\documentclass[traditabstract, longauth]{aa}
\usepackage{graphicx}
\usepackage{subfigure}
\usepackage{txfonts}
\usepackage{xcolor}
\usepackage{longtable}
\usepackage{threeparttable}
\usepackage{natbib,twoopt}
\def\ms{\hbox{\,m\,s$^{-1}$}}         %m.s -1
\def\m2s2{\hbox{\,m$^{2}$\,s$^{-2}$}} %m2.s -2
\def\Msun{\hbox{$\mathrm{M}_{\odot}$}}             %Msun
\def\Rsun{\hbox{$\mathrm{R}_{\odot}$}}
\usepackage{lscape}
\usepackage{pdflscape}

\usepackage{array}

\newcounter{refsysno}
\DeclareRobustCommand{\refsys}[1]{%
   \refstepcounter{refsysno}%
   \therefsysno\label{#1}}

\newcounter{refsysbibno}
\DeclareRobustCommand{\refsysbib}[1]{%
   \refstepcounter{refsysbibno}%
   \therefsysbibno\label{#1}}

\newcounter{reftrno}
\DeclareRobustCommand{\reftr}[1]{%
   \refstepcounter{reftrno}%
   \thereftrno\label{#1}}

\newcounter{reftrbibno}
\DeclareRobustCommand{\reftrbib}[1]{%
   \refstepcounter{reftrbibno}%
   \thereftrbibno\label{#1}}

\begin{document}

\title{Cold Jupiters and improved masses in 38 Kepler and K2 small planet systems from 
3661 HARPS-N radial velocities.}
\subtitle{No excess of cold Jupiters in small planet systems.}
\titlerunning{Cold Jupiters and improved masses in 38 Kepler and K2 small planet systems}
\authorrunning{Bonomo et al.}

\author{A.~S.~Bonomo \inst{1}
\and X.~Dumusque \inst{2} 
\and A.~Massa \inst{3}
\and A.~Mortier \inst{4,5}
\and R.~Bongiolatti \inst{6}
\and L.~Malavolta \inst{7,8}
\and A.~Sozzetti \inst{1}
\and L.~A.~Buchhave \inst{9}
\and M.~Damasso \inst{1}
\and R.~D.~Haywood\inst{10}\fnmsep\thanks{STFC Ernest Rutherford Fellow}
\and A.~Morbidelli \inst{11}
\and D.~W.~Latham\inst{12}
\and E.~Molinari\inst{13}
\and F.~Pepe\inst{2}
\and E.~Poretti\inst{14,15}
\and S.~Udry\inst{2}
\and L.~Affer\inst{16}
\and W.~Boschin\inst{15, 17, 18}
\and D.~Charbonneau\inst{12}
\and R.~Cosentino\inst{15}
\and M.~Cretignier\inst{2}
\and A.~Ghedina\inst{15}
\and E.~Lega\inst{11}
\and M.~L\'{o}pez-Morales\inst{12}
\and M.~Margini\inst{7}
\and A.~F.~Mart\'{i}nez Fiorenzano\inst{15}
\and M.~Mayor\inst{2}
\and G.~Micela\inst{16}
\and M.~Pedani\inst{15}
\and M.~Pinamonti\inst{1}
\and K.~Rice\inst{19,20}
\and D.~Sasselov\inst{12}
\and R.~Tronsgaard\inst{9}
\and A.~Vanderburg\inst{21}
}

\institute{
INAF - Osservatorio Astrofisico di Torino, via Osservatorio 20, 10025 Pino Torinese, Italy  
\and D\'epartement d'astronomie de l'Universit\'e de Gen\`eve, Chemin Pegasi 51, 1290 Versoix, Switzerland
\and Dipartimento di Fisica, Universit\`a degli Studi di Torino, via Pietro Giuria 1, 10125 Torino, Italy
\and School of Physics \& Astronomy, University of Birmingham, Edgbaston, Birmingham, B15 2TT, UK
\and KICC \& Astrophysics Group, Cavendish Laboratory, University of Cambridge, J.J. Thomson Avenue, Cambridge CB3 0HE, UK
\and Dipartimento di Fisica, Universit\`a degli Studi di Milano, Via Celoria 16, I-20133 Milano, Italy
\and Dipartimento di Fisica e Astronomia ``Galileo Galilei'', Universit\`a degli Studi di Padova, Vicolo dell'Osservatorio 3, 35122 Padova, Italy
\and INAF - Osservatorio Astronomico di Padova, vicolo dell'Osservatorio 5, 35122, Padova, Italy
\and DTU Space, National Space Institute, Technical University of Denmark, Elektrovej 328, DK-2800 Kgs. Lyngby, Denmark
\and Astrophysics Group, University of Exeter, Exeter EX4 2QL, UK
\and Universit\'e C\^ote d'Azur, Observatoire de la C\^ote d'Azur, CNRS, Laboratoire Lagrange, Bd de l'Observatoire, CS 34229, 06304 Nice cedex 4, France
\and Center for Astrophysics | Harvard \& Smithsonian, 60 Garden Street, Cambridge, MA 02138, USA
\and INAF - Osservatorio Astronomico di Cagliari, via della Scienza 5, 09047 Selargius, Italy
\and INAF - Osservatorio Astronomico di Brera, Via E. Bianchi 46, 23807 Merate, Italy
\and Fundaci\'on Galileo Galilei - INAF, Rambla Jos\'e Ana Fernandez P\'erez 7, E-38712 Bre\~{n}a Baja, Tenerife, Spain
\and INAF - Osservatorio Astronomico di Palermo, Piazza del Parlamento, 1, 90134 Palermo, Italy
\and Instituto de Astrof\'{\i}sica de Canarias, C/V\'{\i}a L\'actea s/n, E-38205 La Laguna (Tenerife), Canary Islands, Spain
\and Departamento de Astrof\'{\i}sica, Univ. de La Laguna, Av. del Astrof\'{\i}sico Francisco S\'anchez s/n, E-38205 La Laguna (Tenerife), Canary Islands, Spain
\and SUPA, Institute for Astronomy, University of Edinburgh, Blackford Hill, Edinburgh EH9 3HJ, UK
\and Centre for Exoplanet Science, University of Edinburgh, Edinburgh EH9 3FD, UK
\and Department of Physics and Kavli Institute for Astrophysics and Space Research, Massachusetts Institute of Technology, Cambridge, MA 02139, USA
}

\date{Received 21 February 2023 / Accepted 6 April 2023}

\offprints{\email{aldo.bonomo@inaf.it}}

\abstract{
The exoplanet population characterized by relatively short orbital periods ($P<100$~d) around solar-type stars is dominated 
by super-Earths and sub-Neptunes.
However, these planets are missing in our Solar System and the reason behind this absence is still unknown. 
Two theoretical scenarios invoke the role of Jupiter as the possible culprit: Jupiter may have acted as a dynamical barrier to 
the inward migration of sub-Neptunes from beyond the water iceline; alternatively, Jupiter may have considerably reduced 
the inward flux of material (pebbles) required to form super-Earths inside that iceline. 
Both scenarios predict an anti-correlation between the presence of small planets and that of cold Jupiters 
in exoplanetary systems. \\
To test that prediction, we homogeneously analyzed the radial-velocity measurements of 
38 Kepler and K2 transiting small planet systems gathered over nearly ten years with the HARPS-N spectrograph,  
as well as publicly available radial velocities collected with other facilities. 
We used Bayesian differential evolution Markov chain Monte Carlo techniques, which in some cases were coupled with 
Gaussian process regression to model non-stationary variations due to stellar magnetic activity phenomena. 
We detected five cold Jupiters in three systems: two in Kepler-68, two in Kepler-454, and a very eccentric one in K2-312. 
We also found linear trends caused by bound companions in Kepler-93, Kepler-454, and K2-12, with slopes 
that are still compatible with a planetary mass for outer bodies in the Kepler-454 and K2-12 systems. \\
By using binomial statistics and accounting for the survey completeness, we derived an occurrence rate of 
 $9.3^{+7.7}_{-2.9}~\%$ for cold Jupiters with $0.3-13~\rm M_{\rm Jup}$ and $1-10$~AU, which is lower but still compatible at $1.3\sigma$ 
with the value measured from radial-velocity surveys for solar-type stars, regardless of the presence or absence of small planets.
The sample is not large enough to draw a firm conclusion about the predicted anti-correlation between small planets and cold Jupiters; nevertheless, we found no evidence of previous claims of an excess of cold Jupiters in small planet systems. \\
As an important byproduct of our analyses, we homogeneously determined the masses of 64 Kepler and K2 
small planets, reaching a precision better than 5, 7.5, and 10$\sigma$ for 25, 13, and 8 planets, respectively.
Finally, we release  the 3661 HARPS-N radial velocities used in this work to the scientific community. 
These radial-velocity measurements mainly benefit from an improved data reduction software that corrects for  subtle prior systematic effects.}

% 5 {} token are mandatory

\maketitle

\section{Introduction}
\label{introduction}
One of the most striking findings from the detection of almost 4000 transiting planets so far 
is that the most common type of exoplanets in relatively close orbits (orbital periods $P < 100$~d) around solar-type stars are 
small planets (SPs) with radii of $1 < R_{\rm p} < 4~\rm R_\oplus$. These are hosted by 
about half of the solar-type stars in the Milky Way (e.g., \citealt{2015ARA&A..53..409W} and references therein) 
and can be subdivided into two main classes: 
i) high-density super-Earths with $1.0\lesssim R_{\rm p} \lesssim1.7~\rm R_\oplus$ 
and a rocky composition; and 
ii) lower-density sub-Neptunes with $1.7 \lesssim R_{\rm p} \lesssim 4.0~\rm R_\oplus$, 
which are thought to be ice-rich and/or have an atmospheric envelope of hydrogen and helium 
(e.g., \citealt{2019PNAS..116.9723Z}). 
The rocky and ice-rich compositions, if not considerably altered by post-formation processes such as 
core-powered mass loss (e.g., \citealt{2018MNRAS.476..759G, 2019MNRAS.487...24G}) or
atmospheric photo-evaporation for the hottest planets (e.g., \citealt{2014ApJ...792....1L, 2017ApJ...847...29O}), 
would (in principle) reflect different formation locations, 
namely, inside or beyond the water iceline (at $\sim1-3$~AU around a solar-type star), respectively. 
Super-Earths and sub-Neptunes appear to be separated by the so-called radius-valley at $R_{\rm p} \sim 1.7~\rm R_\oplus$ 
in the distribution of planet radii from the \emph{Kepler} space mission \citep{2017AJ....154..109F, 10.1093/mnras/sty1783, 2021ApJ...923..247Z},
even though there may be a certain mixing of the two populations that occurs (e.g., \citealt{2021MNRAS.501.4148L, 2022MNRAS.511.4551L}).

Despite being very abundant overall, super-Earths and sub-Neptunes are absent in our Solar System --
and the reason for that remains an open question. Some theoretical efforts to explain the lack of SPs in the Solar System
have been undertaken, for instance, by \citet{2015ApJ...800L..22I} and \citet{2019A&A...627A..83L}. 
While relying on different frameworks, both works invoke the influence of Jupiter as the possible culprit and 
generalize their outcomes to exoplanetary systems, so as to place the Solar System in the exoplanet context.

The scenario proposed by \citet{2015ApJ...800L..22I} assumes that planet cores form 
preferentially in the proximity of the different icelines (e.g., \citealt{2017A&A...608A..92D}). 
The innermost core is expected to grow faster due to the strong dependence of the accretion timescale 
on orbital radius. If such a core becomes a gas giant, it could then block the migration of the outer cores 
towards the parent star. Occasionally, one core (or more)  could jump over the giant planet and move closer to the star, 
with a ``jumping'' probability depending on the gas-disk profiles and the initial number and total mass of the cores. 
In the simulations carried out by \citet{2015ApJ...800L..22I}, with five cores and a total mass of $\sim 30~\rm M_\oplus$, 
the jumping probability was found to be $\lesssim10-20\%$ (see their Figure 3). 

According to this scenario, the early formation of Jupiter in the Solar System may have prevented the nuclei of Saturn, Uranus, and Neptune 
from migrating towards the Sun and, hence, from becoming a compact system of short-period sub-Neptunes 
such as those observed by the \emph{Kepler}, K2, and TESS space telescopes.

The second framework, described in \citet{2019A&A...627A..83L},  is based on the formation of super-Earths 
inside the water iceline through pebble accretion (e.g., \citealt{2015A&A...578A..36O}). 
With extensive simulations, \citet{2019A&A...627A..83L} 
showed that the outcome in the formation of super-Earths is strongly dependent on the amount of pebble flux 
drifting inwards from the outer regions of the protoplanetary disk. Low pebble fluxes would generate Mars-mass embryos, which 
may then grow to terrestrial planets at orbital distances $a \gtrsim 0.4$~AU through mutual collisions after disk dissipation, 
as has likely occurred in the Solar System (e.g., \citealt{2014prpl.conf..595R}). 
Conversely, higher fluxes of pebbles would produce more massive embryos in shorter time. 
These could migrate towards their parent star if gas was still present in the disk, thereby forming a compact system of close-in super-Earths 
(see their Figure 3). 

When a giant planet forms, it opens a gap in the disk, considerably reducing or even halting the inward flux of pebbles from the regions 
outside its orbit. The formation of Jupiter might thus explain why the Solar System contains no short-period super-Earths, 
but terrestrial planets only: Jupiter may have reduced the flux of material required to form bigger planets within the water iceline.

Both scenarios above predict an anti-correlation between the presence 
of short-period SPs and 
that of cold Jupiters\footnote{We define a cold Jupiter as a planet with mass between 0.3 and 13~$\rm M_{Jup}$ 
and semi-major axis between 1 and 10~AU. Even though some studies have shown that giant planets might have 
masses greater than the deuterium burning limit of 13~$\rm M_{Jup}$ (e.g., \citealt{2014prpl.conf..619C}), 
the above formation models generally refer to cold Jupiters with 
$M_{\rm p} < 13~\rm M_{Jup}$.} (CJs) in exoplanetary systems, which could be tested 
observationally. Previous works by \citet{2018AJ....156...92Z} and \citet{2019AJ....157...52B} seem to contradict this anti-correlation, 
reporting instead an excess of CJs in SP systems. Specifically, by analyzing 65 transiting and 
non-transiting systems with low-mass ($1<M_{\rm p}<10~\rm M_\oplus$) and short-period ($P<100$~d) planets, 
\citet{2019AJ....157...52B} found an occurrence rate of $f_{\rm CJ|SP}=36_{-6}^{+7}~\%$ for gaseous giant planets, with masses of 
$M_{\rm p}=0.5-13~\rm M_{Jup}$ and semi-major axes of $a\sim1-10$~AU. This rate is higher than that found by \citet{2020MNRAS.492..377W}, 
that is, $f_{\rm CJ}=20.2_{-3.4}^{+6.3}\%$\footnote{This value was determined by summing up the occurrence rates 
of cold Jupiters from 300 to 10000~d in Table~3 of \citet{2020MNRAS.492..377W}.}, for $M_{\rm p}=0.3-13~\rm M_{Jup}$ and $a=1-10$~AU, 
from the AAT, HARPS/ESO, HIRES/Keck, and CORALIE radial-velocity (RV) data 
of solar-type stars with time spans longer than eight years, irrespective of the presence or absence of SPs. 

Other works, based on RV surveys only, have attempted to estimate the frequency of CJs in low-mass planet systems and/or 
the frequency of low-mass planets in CJ systems, sometimes with apparently conflicting results. 
For instance, \citet{2018A&A...615A.175B} reported no low-mass planets with $M_{\rm p}=10-30~\rm M_\oplus$ and $P<150$~d 
in 20 CJ systems around solar-type stars observed with HARPS. 
On the contrary, based on the California Legacy Survey conducted with the HIRES/Keck and APF/Lick spectrographs,  
\citet{2022ApJS..262....1R} found that planets with $M_{\rm p} =2-30~\rm M_\oplus$ and $P\lesssim150$~d
may occur approximately twice as frequently around CJ-host stars. 
However, the latter authors observed no significant differences in the occurrence of inner low-mass planets with and without CJ siblings,
when limiting their range in mass to $M_{\rm p}=2-20~\rm M_\oplus$. 

As also noted by \citet{2022ApJS..262....1R}, RV surveys are sensitive to more massive inner planets than transit surveys, 
besides determining minimum masses only. 
Moreover, the adopted ranges in semi-major axes by \citet{2022ApJS..262....1R} for both the inner low-mass planets and the outer CJs 
differ from those in \citet{2019AJ....157...52B}, and they include, for instance, warm Jupiters with $a=0.23-1$~AU. 
This stands in the way of a straightforward comparison of their results with those in \citet{2019AJ....157...52B}.

%\subsection{Aim of the present work}
In the present work, we aim to test the theoretical predictions of \citet{2015ApJ...800L..22I} and \citet{2019A&A...627A..83L} 
by searching for CJs and determining their occurrence rate in 38 transiting systems: 
19 observed by \emph{Kepler} and 19 by K2; 14 of them are in common with the sample studied by \citet{2019AJ....157...52B}.
For this purpose, we used 3661 high-precision HARPS-N radial velocities, 
3471 out of which were collected by the HARPS-N Guaranteed Time of Observations (GTO) consortium, 
and the remaining 190 RVs by other groups, mainly for the purpose of determining the masses and densities of 
\emph{Kepler} and K2 transiting planets. Nonetheless, we monitored these systems over the years 
specifically to look for outer giant planets. An important byproduct of our RV analyses is the improvement in the precision and/or accuracy of planetary masses and densities, 
thanks to the use of a significant number of yet unpublished HARPS-N/GTO RVs as well as the first combination of 
HARPS-N/GTO RVs with literature measurements.

\section{Target selection and radial-velocity data}
\label{target_selection_data}
Among the \emph{Kepler} and K2 systems observed by the HARPS-N/GTO program, we chose those: 
i) hosting small (low-mass) planets with a radius of $1 < R_{\rm p} < 4~\rm R_\oplus$, 
mass of $1<M_{\rm p}< 20~\rm M_\oplus$, and orbital period of $P < 100$~d; and 
ii) having at least 15 HARPS-N RV measurements for a time span longer than $\sim1$~yr. 
This resulted in the selection of the vast majority of \emph{Kepler} and K2 systems monitored by the HARPS-N/GTO program. 
As in the work by \citet{2018AJ....156...92Z}, we adopted a wider range in $M_{\rm p}$ than \citet{2019AJ....157...52B}, 
given that several SPs are known to have $M_{\rm p}> 10~\rm M_\oplus$. 

We also included Kepler-22, even though it meets the second criterion only, 
because the possible presence of CJs may provide valuable information 
on the architecture of a system with a planet in the habitable zone \citep{2012ApJ...745..120B}. 
However, it was not counted in the computation of $f_{\rm CJ|SP}$ in short-period SP systems, 
because Kepler-22b has an orbital period of $289$~d~$>100$~d. 

The HARPS-N radial velocities used in this work 
were extracted with the original HARPS-N Data Reduction Software version 3.7 from the stellar spectra obtained 
before the early failure of the red side of the HARPS-N charge-coupled device (CCD) in late September 2012 (e.g., \citealt{2014A&A...572A...2B}), 
and with the updated DRS version 2.3.5 from the spectra gathered afterwards. 
This latter version of the pipeline, adapted from the ESPRESSO spectrograph to HARPS-N \citep{2021plat.confE.106D},
computes a more stable wavelength solution through a careful selection of the lines of the Thorium-Argon calibration lamp, 
by avoiding saturated Thorium and Argon lines. It also corrects for: i) possible RV long-term variations due to 
changing levels in the flux of the Thorium-Argon calibration lamp with time; and ii) an offset in the DRS v3.7 data, 
occurring at the beginning of June 2020 for the replacement of the Thorium-Argon calibration lamp. 

We observed the majority of the stars in our sample in $\rm OBJ\_AB$ observing mode, 
that is, with fiber A on the target and fiber B on the sky to monitor possible contamination by moonlight. 
For the brightest stars in our sample, namely Kepler-21, Kepler-37, Kepler-68, Kepler-93, Kepler-409, K2-96/HD\,3167, 
K2-167, K2-222, K2-262/Wolf\,503, and K2-312/HD\,80563, we used simultaneous calibration with fiber A on the target and
fiber B on the calibration Thorium-Argon or Fabry-Perot lamp, to achieve 
higher accuracy on the relative RVs. 
We extracted the RVs by cross-correlating the spectra with a stellar template 
close to the stellar spectral type (e.g., \citealt{2002A&A...388..632P}). The only exception is the early-M-late-K-dwarf K2-3, 
for which we used the TERRA software \citep{2012ApJS..200...15A} to overcome the issue of distorted cross-correlation functions (CCFs) 
for cooler stars \citep{2020ExA....49...73R}, thereby achieving a reduced RV scatter \citep{2018A&A...615A..69D}.
For Kepler-10, we performed an additional reduction using the Yarara-v2 tool \citep{2021A&A...653A..43C, 2022A&A...659A..68C}, 
because it proved to slightly enhance the detectability of the planet-induced Doppler signals \citep{Bonomoetalinprep}. 

Possible contamination of the HARPS-N spectra by moonlight was checked following \citet{2017AJ....153..224M}
and corrected by computing the CCF after subtracting the flux of fiber B from the flux of fiber A. 
This procedure led to a reduced RV scatter in a few systems, the most evident cases being Kepler-19, Kepler-107, and K2-110.

Five systems, namely K2-96/HD\,3167, K2-106, K2-111, K2-131, and K2-135/GJ\,9827, were also observed with HARPS-N by other groups. 
To obtain homogeneous HARPS-N datasets across longer time spans and to take advantage of the aforementioned improvements 
of the HARPS-N pipeline, we recomputed all the HARPS-N RVs with the DRS-v2.3.5 from both the spectra acquired by the GTO and 
the other publicly available spectra. 

For each system, we also collected the published RVs gathered with spectrographs other than HARPS-N, 
such as HIRES/Keck, HARPS/ESO, ESPRESSO/VLT, PFS/MagellanII, and APF,  
and analyzed them along with the HARPS-N RVs (see Sect.~\ref{orbital_fitting}). 
This combination is needed to improve the constraints on the presence or lack of CJs, 
determine more precise (and accurate) orbital and physical parameters of the detected CJs, and 
achieve a better precision on the masses and densities of the inner small (low-mass) planets.
A few RV datasets with a limited number of RVs and/or considerably lower precision than 
our HARPS-N RVs were discarded, as they do not yield any improvement in the orbital solution, 
while requiring additional free parameters (the radial-velocity zero point and the uncorrelated jitter term; 
cf. Sect.~\ref{orbital_fitting}). 
We did not use the 71 available HIRES/Keck RVs \citep{2016ApJ...819...83W} 
for the analysis of the Kepler-10 system, because they tend to reduce the RV semi-amplitudes of 
the long-period SPs Kepler-10c and Kepler-10d (Sect.~\ref{nontransiting_planets}) 
compared to those obtained with the 291 HARPS-N RV measurements only \citep{Bonomoetalinprep}, 
even though the simultaneous modeling of both datasets is mainly driven by the much more numerous HARPS-N RVs.
In fact,  \citet{2016ApJ...819...83W} also showed that the signal of Kepler-10c is practically undetected in the HIRES data.

We searched for possible outliers in the RV datasets for each system using the Chauvenet's criterion\footnote{The Chauvenet's criterion states that a value from a set of N measurements can be excluded if its deviation from the mean value is so high that the normal distribution probability that one of the N measurements with an equal or greater deviation may occur is less than 1/2N (see, e.g., \citealt{doi:10.1137/1119078}).}
and removed them. For systems with long-period RV signals, such as 
long-term slopes and/or Keplerians of CJs, 
we applied the Chauvenet's criterion after removing those signals.
Through visual inspection we checked that this criterion efficiently removes all the clear outliers. 

Table~\ref{table_system_parameters} lists the 38 systems in our sample, 
the multiplicity (single planet or multiple planets) of the transiting SPs, the stellar parameters, 
the number of both total RVs and HARPS-N RVs used in this work after the removal of outliers, 
and the total time span of the observations. 
For each target, Table~\ref{table_RV_data} reports the epochs, values, and formal uncertainties of the HARPS-N RVs, 
the activity indicators of the CCF, namely, the full width at half maximum (FWHM), the contrast and bisector span of the CCF, as well as 
the spectroscopic activity indexes $S$ index and $\log{R^{'}_{\rm HK}}$. 
We warn that correlated variations of the FWHM and contrast 
may have occurred at certain epochs due to changes in the focus of HARPS-N, 
but do not affect the RVs because the product $\rm [FWHM \cdot contrast]$ remains constant.

\section{Data analysis}

\subsection{Updated atmospheric and physical stellar parameters}
To maintain uniformity with previous studies of the systems published by the HARPS-N/GTO consortium, 
we derived the atmospheric parameters, that is, effective temperature $T_{\rm eff}$, 
metallicity [Fe/H], and surface gravity $\log{g}$,  
for the host stars Kepler-22, Kepler-109, Kepler-323, Kepler-409, Kepler-1876, 
K2-12, K2-38, K2-106, K2-131, and K2-167 from the HARPS-N spectra. 
To this end, we employed two independent methods, 
ARES+MOOG (e.g., \citealt{2014A&A...572A..95M}) and SPC \citep{2012Natur.486..375B, 2014Natur.509..593B}, 
and adopted as final parameters and uncertainties the average values and error bars 
provided by the two methods (see \citealt{2018MNRAS.481.1839M} for more details). 

To determine the stellar physical parameters, we used the EXOFASTv2 tool
\citep{2017ascl.soft10003E, 2019arXiv190709480E}, which adjusts the stellar radius, mass, and age 
through a Bayesian differential evolution Markov chain Monte Carlo (DE-MCMC) method \citep{TerBraak2006},  
by simultaneously fitting the stellar spectral energy distribution (SED) 
and employing the MESA Isochrones and Stellar Tracks (MIST) \citep{2015ApJS..220...15P}. 
To sample the stellar SED, we utilized the WISE W1, W2, W3, and W4 infrared magnitudes \citep{2014yCat.2328....0C}, 
the 2-MASS near-infrared J, H, and K magnitudes \citep{2003tmc..book.....C}, 
the optical Tycho $B_{\rm T}$ and $V_{\rm T}$ magnitudes \citep{2000A&A...355L..27H}, 
and/or the APASS Johnson B, V and Sloan $g^{'}$, $r^{'}$, $i^{'}$ magnitudes \citep{2016yCat.2336....0H}. 
We imposed Gaussian priors on the $T_{\rm eff}$ and [Fe/H] atmospheric parameters, 
as derived from the analyses of the HARPS-N spectra, 
as well as on the Gaia EDR3 parallax \citep{gaiaedr3}.
A uniform prior was instead used for the V-band extinction, $A_{\rm V}$, with upper limits provided 
by reddening maps \citep{1998ApJ...500..525S, 2011ApJ...737..103S}.

For the remaining 27 systems, we adopted the previously published stellar parameters, giving preference to 
those derived using the Gaia parallaxes and/or asteroseismic analyses of the \emph{Kepler} light curves 
(e.g., Kepler-10, Kepler-454, and Kepler-107). 
In one case (Kepler-20), we redetermined the stellar parameters because the prior on the Gaia parallax 
yields slightly more precise and accurate stellar radius and mass.
The atmospheric and physical parameters of all the host stars in our sample are listed in Table~\ref{table_system_parameters}.

\subsection{Orbital fitting}
\label{orbital_fitting}
We modeled the RV data of all the 38 systems in our sample with non-interacting Keplerian orbits and 
a slope by maximizing a Gaussian likelihood function (e.g., \citealt{2006ApJ...642..505F}) 
through a DE-MCMC technique.  
The parameters of the Keplerian model for each planet in a given system 
are the inferior conjunction time, $T_{\rm c}$, which is equivalent to the transit midtime for 
transiting planets; the orbital period, $P$; the widely adopted parameterization $\sqrt{e}\cos(\omega)$ and $\sqrt{e}\sin(\omega)$ 
of eccentricity, $e,$ and argument of periastron, $\omega$; and the RV semi-amplitude, $K$. 
We included linear slopes to check for significant long-term trends and fit for 
the RV zero point, $\gamma_{\rm i}$, and jitter term, $\sigma_{\rm jit, i}$, for the i-th RV dataset 
gathered with the i-th spectrograph. 
The jitter terms, $\sigma_{\rm jit}$, were summed in quadrature to the formal RV uncertainties, $\sigma_{\rm RV}$, 
to account for additional white noise of unknown (stellar or instrumental) origin (e.g., \citealt{2005ApJ...631.1198G}).

Correlated noise in the RV time series caused by stellar magnetic activity was modeled 
through Gaussian process (GP) regression (e.g., \citealt{2014MNRAS.443.2517H, 2015PhDT.......193H, 2015ApJ...808..127G}) 
within the same DE-MCMC tools, using a covariance matrix described by either 
the quasi-periodic (QP) kernel in the original form of \citet{2006gpml.book.....R}: 

\begin{equation} 
\label{equation_kernel_QP}
k(t, t') = h^{2} \cdot \exp{\left[ - \frac{(t-t')^2}{2\lambda_{1}^2} - \frac{2 \sin^2{\left(\frac{\pi (t-t')}{P_{\rm rot}} \right)}}{\lambda_{2}^2} \right]} 
+ \left[\sigma_{\rm RV}^2(t) + \sigma_{\rm jit}^2 \right] \cdot \delta_{\rm t, t'}
,\end{equation}

\noindent
or the simpler squared-exponential (SE) kernel

\begin{equation} 
\label{equation_kernel_SE}
k(t, t') = h^{2} \cdot \exp{\left[ - \frac{(t-t')^2}{2\lambda_{1}^2}\right]} 
+ \left[\sigma_{\rm RV}^2(t) + \sigma_{\rm jit}^2 \right] \cdot \delta_{\rm t, t'}, 
\end{equation}

\noindent
where $h$ is the semi-amplitude of the correlated noise, $\lambda_1$ is the correlation decay timescale, $P_{\rm rot}$ is the 
period of the quasi-periodic variations, and $\lambda_2$ is the inverse complexity harmonic parameter. 
The hyper-parameters $\lambda_1$,  $P_{\rm rot}$, and $\lambda_2$ can be associated respectively with 
the decay timescale of the active regions, the stellar rotation period and the complexity of the activity signals 
(with $\lambda_2 \sim 3-5$ approaching simpler sinusoidal signals).
We point out that the parameter $\lambda_2$ in Eq.~\ref{equation_kernel_QP} is twice the equivalent parameter $w$ 
in other implementations of the quasi-periodic kernel (e.g., \citealt{2015ApJ...808..127G, 2018A&A...615A..69D}). 

The QP kernel was used to model correlated noise with a quasi-periodic behavior, namely, in the presence of stellar rotation signals. 
The latter were identified when a periodicity in the Generalized Lomb-Scargle (GLS) periodogram of the RVs \citep{2009A&A...496..577Z} 
with a false alarm probability $FAP < 10^{-3}$ was also found in the \emph{Kepler}/K2 light curves 
and/or in the CCF or \emph{S}-index activity indicators. This concerns Kepler-21, Kepler-78, Kepler-102, K2-3, 
K2-36, K2-131, K2-135/GJ\,9827, K2-141, and K2-312/HD\,80653.

The SE kernel was employed for Kepler-93 and K2-2/HIP\,116454 
to account for RV variations on timescales longer than the stellar rotation, 
which are likely due to magnetic activity cycles and/or shorter Rieger-type cycles \citep{1984Natur.312..623R}. 
The DE-MCMC analysis with the SE kernel proved to model such variations efficiently, 
producing flat and Gaussian-distributed residuals. The analysis of the same data with the QP kernel 
yielded very low acceptance rates, which indicates that the GP-QP model is more complex than needed to model long-term variations
in both systems, leaving the QP hyper-parameters $P_{\rm rot}$ and $\lambda_2$ practically unconstrained.

We imposed several priors on the model parameters as well as on the GP hyper-parameters 
in case GP regression was used (i.e., in the presence of correlated noise). 
Specifically, we used Gaussian priors on the transit time, $T_{\rm c}$, and period, $P,$ of the inner transiting SPs, 
as provided by the transit ephemerides derived in previous analyses of the \emph{Kepler} and K2 light curves
(see the second and third columns, and references in Table~\ref{table_planet_parameters}). 
With regard to orbital eccentricities, we adopted: i) circular orbits for the closest SPs, whose orbital circularization times are 
considerably shorter than the stellar age (e.g., \citealt{2008ApJ...686L..29M, 2010ApJ...725.1995M}). 
Null eccentricities for these planets are also consistent with the observation of their secondary eclipses 
at orbital phases $\Delta\phi \sim 0.5$ from transits \citep{2022A&A...658A.132S}; 
ii) half-Gaussian priors with zero mean and $\sigma_{\rm e}=0.098$ for the transiting SPs in multiple systems, 
following the finding of \citet{2019AJ....157...61V}. This prior prevents the fit from converging to spurious eccentricities, 
a well-known critical effect occurring for low signal-to-noise Doppler signals (e.g., \citealt{2011MNRAS.410.1895Z, 2019MNRAS.489..738H}), 
given that the typical RV semi-amplitudes of the transiting SPs in our sample are usually comparable to the RV scatter. 
Spurious high eccentricities may also be unphysical, as they would lead to dynamical instabilities
(e.g., \citealt{2013MNRAS.436.3547G});
iii) uniform priors $U[0, 1[$ for the single-transiting SPs, except for Kepler-22b, Kepler-454b, 
Kepler-409b, and Kepler-1876b, for which we used half-Gaussian priors with zero mean and 
$\sigma_{\rm e}=0.45, 0.35, 0.42, 0.37$, respectively, as derived from asteroseismic and transit light-curve analyses
\citep{2019AJ....157...61V}; and iv) uniform priors for the long-period, non-transiting CJs.
We adopted uninformative priors on the RV semi-amplitudes, $K$, zero points, $\gamma$, 
jitter terms, $\sigma_{\rm jit}$, and slopes, $\dot\gamma$.

%\begin{figure}[t!]
%\centering
%\begin{center}

%\vspace{-0.30cm} 

%\includegraphics[width=7.0cm]{completeness_map_kep_1876.eps}

%\vspace{-0.45cm} 

%\includegraphics[width=7.0cm]{completeness_map_kep_93.eps}

%\vspace{-0.45cm} 

%\includegraphics[width=7.0cm]{average_completeness.eps}

%\vspace{-0.30cm}

%\caption{Recovery rate (completeness) maps of cold Jupiters. 
%\emph{Top panel}: Completeness of Kepler-1876, one of the systems 
%with the worst completeness mainly due to the relatively short temporal baseline of the HARPS-N radial-velocity observations. 
%Black and white indicate 0\% and 100\% recovery rates, respectively. 
%\emph{Middle panel}: Completeness of Kepler-93, showing a very high detection rate, given the long time span of observations.
%\emph{Bottom panel}: Mean survey completeness obtained by averaging out the individual completenesses of the 37 systems.
%}
%\end{center}
%\label{figure_completenesses_Kepler-1876_Kepler-93_average}
%\end{figure}

\begin{figure}[t!]
\vspace{-0.35cm} 
\begin{center}
\includegraphics[width=7.0cm]{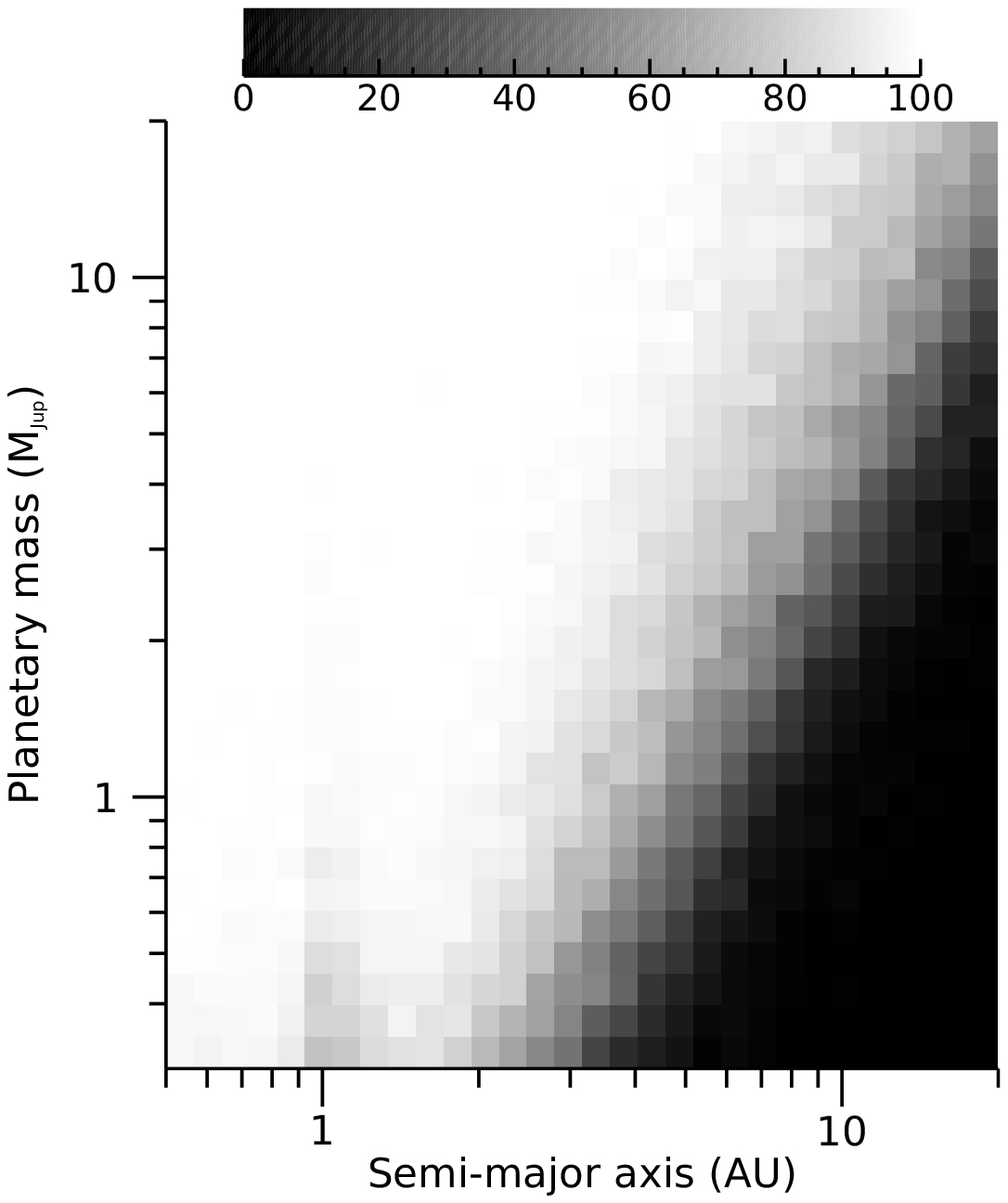}
\end{center}
\vspace{-1.10cm} 
\begin{center}
\includegraphics[width=7.0cm]{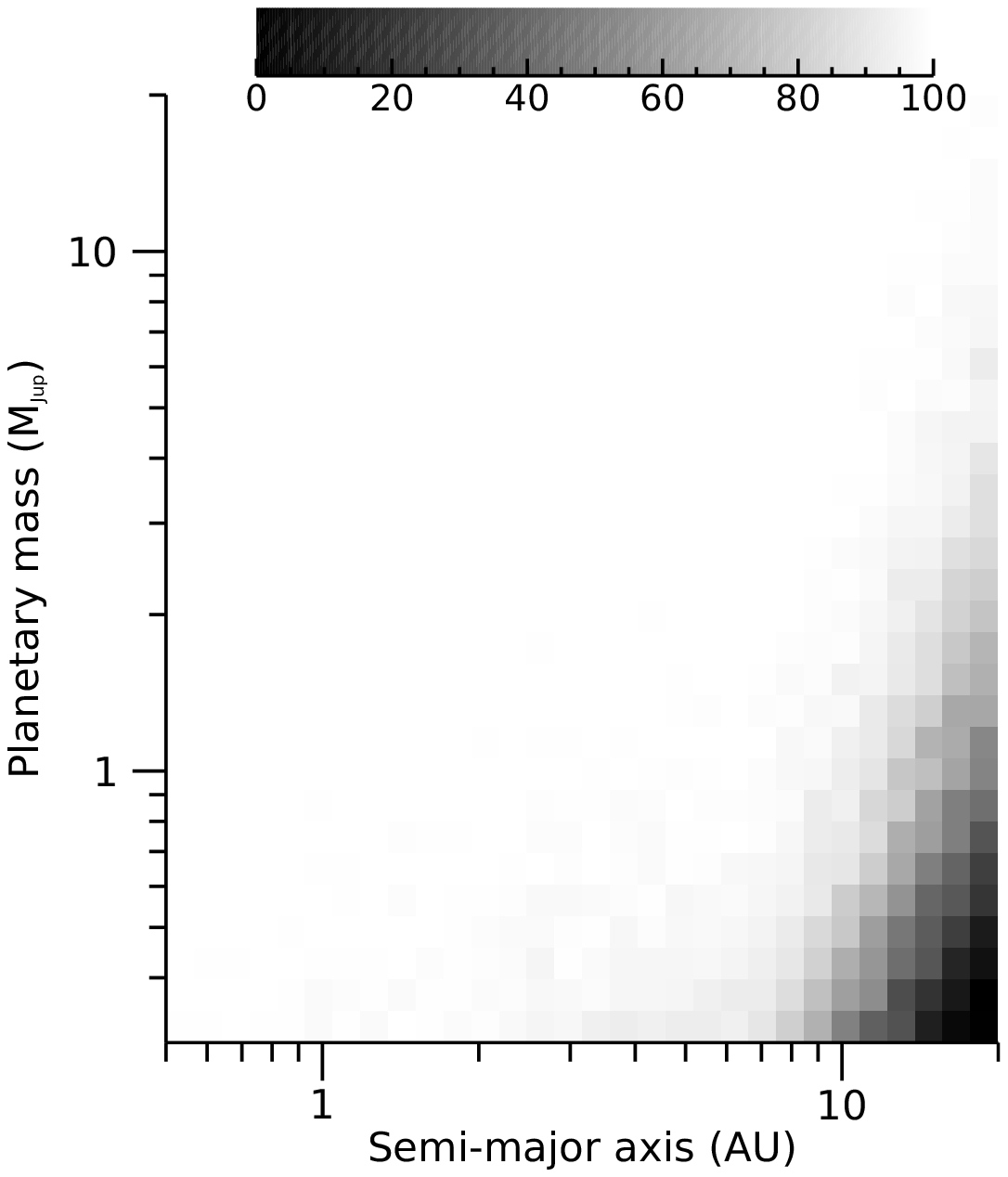}
\end{center}
\vspace{-1.00cm} 
\begin{center}
\includegraphics[width=7.0cm]{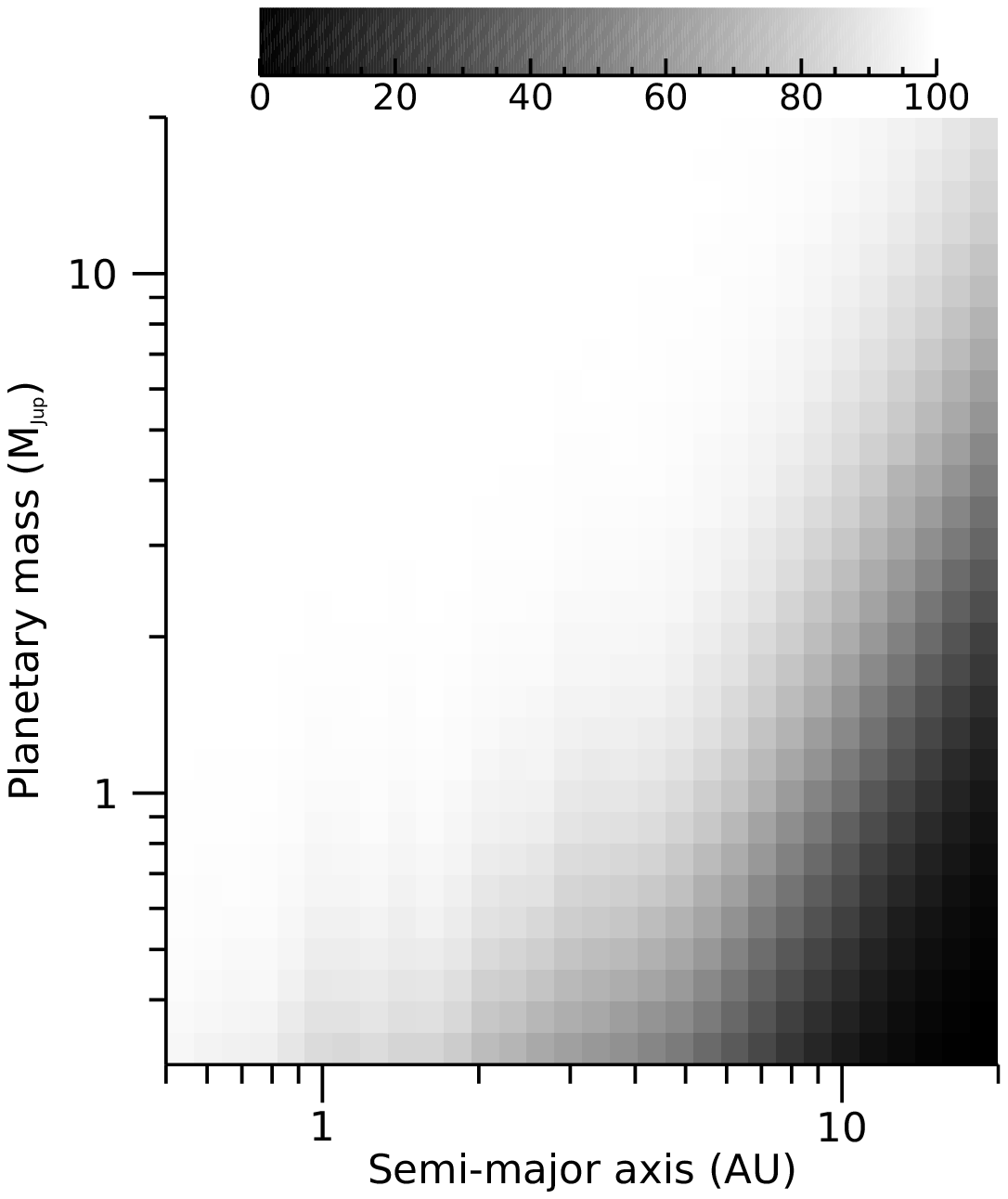}
\end{center}
\vspace{-0.75cm}
\caption{Recovery rate (completeness) maps of cold Jupiters. 
\emph{Top panel}: Completeness of Kepler-1876, one of the systems 
with the worst completeness mainly due to the relatively short temporal baseline of the HARPS-N radial-velocity observations. 
Black and white indicate 0\% and 100\% recovery rates, respectively. 
\emph{Middle panel}: Completeness of Kepler-93, showing a very high detection rate, given the long time span of observations.
\emph{Bottom panel}: Mean survey completeness obtained by averaging out the individual completenesses of the 37 systems.
}
\label{figure_completenesses_Kepler-1876_Kepler-93_average}
\end{figure}

As for the GP hyper-parameters, we imposed uniform priors with bounds wide enough to encompass the expected 
values of $\lambda_1$, $P_{\rm rot}$, and $\lambda_2$, and only a lower bound of 0~$\rm m\,s^{-1}$ for $h$. 
Table~\ref{table_priors_GP_hyperparameters} lists the adopted priors on the GP hyper-parameters for 
the systems that required GP regression to model the non-stationary activity variations along with the 
Keplerian signals. For three systems, namely Kepler-102, K2-141, and K2-132, we further imposed 
$\lambda_{1} > P_{\rm rot}/2$, otherwise $\lambda_{1}$ converged to very low values, on the order of 
a couple of days, making the quasi-periodic term practically irrelevant 
and, hence, unconstrained. Nonetheless, the RV semi-amplitudes  obtained with and without this prior 
(i.e., by also allowing $\lambda_1$ to converge towards very low values) are fully consistent.

Additional signals attributed to non-transiting planets were searched for in the RV residuals with GLS periodograms  
and they were included in the DE-MCMC RV analysis if i) their $FAP < 10^{-3}$; 
ii) their periodicity does not appear in the GLS periodograms of any activity indicator; 
and iii) the difference in the Bayesian information criterion ($BIC$) 
\citep{BurnhamAnderson2004, 2007MNRAS.377L..74L} between 
the model with the additional planet and the model without it is $\Delta BIC>10$ \citep{KassRaftery1995}.
Similarly, possible RV long-term slopes were considered significant if they were detected at more than $3\sigma$ and 
the $\Delta BIC$ in favor of the model with the slope is greater than 10.

We determined the values and the $1\sigma$ uncertainties of the model and derived parameters
from the medians and the 15.87\%-84.14\% quantiles of their posterior distributions. 
For distributions consistent with zero, such as those of eccentricities or RV semi-amplitudes in cases of non-detection 
of the Doppler signals, we provided only the $1\sigma$ upper limits defined as the 0\%-68.27\% quantiles. 
In Table~\ref{table_accelerations} we report for each system the HARPS-N systemic velocities and jitter terms, 
the GP hyper-parameters in the presence of stellar activity signals, and the linear accelerations. In Tables~\ref{table_giant_planets} and \ref{table_nontransiting_planets}, we give 
the parameters of the non-transiting CJs and low-mass planets, respectively. In Table~\ref{table_planet_parameters}, we give the parameters of the 64 Kepler and K2 transiting SPs in our sample
(see Sect.~\ref{results}).

\subsection{Survey completeness}
\label{survey_completeness}
To determine the frequency of CJs in our sample of \emph{Kepler} and K2 systems,
we first need to evaluate the sensitivity of our survey to the presence of such planets. Indeed, 
CJs might not have been detected in some systems because of limited temporal baselines, poor 
temporal sampling, and/or relatively low precision of the RV measurements. 
Therefore, our measure of the occurrence rate of CJs has to take the completeness (or recovery rate) 
of our survey into account. 

The completeness can be estimated with experiments of injection and recovery of planetary signals 
for each system, by considering the real times of the observations and the RV uncertainty of each measurement at time $t$, 
that is, $\sigma(t)=\sqrt{\sigma_{\rm RV}^2(t) + \sigma_{\rm jit}^2}$. Following \citet{2019AJ....157...52B}, 
we simulated signals of CJs in a logarithmic grid of 30x30 cells of planetary mass, $\Delta M_{\rm p}$, versus 
semi-major axis, $\Delta a$, covering the ranges of $0.3-20~\rm M_{Jup}$ in $M_{\rm p}$ and $0.5-20$~AU in $a$. 
For each cell of a given system, we simulated 300 RV signals of CJs at the epochs of our RV observations by 
randomly varying 
i) $M_{\rm p}$ and $a$ within the cell bounds;
ii) $T_{\rm c}$ within the orbital period corresponding to $a$ and the stellar mass $M_\star$ 
(Table~\ref{table_system_parameters}) from Kepler's third law;   
iii) $\cos{i}$ from 0 to 1, where $i$ is the orbital inclination; and
iv) the argument of periastron, $\omega,$ from 0 to $2\pi$, while 
drawing the orbital eccentricity, $e,$ from a beta distribution \citep{2013MNRAS.434L..51K}. 
We then shifted every RV point at time $t$ according to a Gaussian distribution with a mean equal to the RV value 
and a standard deviation $\sigma(t)$. For simplicity, we did not simulate 
stellar magnetic activity signals, assuming that those signals would be efficiently modeled with GP regression, 
as shown in Sect.~\ref{results}.

To establish the recovery rate in every $\Delta M_{\rm p}$-$\Delta a$ cell, we fit the injected signals with 
a slope, a quadratic trend and a Keplerian orbit (with input parameters close to the simulated ones) 
and compared these models with a constant model (i.e., no 
signal) through the $\Delta BIC$ criterion: if $\Delta BIC > 10$ in favor of the model with the planet-induced signal, 
we recorded a detection of the simulated signal, otherwise its non-detection.
Figure~\ref{figure_completenesses_Kepler-1876_Kepler-93_average} shows the completeness of two systems, 
Kepler-1876 (top panel), and Kepler-93 (middle panel) as one of the worst and best cases in our sample, respectively, 
and the average completeness of the 37 systems (bottom panel).

\section{Results}
\label{results}

\subsection{Search for and characterization of cold Jupiters}
We first present the results of the search for CJs in our survey.

\subsubsection{Systems with no long-term trends}
\begin{figure}[t!]
%\centering
\vspace{-0.10cm}
\begin{center}
%\vspace{+0.50cm}
\includegraphics[width=6.0cm, angle=90]{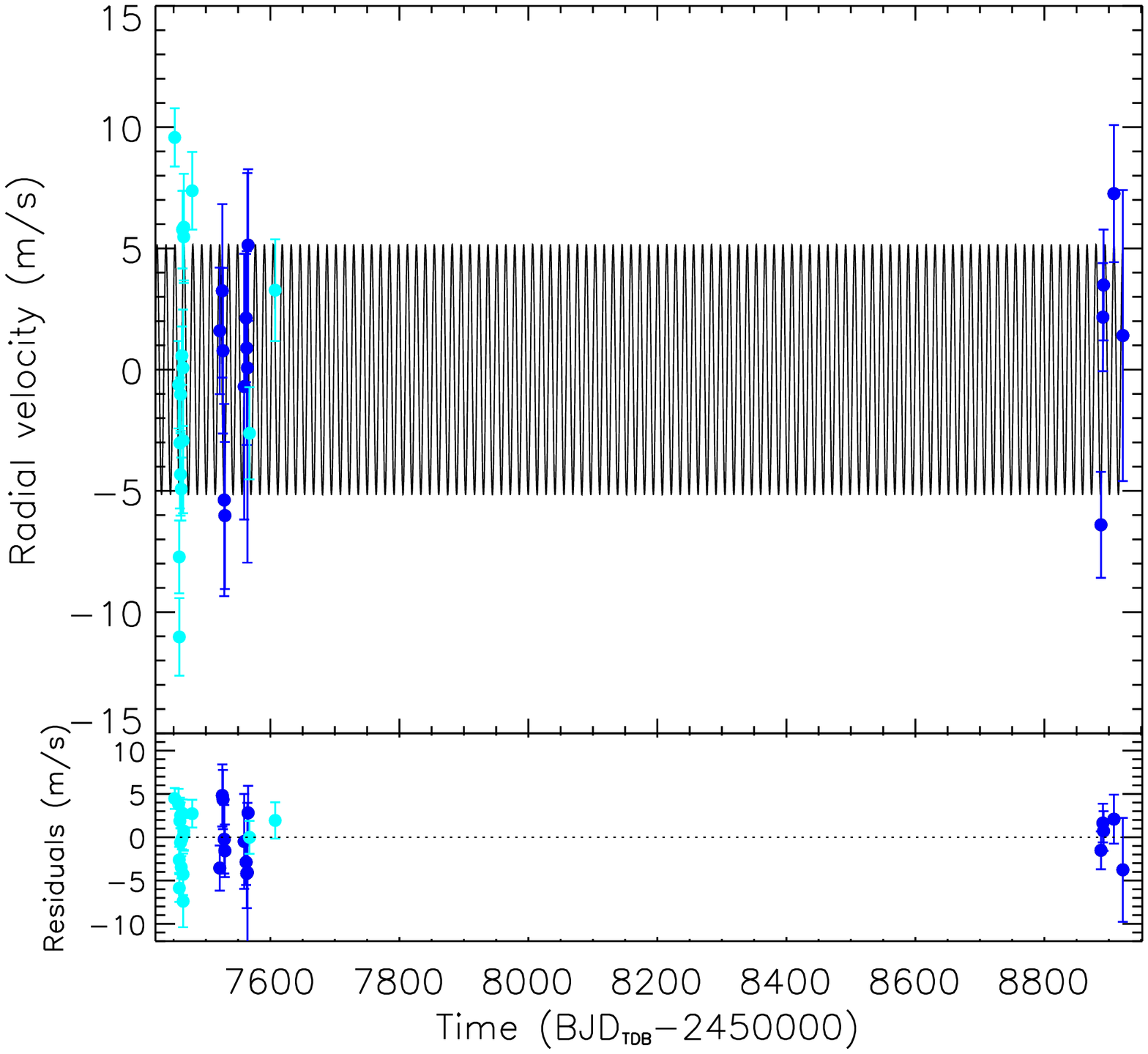} 
\end{center}
%\vspace{+0.30cm}
\begin{center}
\includegraphics[width=6.0cm, angle=90]{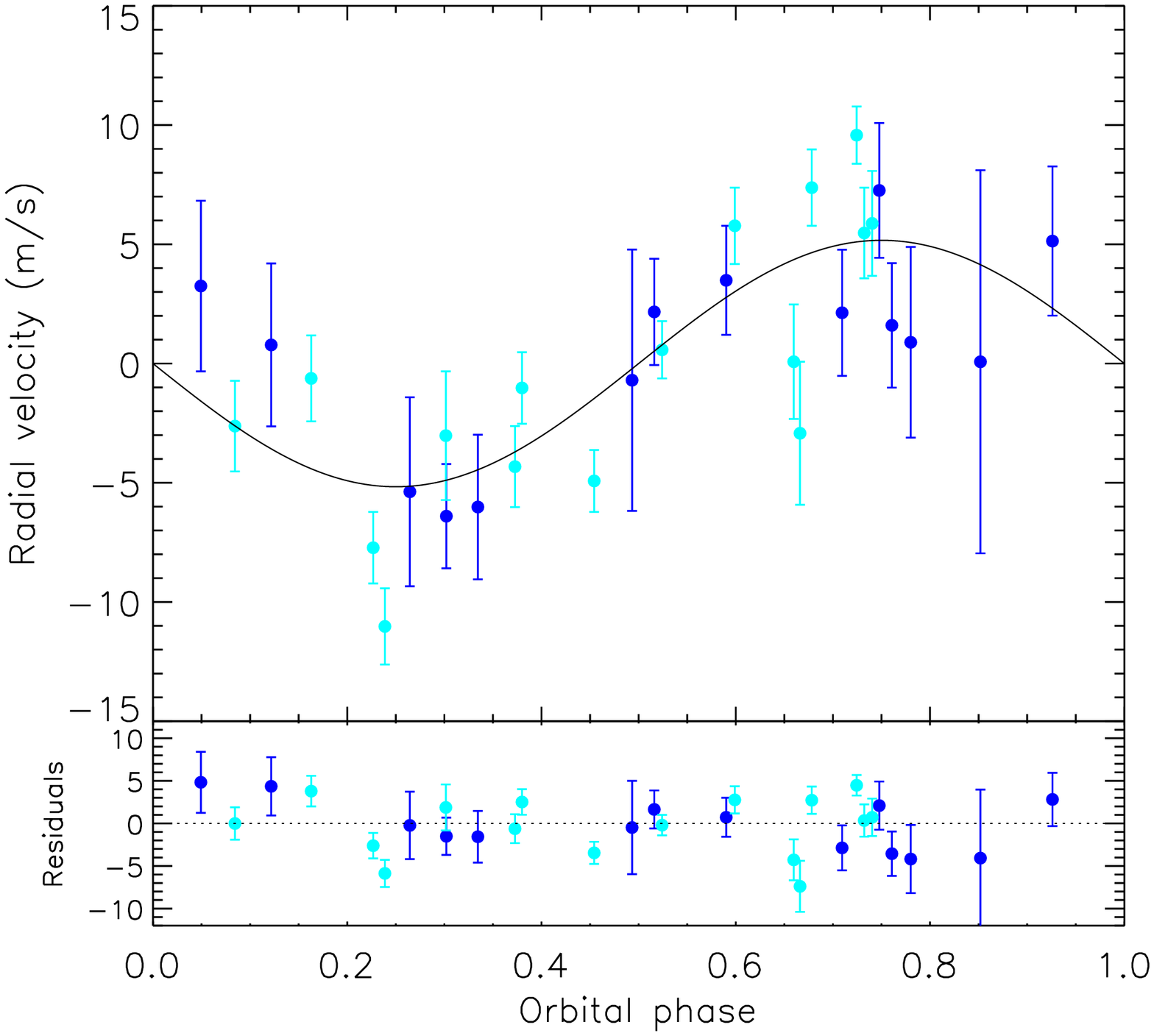} 
\end{center}
%\vspace{+0.10cm}
\caption{Radial-velocity measurements of K2-110. 
\emph{Top panel}: Radial velocities as a function of time, showing no long-term trends. The blue and light blue points display
the measurements obtained with the HARPS-N and HARPS spectrographs, respectively, and the black line indicates the 
Keplerian best-fit model. 
\emph{Bottom panel}: Same radial velocities as in the top panel, but phase folded at the ephemeris of K2-110b ($P=13.86$~d).
}
%\end{center}
\label{figure_RV_K2-110}
\end{figure}

\begin{figure}[h!]
%\vspace{-0.20cm}
\begin{center}
\includegraphics[width=6.0cm, angle=90]{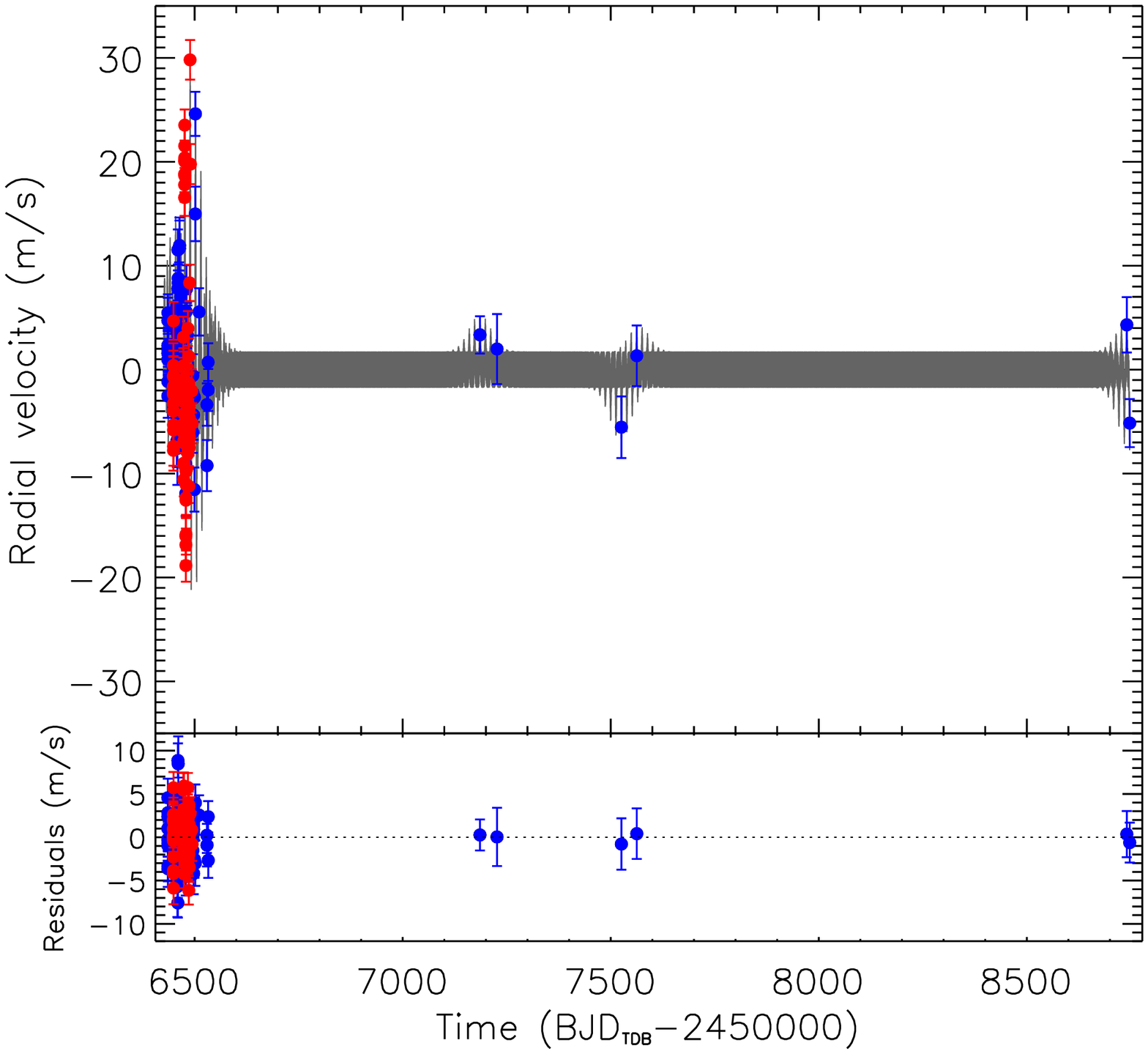}
\end{center}
%\vspace{0.50cm}
\begin{center}
\includegraphics[width=6.0cm, angle=90]{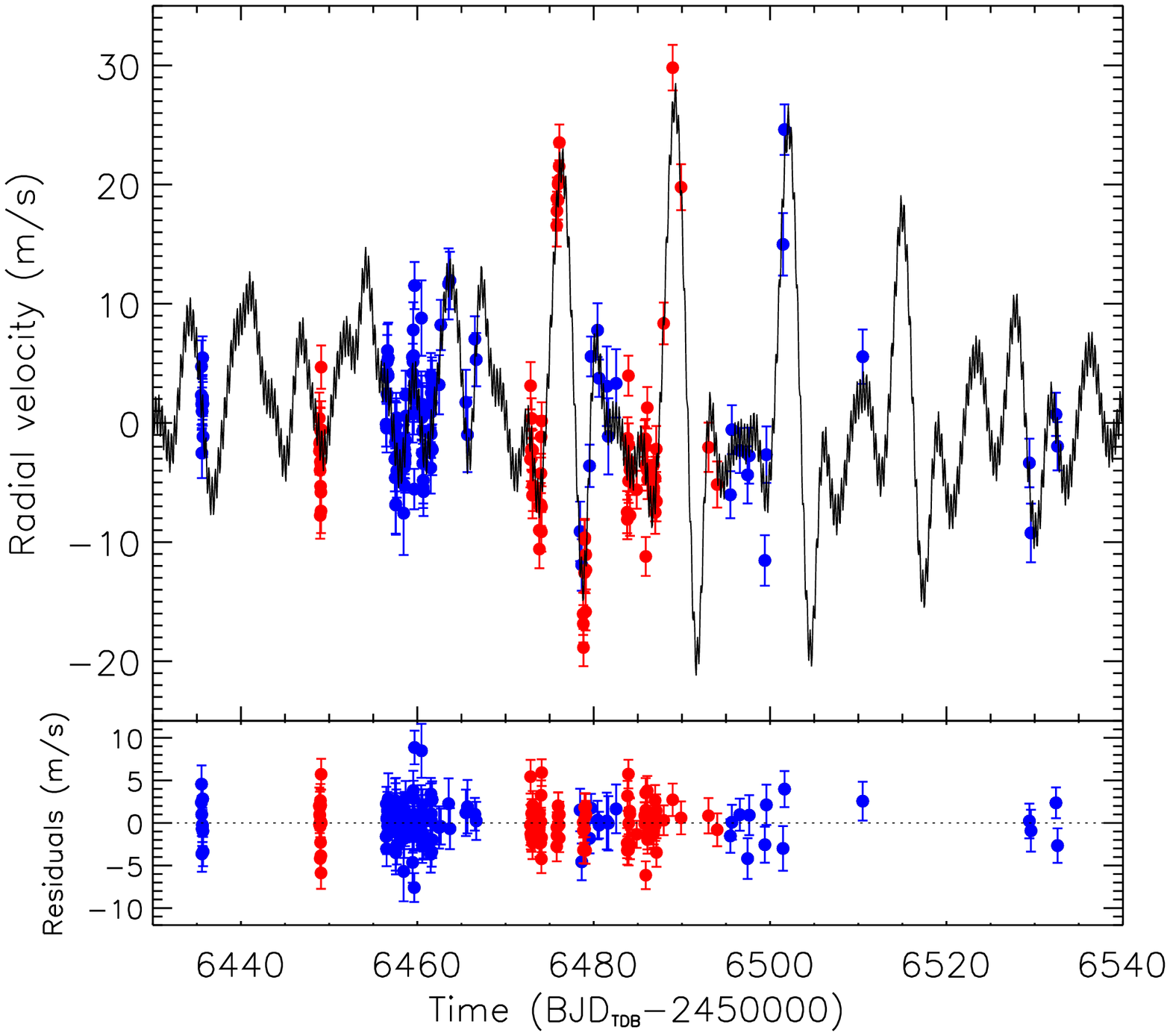}
\end{center}
\begin{center}
\includegraphics[width=6.0cm, angle=90]{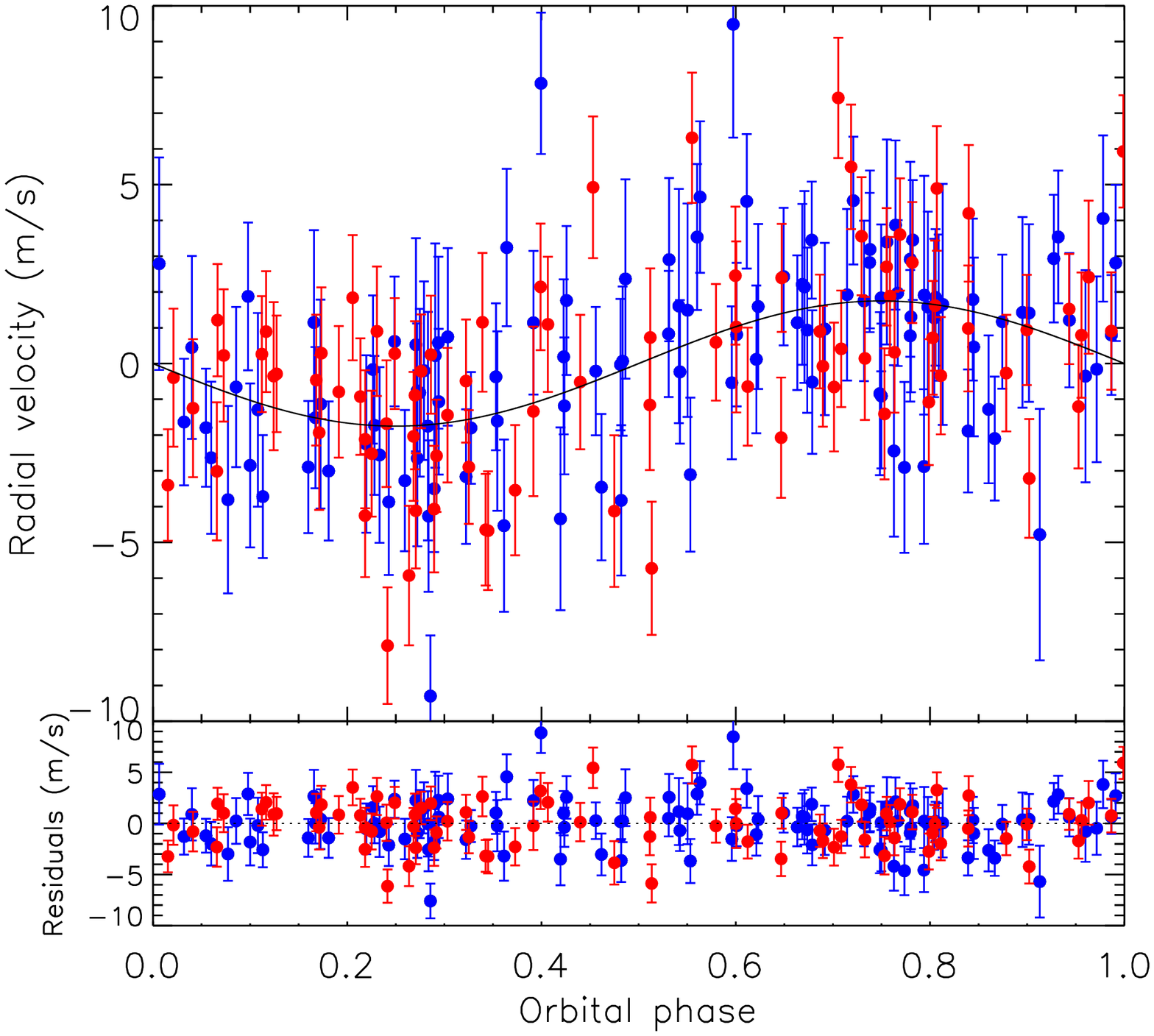}
\end{center}
%\vspace{0.20cm}
\caption{Radial-velocity measurements of Kepler-78. 
\emph{Top panel}: Radial velocities as a function of time, showing variations due to both stellar activity and the ultra-short-period planet Kepler-78b, 
but no significant long-term trends. The blue and red circles indicate the radial-velocity points gathered with the HARPS-N and
HIRES spectrographs, respectively.
\emph{Middle panel}: Zoom on the first 100 days, mainly showing the non-stationary 
stellar activity signal, which was modeled with Gaussian process regression and a quasi-periodic kernel (black solid line).
\emph{Bottom panel}: Radial velocities phase-folded at the ephemeris of Kepler-78b ($P=0.35$~d), 
after removing the Gaussian process model for the stellar activity variations.
}
\label{figure_RV_Kepler-78}
\end{figure}

Altogether 31 out of 38 systems do not show any significant long-term trend as caused by sufficiently massive outer companions 
within the completeness limits (Table~\ref{table_accelerations}). 
The RVs of two of these systems, K2-110 \citep{2017A&A...604A..19O} and Kepler-78 \citep{2013Natur.503..377P, 2013Natur.503..381H}, 
are shown in Figures~\ref{figure_RV_K2-110} and \ref{figure_RV_Kepler-78} as representative cases. 
Possible extra noise in the K2-110 system was taken into account through the white noise term, $\sigma_{\rm jit}$, only, because 
the host star is not magnetically active. On the contrary, GP regression with a quasi-periodic kernel was needed to model 
the strongly correlated variations of the active star Kepler-78 (Fig.~\ref{figure_RV_Kepler-78}; see also \citealt{2015ApJ...808..127G}).

%\begin{figure*}[t!]
%\centering
%\includegraphics[width=6.0cm, angle=90]{plot_time_K2-110b.ps}
%\includegraphics[width=6.0cm, angle=90]{plot_phase_K2-110b.ps}
%\vspace{0.20cm}
%\caption{Radial-velocity measurements of K2-110. 
%\emph{Left panel}: Radial velocities as a function of time, showing no long-term trends. The blue and light blue points display
%the measurements obtained with the HARPS-N and HARPS spectrographs, respectively, and the black line indicates the 
%Keplerian best-fit model. 
%\emph{Right panel}: Same radial velocities as in the left panel, but phase folded at the ephemeris of K2-110b ($P=13.86$~d).
%}
%\label{figure_RV_K2-110}
%\end{figure*}

\subsubsection{Systems with long-term linear trends}
\begin{figure*}[h!]
%\centering
\begin{subfigure}
%\vspace{-0.20 cm}
\centering
\hspace{1cm}
\includegraphics[width=6.0cm, angle=90]{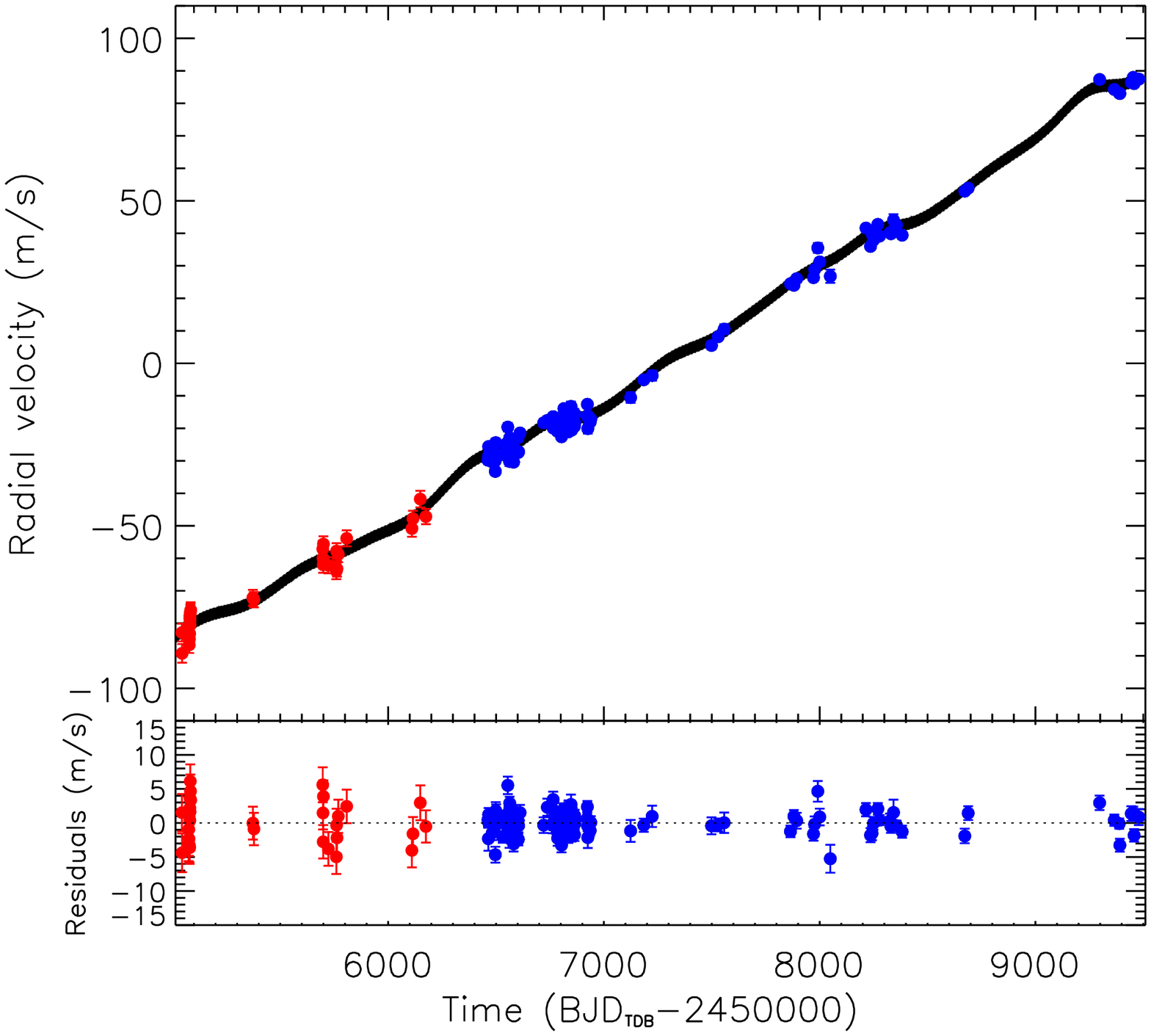}
\includegraphics[width=6.0cm, angle=90]{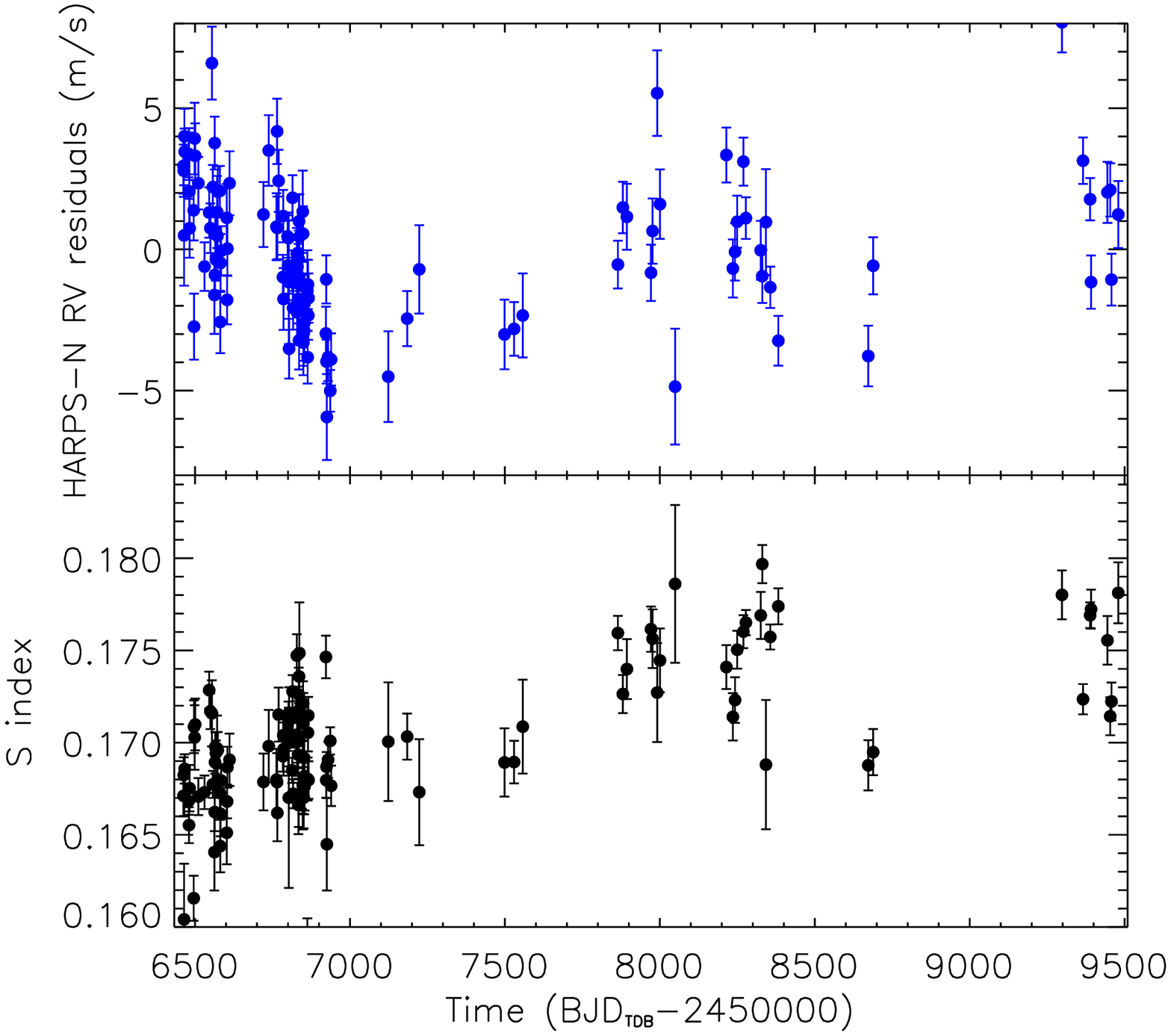}
\vspace{0.10cm}
\caption{Radial-velocity and CaII \emph{S}-index measurements of Kepler-93.
\emph{Left panel:} HIRES (red circles) and HARPS-N (blue circles) radial velocities. The black line shows  
the best-fit model with a long-term linear slope caused by a non-planetary companion, 
a Keplerian orbit for the transiting planet Kepler-93b ($P=4.73$~d), and Gaussian process regression with a squared exponential kernel 
to account for long-term stellar activity variations.
\emph{Right panel}: Residuals of the HARPS-N radial velocities after subtracting the linear long-term trend 
and the signal of Kepler-93b (top), and measurements of the stellar activity $S$ index (bottom). The increase in both the RVs and 
the $S$ index from $\sim7000$ to $\sim8300$~$\rm BJD_{TDB}-2450000$ indicates that the small-amplitude long-term variations 
superposed to the linear slope seen in the left panel are likely due to stellar magnetic activity.
}
\label{figure_RV_Kepler-93}
\end{subfigure}
%\end{figure*}

%\begin{figure*}[h!]
\begin{subfigure}
\centering
\hspace{1cm}
\includegraphics[width=6.0cm, angle=90]{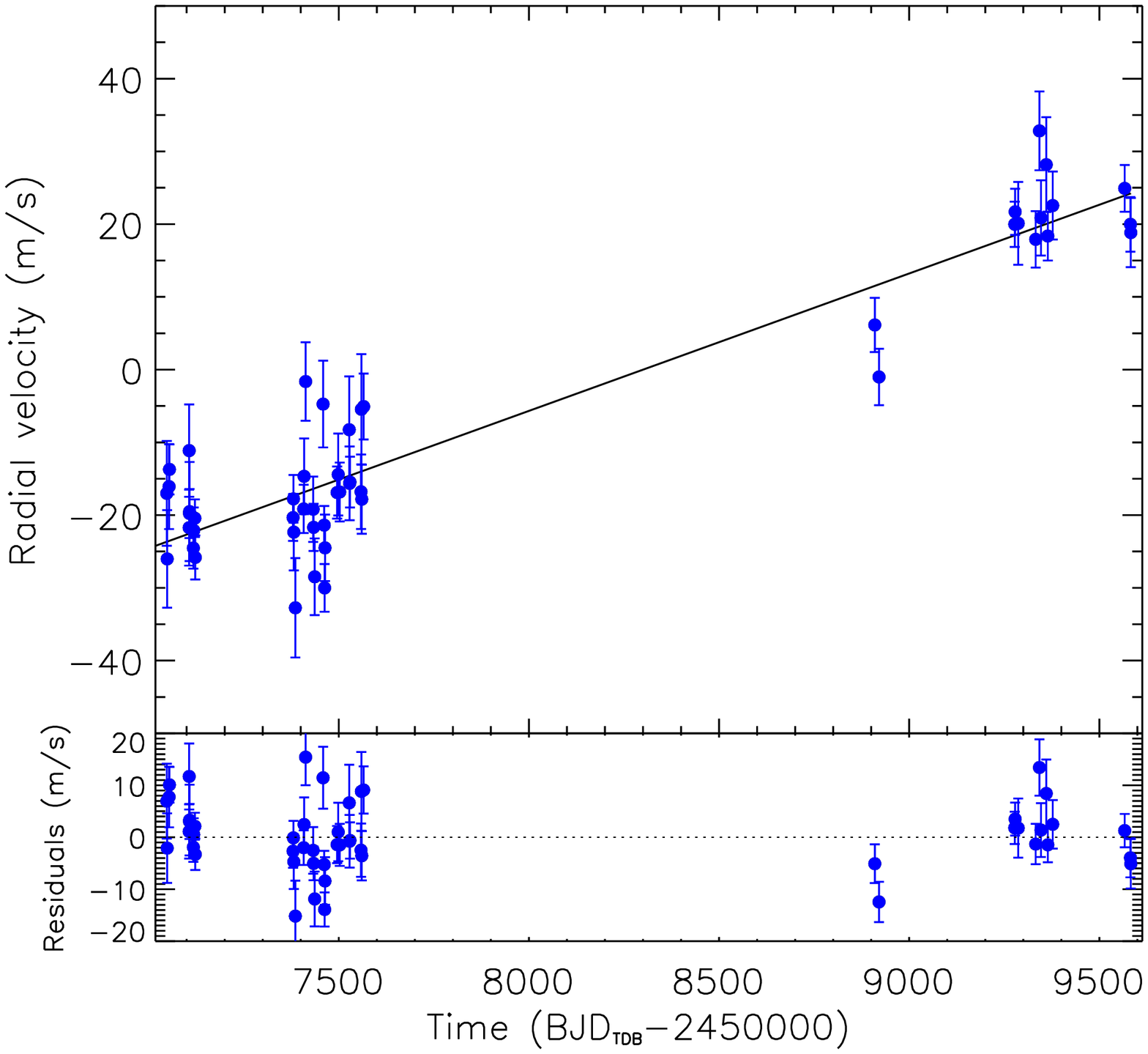}
\includegraphics[width=6.0cm, angle=90]{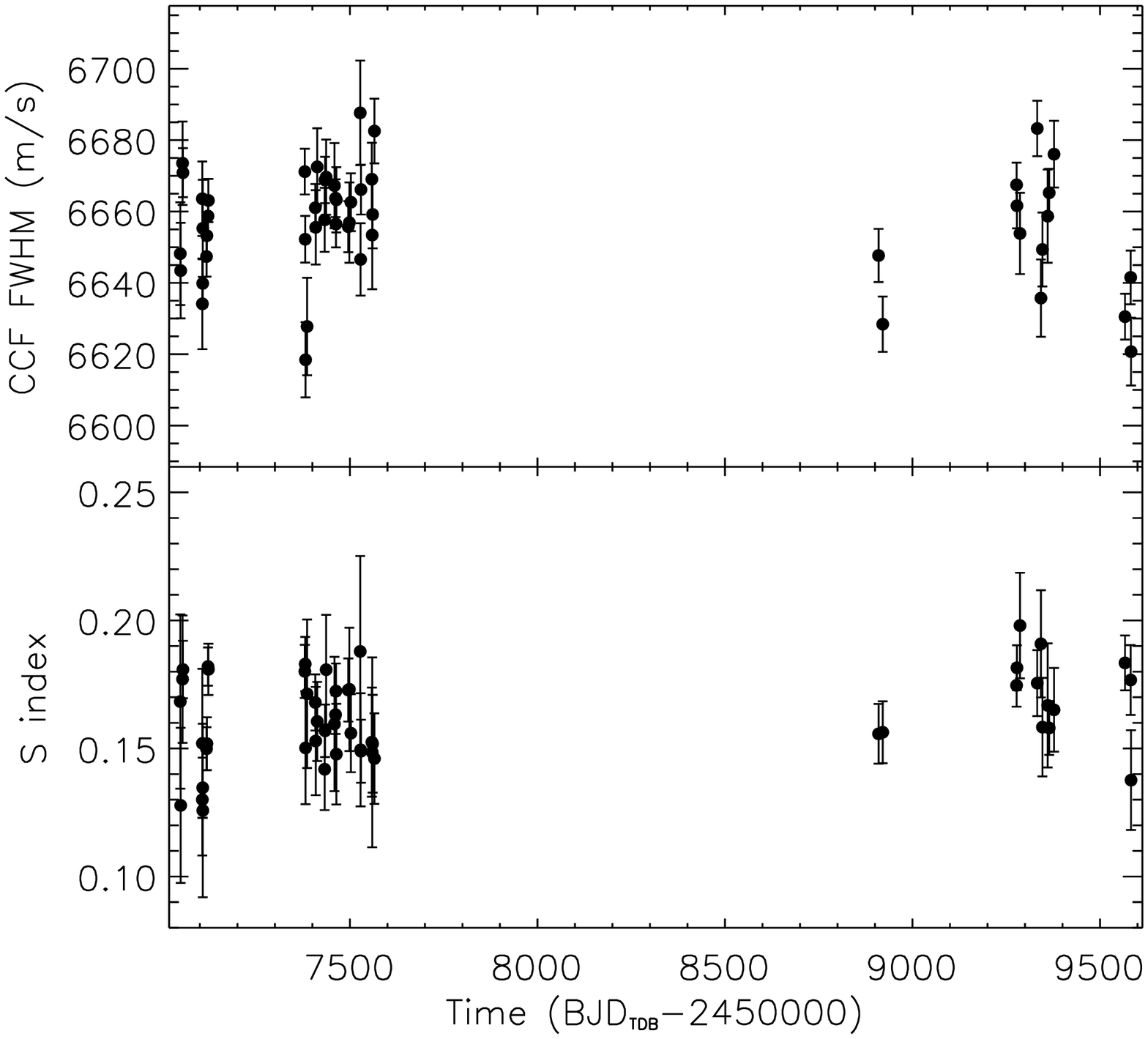}
\vspace{0.10cm}
\caption{Radial-velocity and activity index measurements of K2-12.
\emph{Left panel:} HARPS-N radial velocities showing the long-term linear slope due to a companion of yet unknown nature. 
The Doppler signal of the transiting planet K2-12b ($P=8.28$~d) is undetected, hence, only an upper limit to its mass could be given.
\emph{Right panel}: FWHM of the cross-correlation function (top) and CaII \emph{S}-index (bottom) time series showing no significant linear trends. This would indicate that the radial-velocity linear trend in the left panel is caused by a physically bound companion.
}
\label{figure_RV_vs_time_K2-12}
\end{subfigure}
%\end{figure*}

%\begin{figure*}[h!]
\begin{subfigure}
\centering
\hspace{1cm}
\includegraphics[width=6.0cm, angle=90]{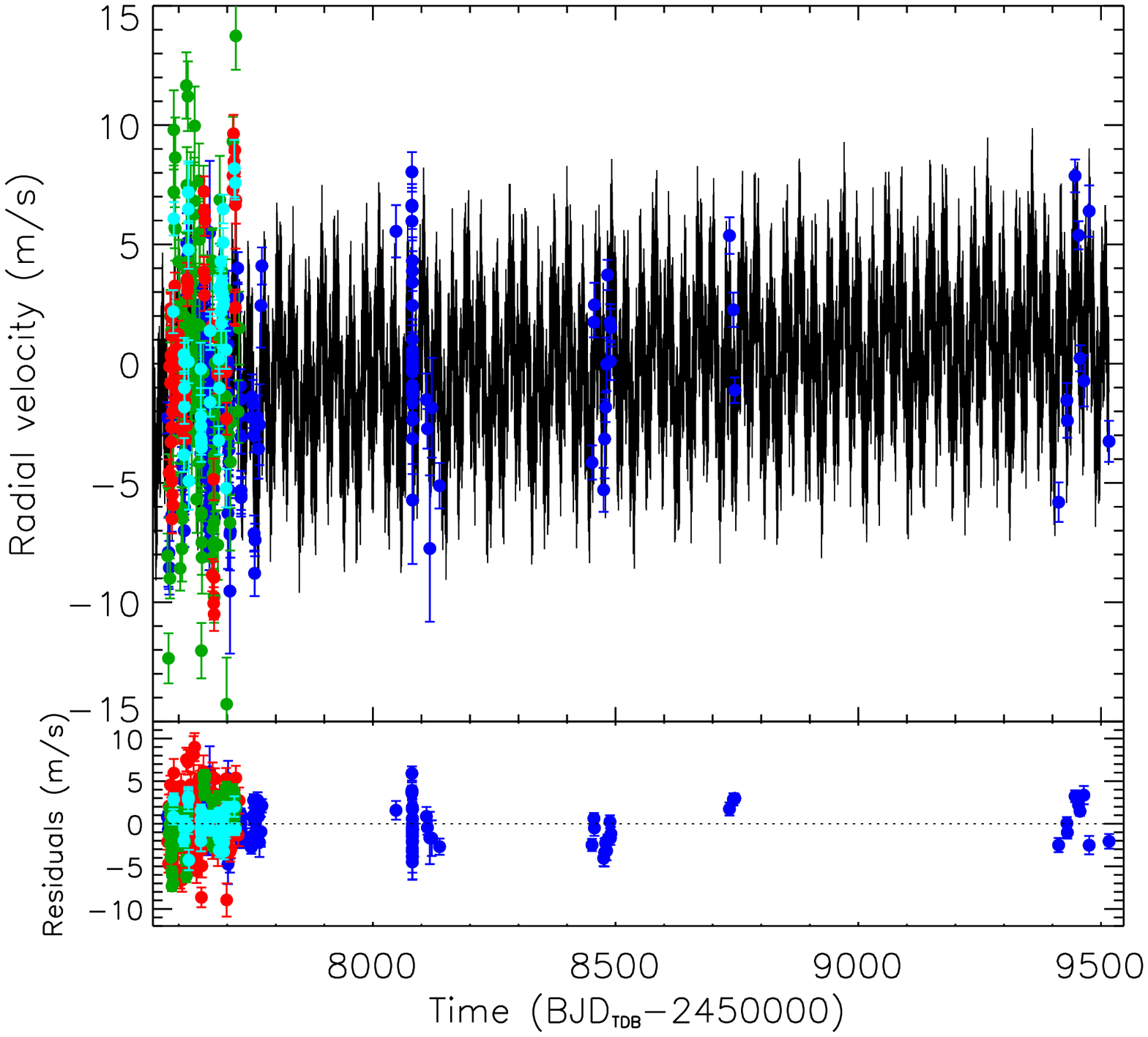}
\includegraphics[width=6.0cm, angle=90]{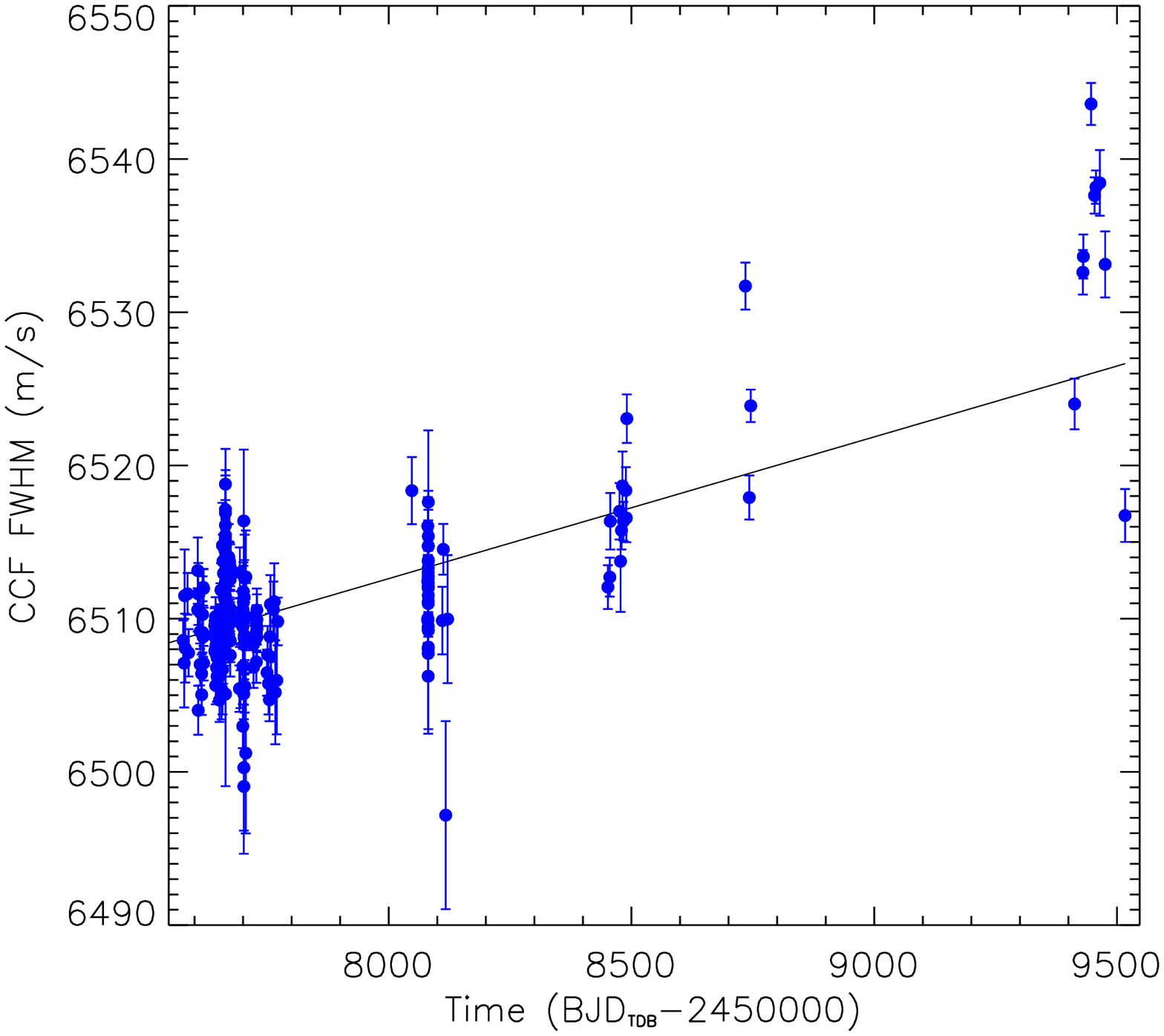}
\vspace{0.10cm}
\caption{Radial-velocity and CCF FWHM measurements of K2-96/HD\,3167.
\emph{Left panel}: Radial velocities showing a long-term linear trend: red, green, blue, and light blue points 
show the measurements obtained with the HIRES, APF, HARPS-N, and HARPS spectrographs, respectively. 
\emph{Right panel}: Linear trend observed in the FWHM of the HARPS-N cross-correlation functions. 
A similar behavior is observed in the CaII $S$ index. This implies that the slope observed in the RVs (left panel) 
is produced by stellar activity, and not by a physical companion.
}
\label{figure_RV_FWHM_HD3167}
\end{subfigure}
\end{figure*}

Five systems show long-term linear trends, namely Kepler-93 (\citealt{2015ApJ...800..135D}; Fig.~\ref{figure_RV_Kepler-93}), 
Kepler-454 (\citealt{2016ApJ...816...95G}; Fig.~\ref{figure_RV_Kepler-454}),  
K2-12 (\citealt{2018AJ....155..136M}; Fig.~\ref{figure_RV_vs_time_K2-12}), 
K2-96/HD\,3167 (\citealt{2017AJ....154..122C, 2017AJ....154..123G}; Fig.~\ref{figure_RV_FWHM_HD3167}), 
and K2-262/Wolf\,503 \citep{2021AJ....162..238P}.
The trends in Kepler-93, Kepler-454 and K2-12 are caused by bound companions, 
while those in K2-96/HD\,3167 and K2-262/Wolf\,503 are due to stellar magnetic activity, 
because similar trends are also seen in the activity indicators
(see Fig.~\ref{figure_RV_FWHM_HD3167} for the case of K2-96/HD\,3167; cf. also \citealt{2022A&A...668A..31B}). 
Gaussian processes with a SE kernel were used to model the RVs of Kepler-93 
to account for long-term variations overlapping with the linear trend and the signal of the planet Kepler-93b 
(Fig.~\ref{figure_RV_Kepler-93}). 
Such variations are likely due to stellar activity, as they seem to follow those of the \emph{S}-index activity indicator
after 7000~$\rm BJD_{TDB}-2450000$ with a Pearson correlation coefficient of $0.26\pm0.03$ (Fig.~\ref{figure_RV_Kepler-93}, right panel).

From the temporal baseline, $\Delta T$, of the RV time series and the amplitude of the trend, $\Delta K = \dot\gamma \cdot \Delta T$, 
we could estimate the lower limits of orbital period, $P,$ and minimum mass, $M_{\rm p} \sin{i}$, of the companions 
producing the trends, by assuming circular or quasi-circular orbits, 
namely: $P \geq 4\cdot \Delta T$ and $M_{\rm p} \sin{i} \geq (\Delta K / 28.4~{\rm m\,s^{-1}}) \cdot (M_{\rm s} / {\rm M_\odot})^{2/3} \cdot (P / {\rm yr})^{1/3}$. 
Table~\ref{table_companion_trends} reports the lower limits of the orbital parameters and $M_{\rm p} \sin{i}$ of the outer companions 
in the Kepler-93, Kepler-454, and K2-12 systems. The companion generating the linear trend in Kepler-93 is a brown dwarf or a low-mass 
star, having $M_{\rm p} \sin{i} \geq 21~\rm M_{Jup}$, 
while the masses of the companions producing the slopes in the Kepler-454 and K2-12 systems are currently compatible with a 
planetary companion, but further monitoring is needed to unveil their nature.

\subsubsection{Systems with Keplerian signals of cold Jupiters}
Resolved Keplerian orbits of CJs are observed only in three of the thirty-eight systems, 
namely: Kepler-68 \citep{2013ApJ...766...40G}, Kepler-454 \citep{2016ApJ...816...95G}, and K2-312/HD\,80653 \citep{2020A&A...633A.133F}.

\begin{figure}[h!]
%\vspace{-0.2cm}
\begin{center}
\includegraphics[width=6.0cm, angle=90]{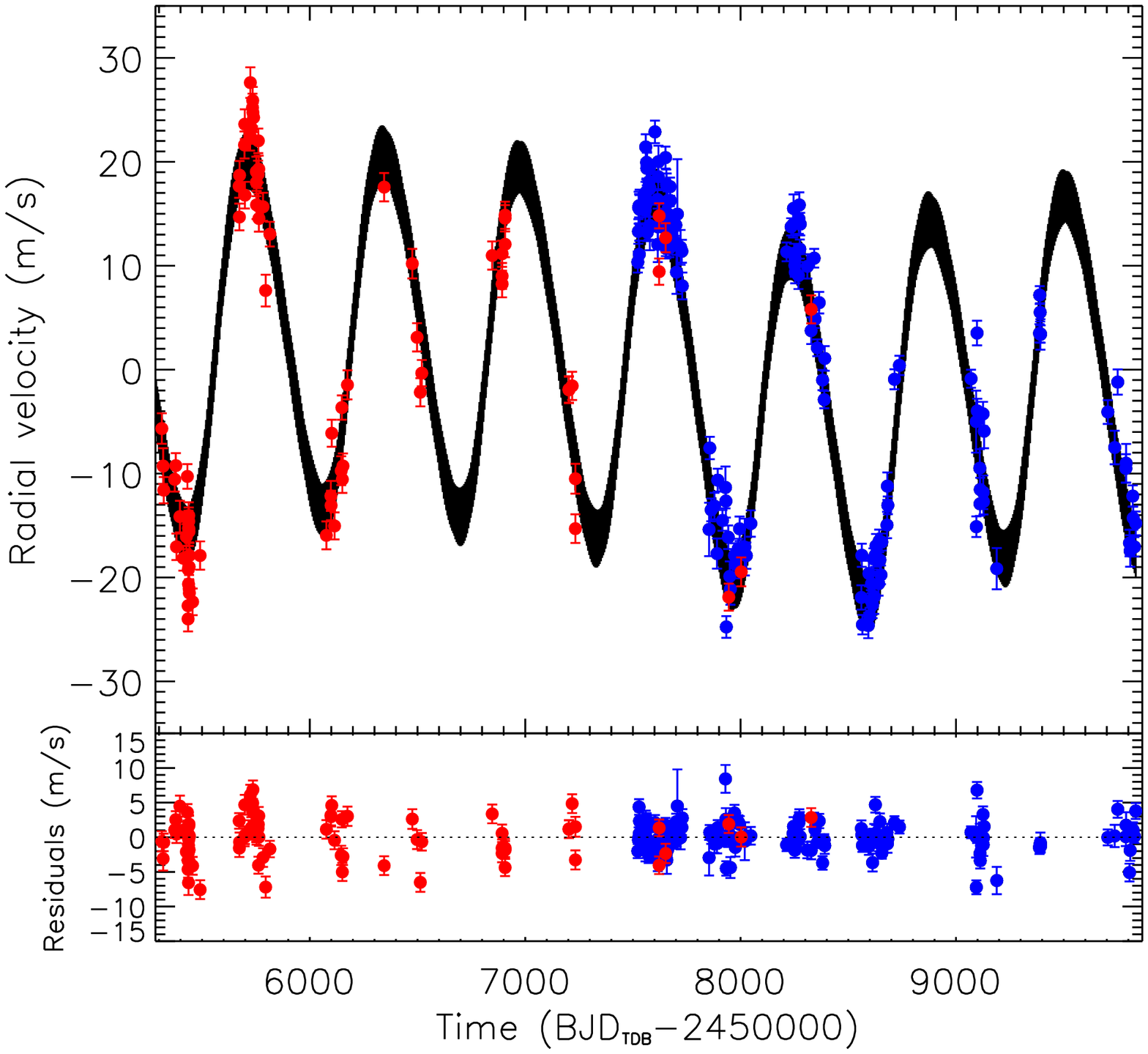}
\end{center}
\begin{center}
\includegraphics[width=6.0cm, angle=90]{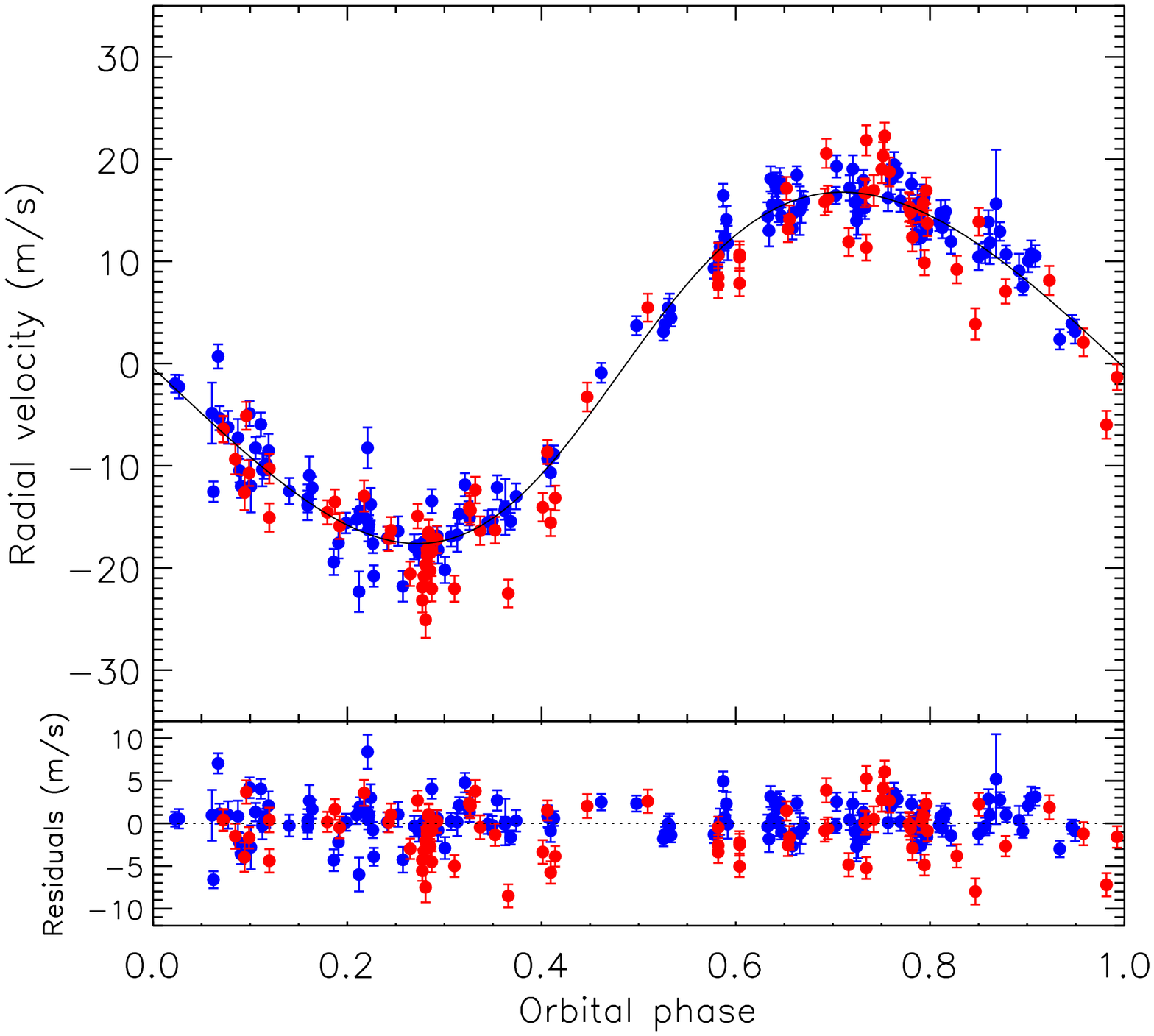}
\end{center}
\begin{center}
\includegraphics[width=6.0cm, angle=90]{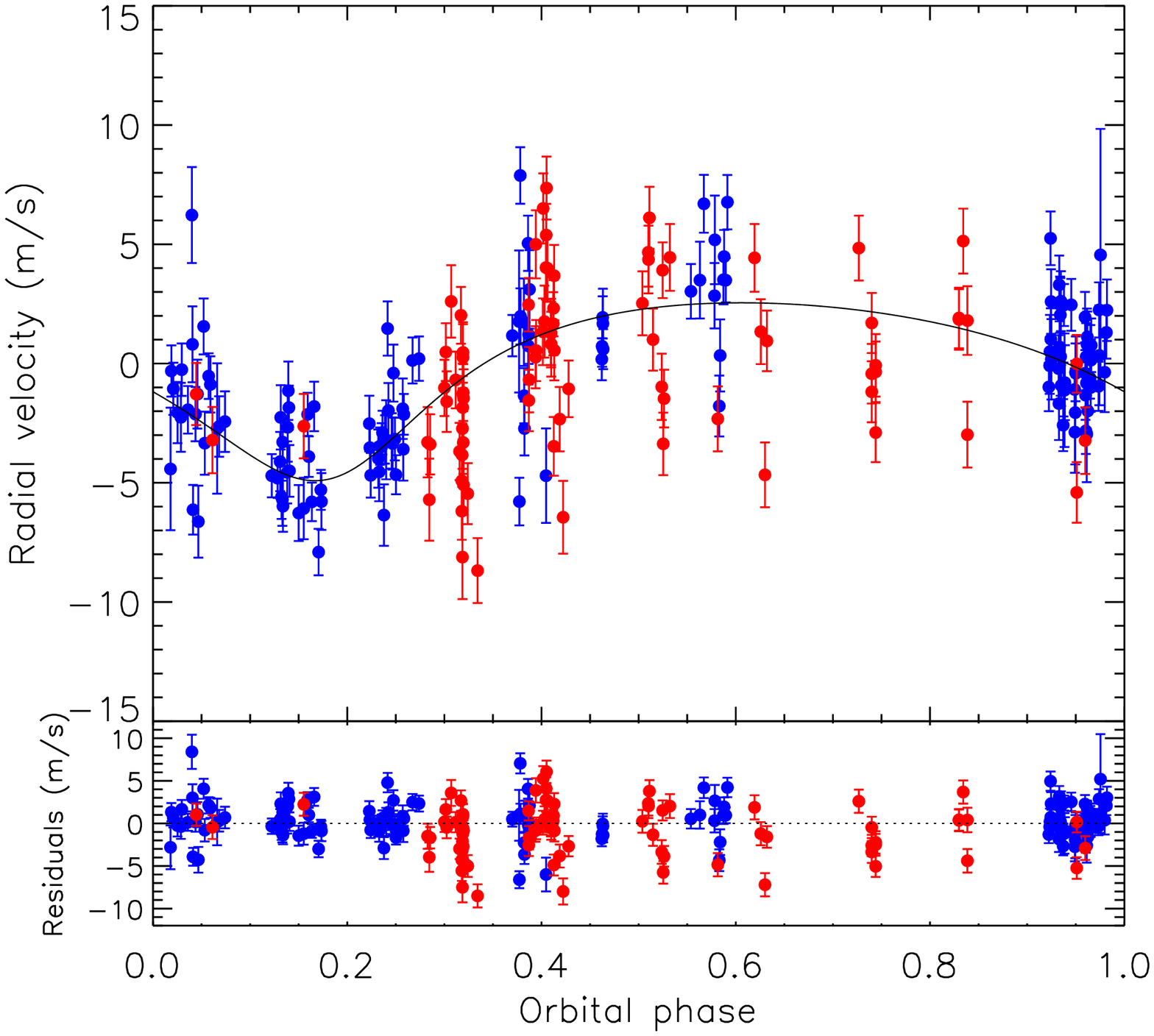}
\end{center}
\caption{Radial-velocity measurements of Kepler-68.
\emph{Top panel:} HIRES (red circles) and HARPS-N (blue circles) radial velocities and the best-fit model (black line) with four Keplerians corresponding to the two transiting planets Kepler-68b ($P=5.40$~d) and Kepler-68c ($P=9.60$~d), and the two cold giant planets Kepler-68d ($P=633$~d $=1.7$~yr) 
and Kepler-68e ($P\sim3450$~d $=9.4$~yr). 
\emph{Middle and bottom panels:} Phase-folded radial-velocity signals of the cold giant planets Kepler-68d and Kepler-68e. Note: orbital phases equal to 0 and 1 correspond to inferior conjunction.
}
\label{figure_RV_Kepler-68}
\end{figure}

\begin{figure*}[h!]
\centering
\includegraphics[width=6.0cm, angle=90]{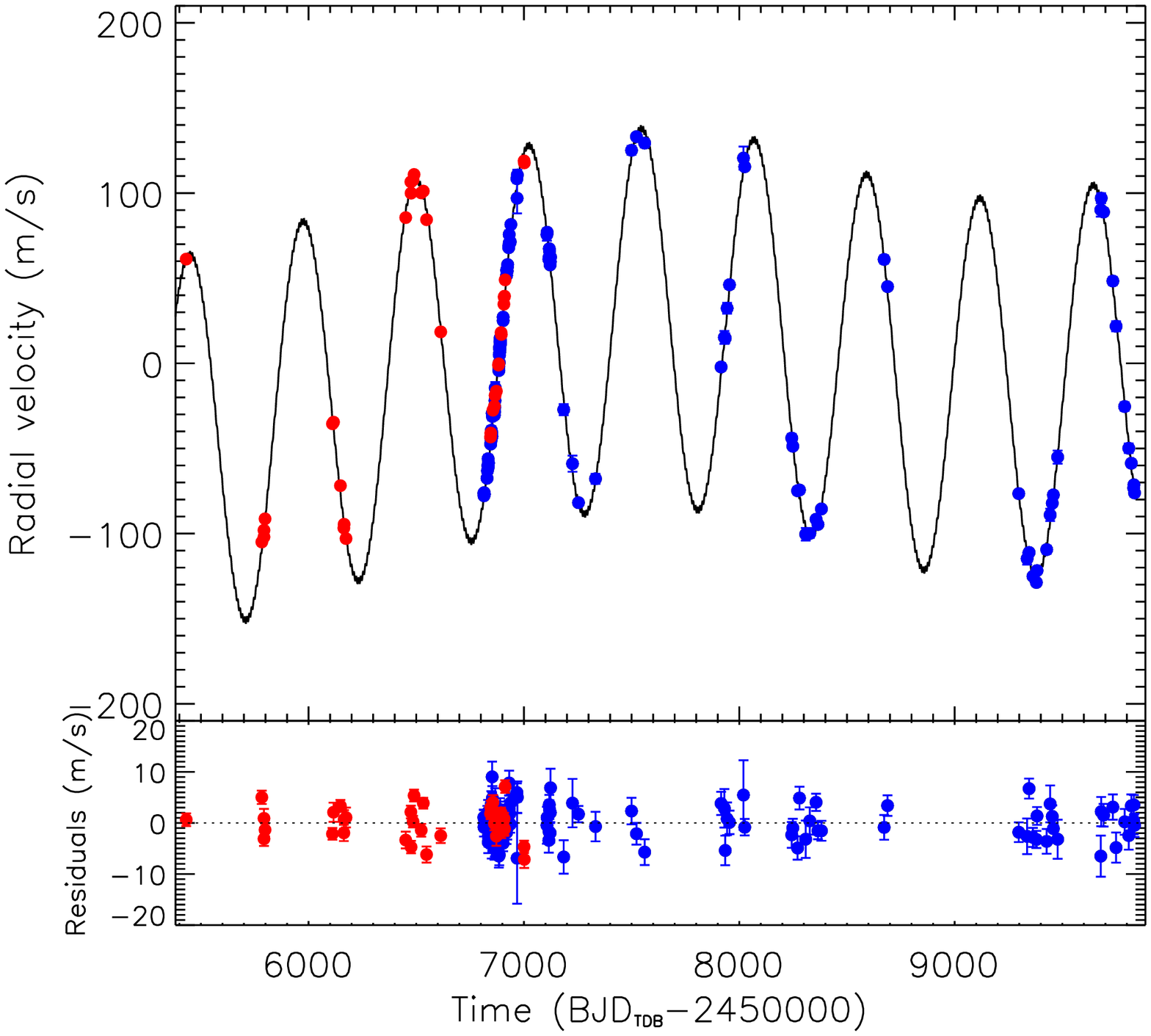}
\includegraphics[width=6.0cm, angle=90]{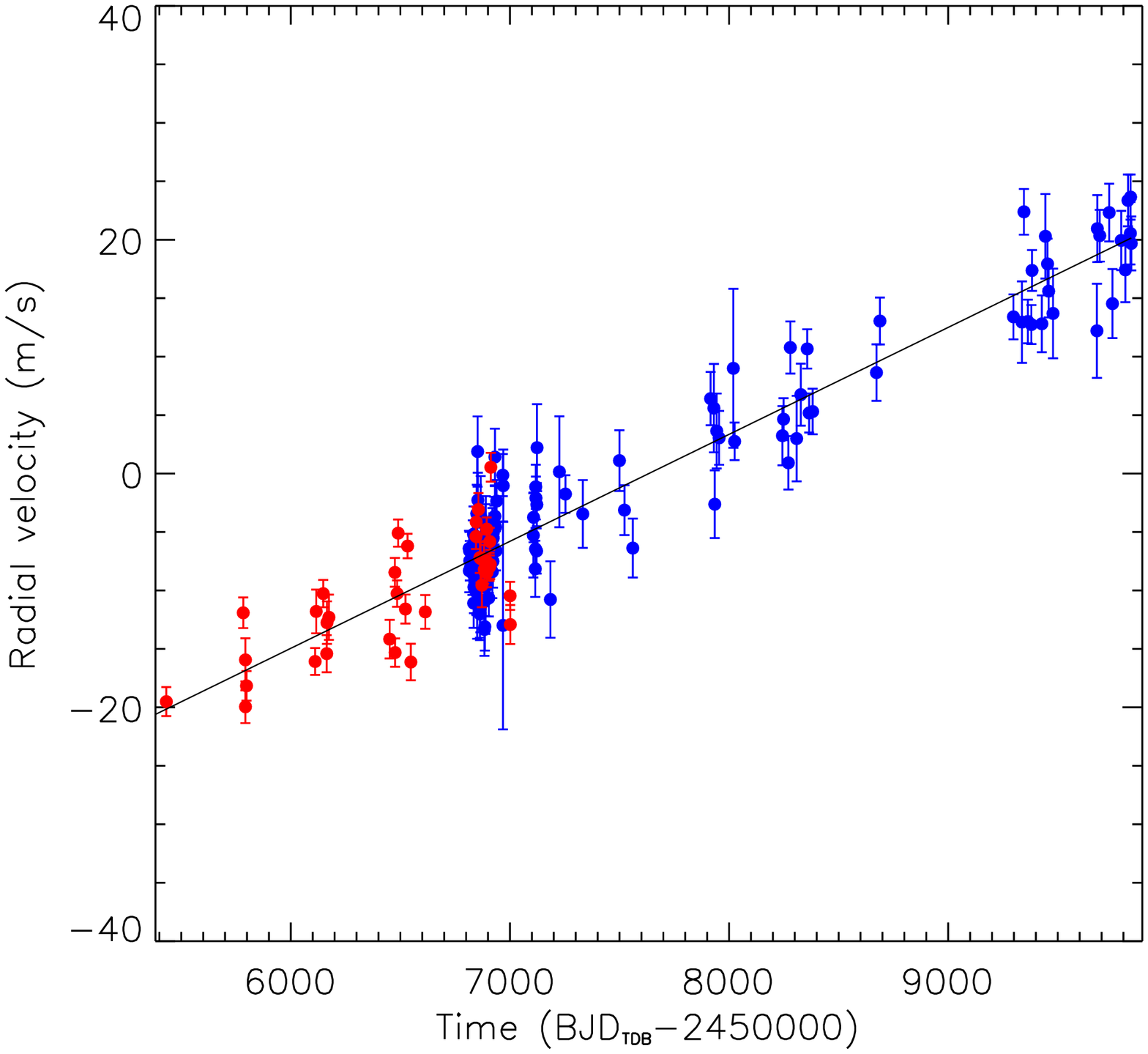}
\par\bigskip
\includegraphics[width=6.0cm, angle=90]{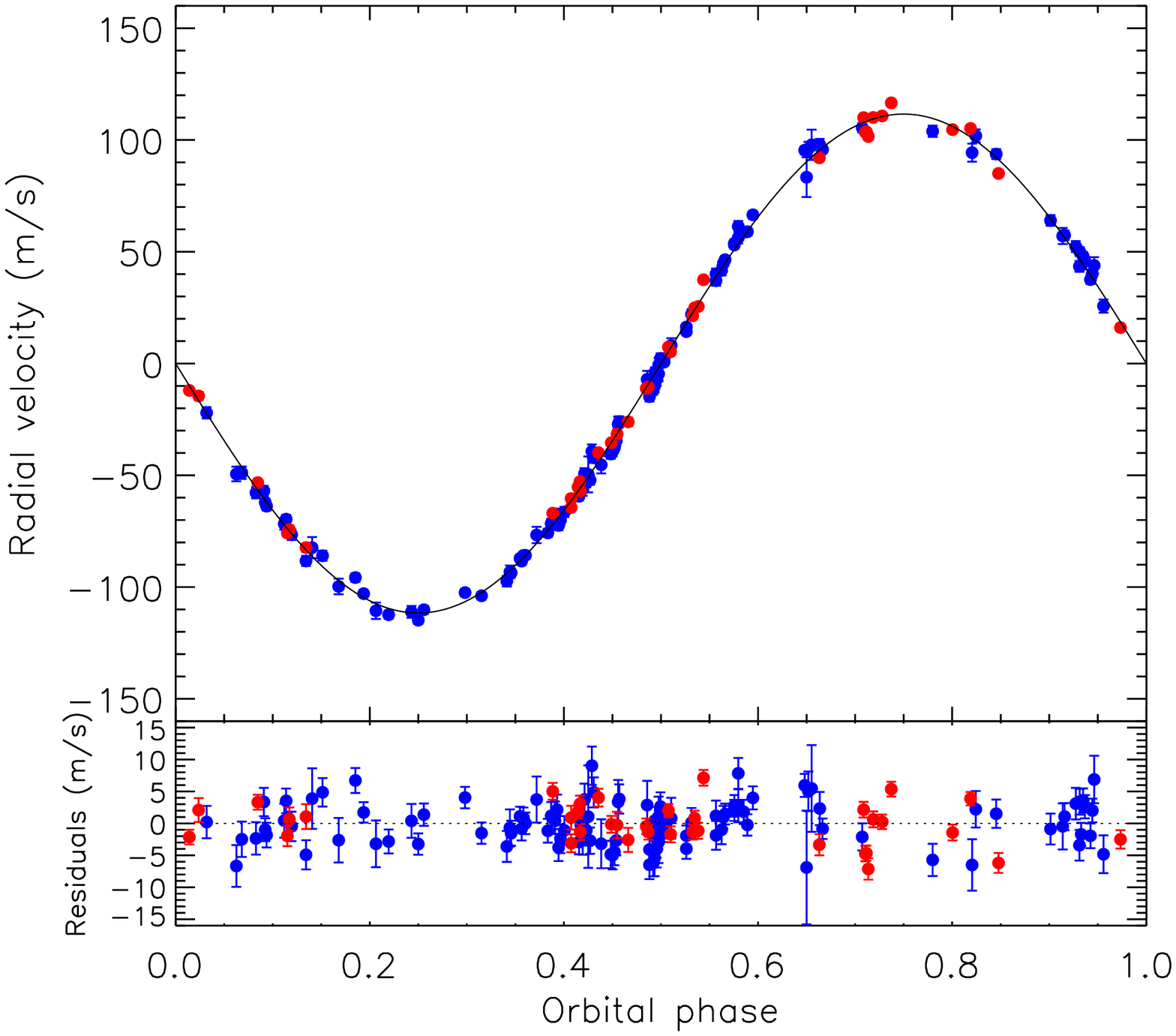}
\includegraphics[width=6.0cm, angle=90]{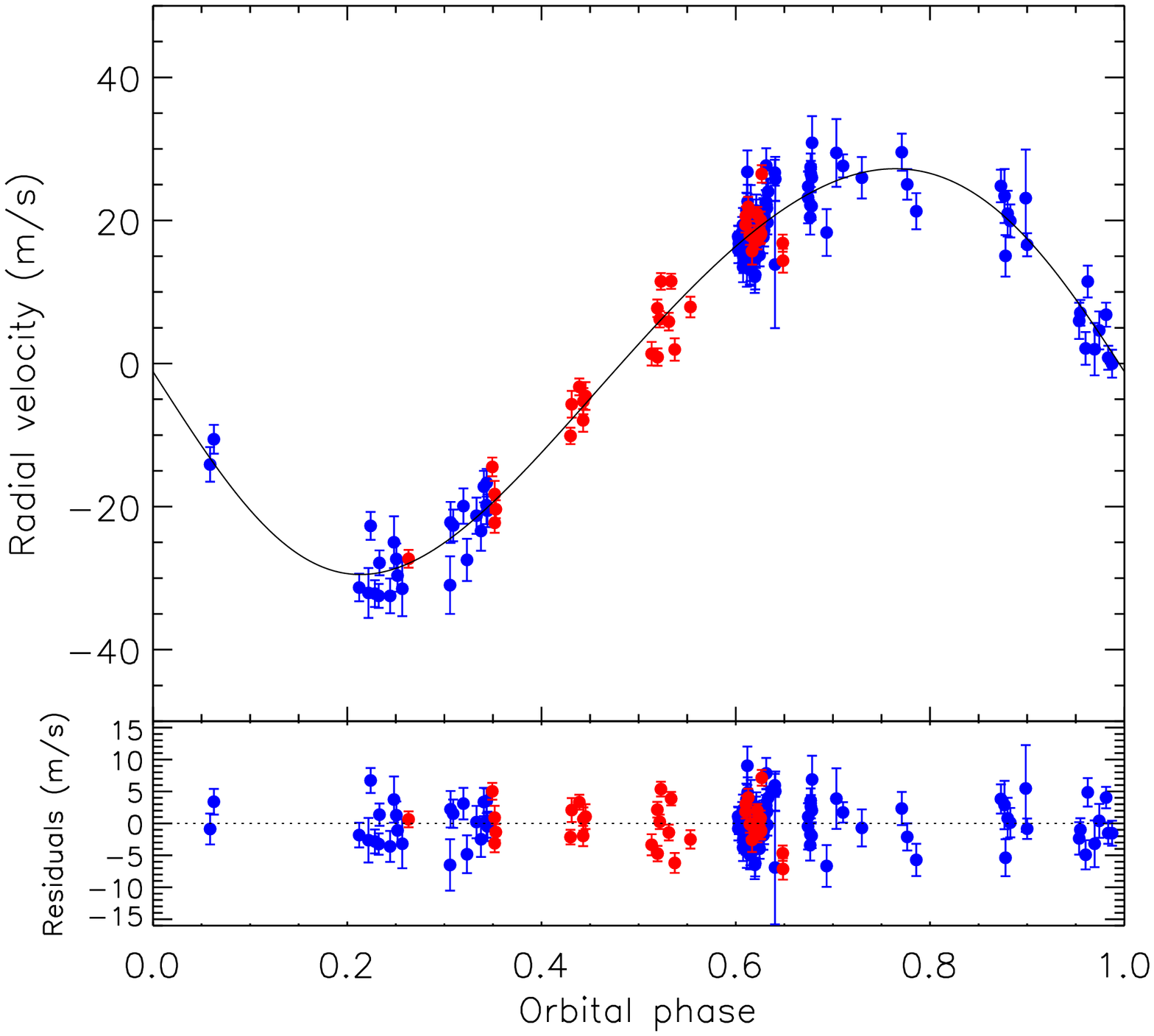}
\vspace{0.20cm}
\caption{Radial-velocity measurements of Kepler-454.
\emph{Top left panel:} HIRES (red circles) and HARPS-N (blue circles) radial velocities and the best-fit model (black line) with a long-term linear trend and three Keplerians corresponding to the transiting planet Kepler-454b ($P=10.57$~d) and the two cold giant planets Kepler-454c ($P=524$~d $=1.4$~yr) and Kepler-454d ($P\sim4070$~d $=11.1$~yr). 
\emph{Top right panel:} Linear trend after removing the three Keplerian signals, 
which is caused by an additional fourth companion of yet unknown nature. 
\emph{Bottom panels:} Phase-folded radial-velocity signals of the cold giant planets Kepler-454c (left) and Kepler-454d (right). Note: orbital phases equal to 0 and 1 correspond to inferior conjunctions. }
\label{figure_RV_Kepler-454}
\end{figure*}

The two CJs, Kepler-68d and Kepler-454c, with $P=633$ and 524~d and $M_{\rm p} \sin{i}=0.75$ and 4.51~$\rm M_{Jup}$,
were previously discovered by \citet{2013ApJ...766...40G} and \citet{2016ApJ...816...95G}. 
Moreover, a long-term quadratic trend in Kepler-68 \citep{2019AJ....157..145M} and a slope in Kepler-454 \citep{2016ApJ...816...95G}
were also found in the RV data,  
revealing the presence of additional outer companions in both systems, 
given that no trends were seen in the activity indicators. 
Our analysis not only allowed us to refine the parameters of the Kepler-68d and Kepler-454c giant planets, 
but also unveiled two additional CJs, namely: Kepler-68e and Kepler-454d, with $P\sim3450$ and 4070~d and 
minimum masses of 0.27 and 2.31~$\rm M_{Jup}$, respectively (see also \citealt{Marginietalinprep} for Kepler-68). 
The RVs of both systems as a function of time and the phase-folded RV signals of the CJs
Kepler-68d, Kepler-68e, Kepler-454c, and Kepler-454d are shown in Figs.~\ref{figure_RV_Kepler-68} and \ref{figure_RV_Kepler-454}.

\begin{figure}[h!]
\centering
\includegraphics[width=6.0cm, angle=90]{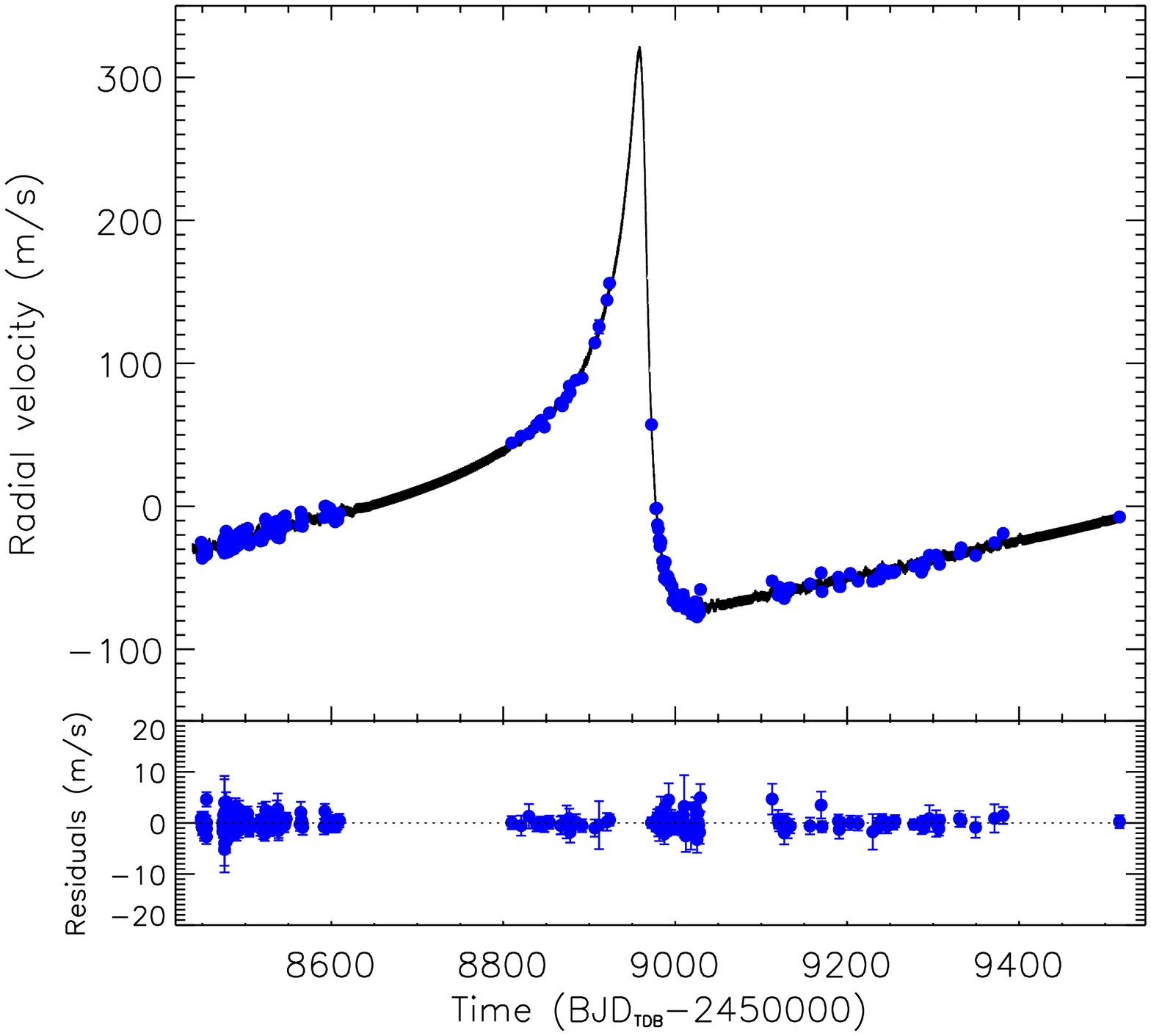}
\vspace{0.20cm}
\caption{HARPS-N radial velocities of K2-312\,/\,HD\,80653 and the best-fit model (black line) with two Keplerians, which correspond to the 
ultra-short-period transiting planet K2-312b ($P=0.72$~d) and the eccentric cold giant planet K2-312c ($P\sim921$~d $=2.5$~yr; $e=0.85$), and Gaussian process regression with a quasi-periodic kernel.}
\label{figure_RV_vs_time_K2-312}
\end{figure}

The RV monitoring of the K2-312 system revealed that the linear trend observed by \citet{2020A&A...633A.133F} is due 
to a very eccentric CJ, namely, K2-312c, with $P=921$~d, $M_{\rm p} \sin{i}=5.41~\rm M_{Jup}$, and $e=0.85$ 
(Fig.~\ref{figure_RV_vs_time_K2-312}; \citealt{Porettietalinprep}). 
The parameters of the five CJs with resolved Keplerian orbits and their $1\sigma$ uncertainties are given in Table~\ref{table_giant_planets}.

\subsection{Occurrence rate of cold Jupiters}
By considering the three systems with resolved Keplerian orbits (Kepler-68, Kepler-454, and K2-312) and 
possibly the K2-12 system, which shows a trend that is currently compatible with a planetary companion at orbital distance $a < 10$~AU, 
we can derive the occurrence rate of CJs, $f_{\rm CJ|SP}$, in our sample. 
From this sample, we have to remove Kepler-22 for the reason explained in Sect.~\ref{target_selection_data}, which
yields a total of 37 \emph{Kepler} and K2 systems. 
We used binomial statistics, namely:
\begin{equation}
b(d|N_{\star,\rm eff},f_{\rm CJ|SP})=\frac{N_{\star,\rm eff}!}{d!(N_{\star,\rm eff}-d)!}f_{\rm CJ|SP}^d(1-f_{\rm CJ|SP})^{N_{\star,\rm eff}-d},
\label{binom_equation}
\end{equation}

\noindent
where $d$ is the number of systems with detected CJs, that is, $d=3$ or 4, 
depending on whether K2-12 is included or not; and $N_{\star,\rm  eff}$ is not just the number of systems 
in our sample $N_{\star}=37$, but the ``effective'' number of stars
$N_{\star,\rm  eff} = N_\star \cdot C$, where $C$ is the average completeness obtained by computing the mean of the 
completeness maps of the 37 systems (Fig.~\ref{figure_completenesses_Kepler-1876_Kepler-93_average}, bottom panel).

To compare our results with those of \citet{2020MNRAS.492..377W}, we computed $C$ and $f_{\rm CJ|SP}$ for 
the range $\Delta M_{\rm p}=0.3-13~\rm M_{Jup}$ in planetary mass, and 
different intervals $\Delta a=1-2$, $2-4$, $4-10$, and $1-10$~AU in semi-major axis
(see Table~\ref{completeness_frequency}). 
Similarly, we also derived $C$ and $f_{\rm CJ|SP}$ for $\Delta M_{\rm p}=0.5-13$, $0.5-20~\rm M_{Jup}$, 
and $\Delta a=1-10$ and $1-20$~AU, for comparison with the $f_{\rm CJ|SP}$ found by \citet{2019AJ....157...52B}.
We report our values and $1\sigma$ error bars of the occurrence rates of CJs as well as 
those obtained by \citet{2020MNRAS.492..377W} and \citet{2019AJ....157...52B} in Table~\ref{frequency_comparison}. 
In summary, we found $f_{\rm CJ|SP}=9.3^{+7.7}_{-2.9}~\%$ for d=3 and $f_{\rm CJ|SP}=12.3^{+8.1}_{-3.7}~\%$ for d=4,  
namely, by considering the long-term trend in K2-12 as planetary in origin, 
for $\Delta M_{\rm p}=0.3-13~\rm M_{Jup}$ and $\Delta a=1-10$~AU. 
The former value is lower than that found by \citet{2020MNRAS.492..377W}, namely, $f_{\rm CJ}=20.2^{+6.3}_{-3.4}~\%$, 
by a factor of 2; however, given the large uncertainties, the two measurements are compatible at 1.3$\sigma$.
Our $f_{\rm CJ|SP}$ is four times lower than that derived by \citet{2019AJ....157...52B}.

\subsection{Non-transiting low-mass planets}
\label{nontransiting_planets}
By employing GLS periodograms of the RV residuals and Bayesian model comparison through the $BIC$ criterion
(Sect.~\ref{target_selection_data}), we confirmed the presence of additional RV signals that can be attributed to the non-transiting planets 
Kepler-10d ($P=151.0$~d, $M_{\rm p} \sin{i}=12.8~\rm M_{\oplus}$; \citealt{Bonomoetalinprep}), 
K2-96b/HD\,3167b ($P=8.4$~d, $M_{\rm p} \sin{i}=4.3~\rm M_{\oplus}$; \citealt{2017AJ....154..122C}),
K2-96e/HD\,3167e ($P=96.6$~d, $M_{\rm p} \sin{i}=8.4~\rm M_{\oplus}$; \citealt{2022A&A...668A..31B}), and 
K2-111c ($P=15.7$~d, $M_{\rm p} \sin{i}=11.1~\rm M_{\oplus}$; \citealt{2020MNRAS.499.5004M}); 
their orbital parameters and minimum masses are given in Table~\ref{table_nontransiting_planets}. 
Concerning K2-96e/HD\,3167e, we found a slightly different $P$ than \citet{2022A&A...668A..31B}, namely, $P_{\rm e}=102.09\pm0.52$~d, 
which is likely due to a different treatment of the activity signal: we fit a slope to all the RVs 
(see Fig.~\ref{figure_RV_FWHM_HD3167}), while \citet{2022A&A...668A..31B} included in their MCMC analysis 
two activity-decorrelation terms for the HARPS and HARPS-N spectrographs. In any case, as noted by \citet{2022A&A...668A..31B}, 
the observing spectral window allows for different solutions of $P_{\rm e}$ with multiple peaks in the posterior distribution 
(see their Fig.~12). 

\begin{figure}[h!]
\centering
\includegraphics[width=6.0cm, angle=90]{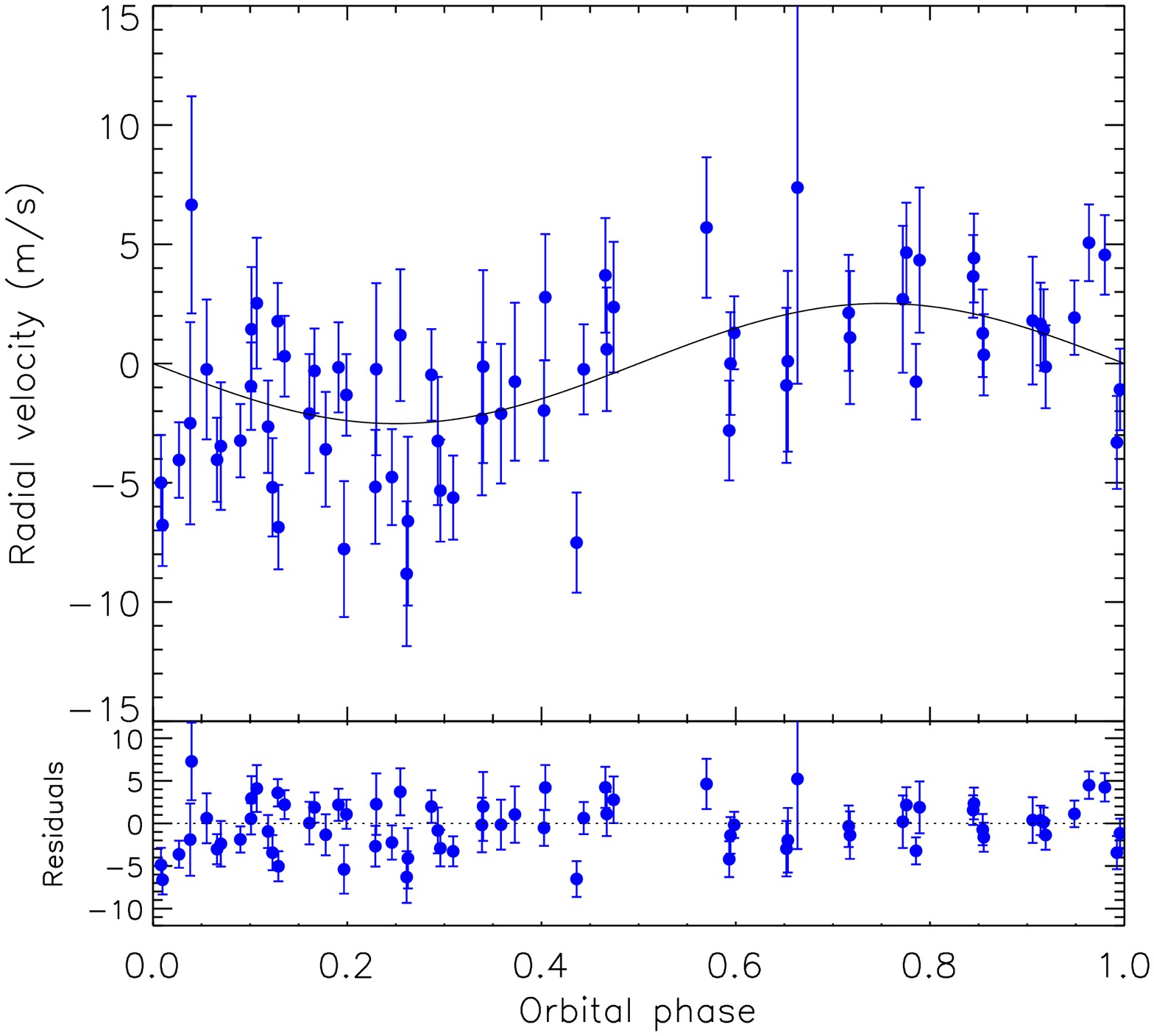}
\vspace{0.20cm}
\caption{Phase-folded HARPS-N radial velocities (blue dots) of Kepler-1876 at the period $P=15.8$~d of the possible planet candidate Kepler-1876c. The black solid line shows the best-fit model.}
\label{figure_RV_phase_Kepler-1876}
\end{figure}

We report a new planet candidate, Kepler-1876c ($P=15.8$~d, $M_{\rm p} \sin{i}=11.0~\rm M_{\oplus}$), in the single
transiting system Kepler-1876 (Fig.~\ref{figure_RV_phase_Kepler-1876}). The GLS periodogram of the HARPS-N RVs of Kepler-1876 
shows a significant periodicity at $P=15.8$~d with FAP of $5.6 \cdot 10^{-5}$, 
which does not appear in any of the activity indicators. 
Nonetheless, the $\Delta BIC$ in favor of the two-planet model is currently $\Delta BIC=5.3 < 10$; thus, more RVs are needed to confirm the planetary nature of this signal, 
also by  checking that its phase and amplitude do not change with time.

We cannot confirm the signals of the non-transiting planet Kepler-20g \citep{2016AJ....152..160B}
or the non-transiting candidate K2-2c \citep{2015ApJ...800...59V} with respective periods of $P=35$ and 45~d and RV semi-amplitudes of
$K=4.1$ and $\sim 2~\rm m\,s^{-1}$. The former signal is no longer present after 
the improvement of the DRS software to extract the HARPS-N RVs (Sect.~\ref{target_selection_data}). 
The latter was likely due to a combination of the spectral window of the original RVs in \citet{2015ApJ...800...59V} 
and an activity signal with periodicity of $\sim 270$~d, which is also seen in the 
GLS periodograms of the \emph{S}-index and FWHM activity indicators,
and was modeled with a GP-SE approach (see Table~\ref{table_accelerations}). 
The fact that the 45~d signal could have originated from stellar activity variations 
was already discussed and considered by \citet{2015ApJ...800...59V}.

As specified in Sect.~\ref{orbital_fitting}, we used non-interacting Keplerians to model 
RV planetary signals and GLS periodograms to search for non-transiting planets in the RV residuals. 
This implies that additional planets revealed by transit timing variations (TTVs), 
with undetectable signal in the RVs only (according to the criteria given in Sect.~\ref{orbital_fitting}), 
such as Kepler-19c and d \citep{2017AJ....153..224M},
were not modeled in this study for uniformity.

\subsection{Improved physical and orbital parameters of 64 Kepler and K2 small planets}
As mentioned in Sect.~\ref{introduction}, the DE-MCMC analyses of new HARPS-N RVs, 
combined (in some cases for the first time) with the literature RVs obtained with other instruments, 
allowed us to update the orbital and physical parameters of the 64 transiting planets. 
In particular, from the stellar parameters given in Table~\ref{table_system_parameters}, 
the transit parameters ($P$, $R_{\rm p}$ and $i$) from the 
literature\footnote{For the systems with updated $R_\star$ values in Table~\ref{table_system_parameters}, 
we recomputed $R_{\rm p}$ from the $R_{\rm p}/R_\star$ transit parameter and $R_\star$. 
In cases where the impact parameter $b$ only is provided in literature assuming circular orbits, 
we derived the orbital inclination as $i=\arccos{(b \cdot R_\star/a)}$.}, 
and our updated RV semi-amplitudes $K$, 
we re-derived the masses, densities, and surface gravities for all the planets as in Table~\ref{table_planet_parameters}. 
We also report in the same table the planet equilibrium temperature, $T_{\rm eq}$, assuming a null Bond albedo and full 
redistribution of heat from the dayside to the nightside (e.g., \citealt{2007ApJ...667L.191L}), and 
the incident flux, $F_{\rm p}$. Both $T_{\rm eq}$ and $F_{\rm p}$ 
were updated in case the stellar $T_{\rm eff}$ needed to be re-determined from our HARPS-N spectra (Table~\ref{table_system_parameters}).

Table~\ref{table_planet_parameters} also emphasizes the fundamental contribution of HARPS-N in 
determining precise masses and densities to infer the composition of SPs. 
Specifically, the masses of 25 planets are determined with a precision higher than $5\sigma$ ($M_{\rm p}/\sigma_{M_{\rm p}} > 5$), 
13 of which with $M_{\rm p}/\sigma_{M_{\rm p}} > 7.5$, and 8 with $M_{\rm p}/\sigma_{M_{\rm p}} > 10$. 
Only upper limits on $M_{\rm p}$ are given for 20 planets whose induced Doppler signal was not detected. 
We note that a discussion of the planetary compositions from the measure of planetary masses and densities
is beyond the scope of this work\footnote{In particular, we refer the reader to companion papers, 
which discuss in detail the updated compositions of Kepler-10b and c \citep{Bonomoetalinprep}, 
Kepler-68b and c \citep{Marginietalinprep} and K2-418b/EPIC-229004835b \citep{Tronsgaardetalinprep}}.

\newpage
\section{Discussion and conclusions}
The value of the occurrence rate of CJs from the sample of 37 Kepler and K2 systems
monitored by the HARPS-N/GTO consortium over the long term, namely, $f_{\rm CJ|SP}=9.3^{+7.7}_{-2.9}\%$,
is lower than the frequency of $f_{\rm CJ}=20.2^{+6.3}_{-3.4}\%$ derived by \citet{2020MNRAS.492..377W} 
for solar-type stars from RV surveys, 
regardless of the possible presence of inner SPs (Table~\ref{frequency_comparison}). 
This might hint at the theoretical anti-correlation between the presence of SPs 
and CJs predicted by \citet{2015ApJ...800L..22I} and \citet{2019A&A...627A..83L}.
However, the large uncertainty on our $f_{\rm CJ|SP}$ associated with the inevitably limited target sample 
does not allow us to draw a firm conclusion. 
Moreover, the sample considered by \citet{2020MNRAS.492..377W} likely contains a 
certain fraction of stars hosting SPs, which went undetected through the Doppler method 
and, thus, a comparison of cold-Jupiter occurrence rates for stars with and without SPs is not straightforward. 
Nonetheless, assuming that the aforementioned anti-correlation does hold, 
the possible ``contamination'' of the sample of \citet{2020MNRAS.492..377W} 
by SP systems would tend to make $f_{\rm CJ}$ lower than it really is  
(in other words, $f_{\rm CJ}$ might be higher than $\sim 20\%$ for solar-type stars hosting no SPs).

Our results do not support the claim by \citet{2019AJ....157...52B} regarding an excess of CJs in 
SP systems (see Table~\ref{frequency_comparison}): 
according to their $f_{\rm CJ|SP}$, we should have discovered CJs in $12\pm2$ of our systems, 
while we only found them in 3 systems.
The reason for such a discrepancy lies at least in part in an incorrect interpretation of the planetary origin 
of some linear trends by \citet{2019AJ....157...52B} 
(see their Fig. 3): for instance, the trend in the GJ\,273 system is due to secular acceleration, 
the slope in HD\,3167 to stellar activity (Fig.~\ref{figure_RV_FWHM_HD3167}), 
and those in Kepler-93b and Kepler-407b are compatible with being generated by brown dwarfs or low-mass stellar companions. 
In addition, $f_{\rm CJ|SP}$ was computed by \citet{2019AJ....157...52B} in a different way, 
namely, by assuming a double power law $f_{\rm CJ|SP} = A \cdot M_{\rm p}^{\alpha} \cdot a^{\beta}$, 
deriving the coefficients $A$, $\alpha,$ and $\beta$ with a likelihood approach, 
and integrating $f_{\rm CJ|SP}$ over the parameter space. 
Nonetheless, the exclusion of approximately half of the long-term trends found by \citet{2019AJ....157...52B},
since they were not due to CJs, would  considerably reduce their $f_{\rm CJ|SP}$ and make it more compatible with our estimate.

The present work will be extended to a larger sample, including the TESS SPs  
monitored with HARPS-N since 2019 (e.g., \citealt{2021AJ....162...79C, 2021MNRAS.501.4148L, 2022A&A...667A...8N}), 
as well as the K2 and TESS SP systems observed from the Southern hemisphere 
with other spectrographs such as ESPRESSO and HARPS. 
Performing the same analyses of RV data in Sect.~\ref{orbital_fitting} on a sample at least three times as large 
will allow us to
\begin{itemize}
\item[(i)] derive a more precise $f_{\rm CJ|SP}$ in SP systems. 
\item[(ii)] compute $f_{\rm CJ|SP}$ as a function of  the multiplicity of SPs. Indeed, the predicted anti-correlation between CJs and SPs 
is expected to be more pronounced for systems with a higher level of multiplicity of SPs, because single cores have a higher
probability of "jumping" inside the orbit of the cold gas giant during their inward migration 
(see Fig. 3 in \citealt{2015ApJ...800L..22I}).
Detecting non-transiting planets with RVs (Sect.~\ref{nontransiting_planets}), especially in single transiting systems, 
is therefore crucial for more detailed studies of the possible dependence of $f_{\rm CJ|SP}$ on the multiplicity of SPs.
\item[(iii)] compute $f_{\rm CJ|SP}$ as a function of the cold-Jupiter multiplicity: if multiple CJs were formed, it would be (in principle)
even more difficult for the farther icy cores to migrate inward because the multiple CJs would act as a stronger dynamical barrier than 
a single CJ. The Jupiter-Saturn pair in the Solar System might have prevented the Uranus and Neptune cores from 
migrating towards the Sun. Nevertheless, the CJ pairs we discovered in the Kepler-68 and Kepler-454 systems did not hinder 
the formation of the inner SPs Kepler-68b/c and Kepler-454b.
\item[(iv)] determine $f_{\rm CJ|SP}$ as a function of the  composition of SPs, which would be mainly ice-rich (sub-Neptunes) in the scenario of \citet{2015ApJ...800L..22I} (cf. also \citealt{2019PNAS..116.9723Z})
or rocky (super-Earths) in the scenario of \citet{2019A&A...627A..83L}; this is because they are expected to form 
beyond or inside the water iceline, respectively.
Unveiling a possible anti-correlation between CJs and predominantly ice-rich or rocky SPs
may thus yield important clues on the mechanisms of formation of short-period SPs. 
An anti-correlation between rocky super-Earths and CJs would not be expected instead, if the former mainly
originate inside rings of silicate-rich planetesimals at approximately 1~AU and then migrate inward \citep{2023NatAs...7..330B}. 
Actually, a positive correlation could exist if the mass initially in the ring of silicate-rich planetesimals, leading to rocky super-Earths, 
is correlated with the mass of the ice-rich planetesimals, leading to the formation of the cores of outer giant planets. 
In this regard, the improvement in the mass determination of several of the 64 Kepler and K2 transiting SPs 
from our RV analyses (Table~\ref{table_planet_parameters})
is useful to distinguish between rocky and non-rocky compositions.
\end{itemize}

\noindent
For these purposes, we recommend continuing to follow up on SP systems, even after achieving the desired precision on planetary masses 
for the investigation of SP compositions. 
This can be done at a very low cost, given that just a few RV measurements over the years would, in principle, be sufficient to search for outer giant planets, 
provided that the spectrograph is stable. For the brightest targets, additional, complementary information on the presence of CJs will also be provided by the Gaia mission (e.g., \citealt{2022arXiv220605439H}), starting with the publication of Data Release 4, slated for the end of 2025.

The 3661 HARPS-N RVs and activity indicators used for this work are released to the scientific community 
via the CDS\footnote{\url{http://vizier.u-strasbg.fr/cgi-bin/VizieR}} and DACE\footnote{\url{https://dace.unige.ch/dashboard/}}
databases to serve as the basis for further studies.

%\clearpage
%\newpage

%\bibliographystyle{aa} % style aa.bst
%\bibliography{Bonomoetal_2023_astroph} % your references Yourfile.bib

\begin{acknowledgements}
This work is based on observations made with the Italian Telescopio Nazionale Galileo (TNG)
operated on the island of La Palma by the Fundaci\'{o}n Galileo Galilei of the INAF at the Spanish Observatorio del Roque de los Muchachos
of the Instituto de Astrofisica de Canarias (GTO programme). 
The HARPS-N project was funded by the Prodex Program of the Swiss Space Office (SSO), the Harvard-University Origin of Life Initiative (HUOLI), the Scottish Universities Physics Alliance (SUPA), the University of Geneva, the Smithsonian Astrophysical Observatory (SAO), the Italian National Astrophysical Institute (INAF), the University of St. Andrews, Queen's University Belfast and the University of Edinburgh.
%, the University of St. Andrews, QueenÕs University Belfast and the University of Edinburgh. 
This paper is based on small-size planetary systems discovered by the \emph{Kepler} mission. 
Funding for the Kepler mission was provided by the NASA Science Mission directorate.
This work has made use of data from the European Space Agency (ESA) mission Gaia (\url{https://www.cosmos.esa.int/gaia}), 
processed by the Gaia Data Processing and Analysis Consortium (DPAC, \url{https://www.cosmos.esa.int/web/gaia/dpac/consortium}). 
Funding for the DPAC has been provided by national institutions, in particular the institutions participating in the Gaia Multilateral Agreement.
This project has received funding from the European Research Council (ERC) under the European Union's Horizon 2020 research and innovation program (grant agreement SCORE No 851555). This work has been carried out within the framework of the National Centre of Competence in Research PlanetS supported by the Swiss National Science Foundation under grants 51NF40\_182901 and 51NF40\_205606. The authors acknowledge the financial support of the SNSF. 
We acknowledge financial contribution from the agreement ASI-INAF n.2018-16-HH.0.
F.P.E. would like to acknowledge the Swiss National Science Foundation (SNSF) for supporting research with HARPS-N through the SNSF grants nr. 140649, 152721, 166227 and 184618. 
R.D.H. is funded by the UK Science and Technology Facilities Council (STFC)'s Ernest Rutherford Fellowship (grant number ST/V004735/1).
\end{acknowledgements}

\bibliographystyle{aa} % style aa.bst
\bibliography{46211} % your references Yourfile.bib

\clearpage
\onecolumn

%%%%%%%%%%%%%%%%%%%%%%%%%%%%%%%%%%%%%%%
%%%%%% TABLE OF STELLAR PARAMETERS AND RV DATASETS%%%%%%%%
%%%%%%%%%%%%%%%%%%%%%%%%%%%%%%%%%%%%%%%

\tiny

\begin{landscape} 

\renewcommand{\arraystretch}{1.2}

\begin{longtable}{l c c c c c c c c c c c} 
\caption{Kepler and K2 systems in our sample.}\\
%From left to right the columns report the name of the system, the multiplicity of transiting planets, the stellar parameters (mass, radius, effective temperature, metallicity and isochronological age), the number of radial velocities from all surveys ($N_{\rm RV}$~tot), the number of HARPS-N radial velocities ($N_{\rm RV}$~HN), the number of radial-velocity datasets ($N_{\rm Dat}$), the total duration of the radial-velocity time series, and the literature references for both the stellar parameters and the radial-velocity measurements.}\\
\hline
System & Multiple/Single     & $M_{\rm s}$ & $R_{\rm s}$ & $T_{\rm eff}$ & $\rm [Fe/H]$ & Age   & $N_{\rm RV}$   & $N_{\rm RV}$  & $N_{\rm Dat}$ & Duration        & System and/or literature   \\
             & Transiting System  & [$\Msun$]     & [$\Rsun$]     & [K]               &   [dex]             & [Gyr]  &  tot                      &  HN                     &                               & [day]                 & data reference  \\
\hline
\endfirsthead
\caption{continued.}\\
\hline 
System & Multiple/Single      & $M_{\rm s}$ & $R_{\rm s}$ & $T_{\rm eff}$ & $\rm [Fe/H]$ & Age   & $N_{\rm RV}$   & $N_{\rm RV}$  & $N_{\rm Dat}$ & Duration        & System and Data   \\
             & Transiting System & [$\Msun$]     & [$\Rsun$]     & [K]               &   [dex]             & [Gyr]  &  tot                       &  HN                     &                               & day           & Reference  \\
\hline
\endhead
\hline
\endfoot
Kepler-10 & m & $0.910\pm0.021$ & $1.065\pm0.009$ & $5708\pm28$ & $-0.15\pm0.04$ & $10.6_{-1.3}^{+1.5}$ & 291 & 291& 2$^1$ & 4021 & \refsys{2014ApJ...789..154D} \\ 
Kepler-19 & s & $0.936\pm0.040$ & $0.859\pm0.018$ & $5541\pm60$ & $-0.13\pm0.06$ & $1.9 \pm 1.7$ & 104 & 104& 2$^1$ & 2912 & \refsys{2011ApJ...743..200B}, \refsys{2017AJ....153..224M} \\ 
Kepler-20 & m & $0.929\pm0.053$ & $0.9164_{-0.0077}^{+0.0087}$ & $5495\pm50$ & $0.07\pm0.08$ & $5.6_{-3.5}^{+4.5}$ & 161 & 131& 2 & 3669 & \refsys{2016AJ....152..160B}, this work$^2$ \\ 
Kepler-21 & s &$1.408_{-0.030}^{+0.021}$ & $1.902_{-0.012}^{+0.018}$ & $6305\pm50$ & $-0.03\pm0.10$ & $2.84 \pm 0.35$ & 98 & 98& 1 & 1972 & \refsys{2012ApJ...746..123H}, \refsys{2016AJ....152..204L} \\ 
Kepler-22 & s & $0.857_{-0.043}^{+0.051}$ & $0.869 \pm 0.011$ & $5596\pm61$ & $-0.255\pm0.065$ & $7.0_{-4.2}^{+4.0}$ & 70 & 55& 3$^1$ & 3913 & \refsys{2019AJ....157...52B}, this work$^2$ \\ 
Kepler-37 & m & $0.790_{-0.030}^{+0.033}$ & $0.7890_{-0.0056}^{+0.0064}$ & $5357\pm68$ & $-0.36\pm0.05$ & $7.6_{-3.1}^{+3.4}$ & 145 & 114& 2 & 3424 & 7, \refsys{2021MNRAS.507.1847R} \\ 
Kepler-68 & m & $1.057_{-0.020}^{+0.022}$ & $1.2564 \pm 0.0084$ & $5847\pm75$ & $0.11\pm0.06$ & $6.84_{-1.04}^{+0.90}$ & 225 & 143 & 2 & 4521 &  \refsys{2019AJ....157..145M}, \refsys{Marginietalinprep} \\
Kepler-78 & s & $0.779_{-0.046}^{+0.032}$ & $0.7475_{-0.0078}^{+0.0077} $ & $5058\pm50$ & $-0.18\pm0.08$ & N.A. & 201 & 117 & 2 & 2312 &  \refsys{2013Natur.503..377P}, \refsys{2013Natur.503..381H}, \refsys{2019ApJ...883...79D} \\
Kepler-93 & s & $0.911\pm 0.033$ & $0.919\pm0.011 $ & $5669\pm75$ & $-0.18\pm0.10$ & $6.6\pm0.9$ & 153 & 121 & 2 & 4435 & \refsys{2014ApJS..210...20M}, \refsys{2015ApJ...800..135D}\\
Kepler-102 & m & $0.803 \pm 0.021$ & $0.724\pm0.018 $ & $4909\pm98$ & $0.11\pm0.04$ & $1.1_{-0.5}^{+3.6}$ & 146 & 74 & 3$^1$ & 3746 &  \refsys{2023AJ....165...74B} \\
Kepler-103 & m & $1.212_{-0.033}^{+0.024}$ & $1.492_{-0.022}^{+0.024}$ & $6009\pm64$ & $0.16\pm0.04$ & N.A. & 60 & 60 & 1 & 1594 &  \refsys{2019MNRAS.490.5103D} \\
Kepler-107 & m & $1.238\pm0.029$ & $1.447\pm0.014$ & $5854\pm61$ & $0.321\pm0.065$ & $4.29_{-0.56}^{+0.70}$ & 121 & 121 & 1 & 2180 &  \refsys{2019NatAs...3..416B} \\
Kepler-109 & m & $1.094_{-0.078}^{+0.086}$ & $1.387\pm0.021$ & $5950\pm62$ & $-0.020\pm0.065$ & $6.2_{-2.2}^{+2.7}$ & 66 & 51 & 2 & 3663 &  7, this work$^2$ \\
Kepler-323 & m & $1.015_{-0.070}^{+0.072}$ & $1.120_{-0.018}^{+0.020}$ & $6004\pm71$ & $-0.14\pm0.07$ & $5.4_{-2.8}^{+3.4}$ & 48 & 48 & 1 & 2237 &  this work$^2$ \\
Kepler-409 & s & $0.913_{-0.048}^{+0.050}$ & $0.897_{-0.011}^{+0.012}$ & $5421\pm64$ & $0.05\pm0.07$ & $5.8_{-3.6}^{+4.5}$ & 65 & 40 & 2 & 3362 &  7, this work$^2$ \\
Kepler-454 & s & $1.03_{-0.03}^{+0.04}$ & $1.066\pm0.012$ & $5687\pm49$ & $0.32\pm0.08$ & $5.25_{-1.39}^{+1.41}$ & 147 & 111 & 2 & 4404 & \refsys{2016ApJ...816...95G}  \\
Kepler-538 & s & $0.892_{-0.035}^{+0.051}$ & $0.8717_{-0.0061}^{+0.0064}$ & $5534\pm61$ & $-0.09\pm0.065$ & $5.3_{-3.0}^{+2.4}$ & 111 & 85 & 2 & 3336 & \refsys{2019AJ....158..165M}  \\
Kepler-1655 & s & $1.03\pm0.04$ & $1.03\pm0.02$ & $6148\pm71$ & $-0.24\pm0.05$ & $2.56\pm1.06$ & 97 & 97 & 1 & 1566 & \refsys{2018AJ....155..203H}  \\
Kepler-1876 & s & $1.187_{-0.088}^{+0.075}$ & $1.477_{-0.022}^{+0.021}$ & $6104\pm57$ & $0.01\pm0.06$ & $4.4_{-1.4}^{+2.1}$ & 70 & 70 & 1 & 464 & this work$^2$ \\
K2-2\,/\,HIP\,116454 & s & $0.755\pm0.027$ & $0.716\pm0.024$ & $5089\pm50$ & $-0.16\pm0.08$ & N.A. & 108 & 108 & 2 & 2367 & \refsys{2015ApJ...800...59V}  \\
K2-3 & m & $0.62\pm0.06$ & $0.60\pm0.06$ & $3835\pm70$ & $-0.01\pm0.09$ & $ > 1$ & 327 & 195 & 3 & 898 & \refsys{2018A&A...615A..69D}  \\
K2-12 & s & $0.965_{-0.048}^{+0.064}$ & $1.117\pm0.014$ & $5672\pm63$ & $0.00\pm0.07$ & $8.9_{-3.3}^{+3.0}$ & 50 & 50 & 1 & 2535 & this work$^2$ \\
K2-36 & m & $0.79\pm0.01$ & $0.718_{-0.006}^{+0.008}$ & $4916\pm37$ & $-0.09_{-0.04}^{+0.06}$ & N.A. & 86 & 86 & 1 & 1865 & \refsys{2019A&A...624A..38D}  \\
K2-38 & m &$1.054_{-0.063}^{+0.070}$ & $1.141\pm0.012$ & $5705\pm66$ & $0.235\pm0.070$ & $6.1_{-3.1}^{+3.4}$ & 97 & 40 & 4 & 1576 & \refsys{2020A&A...641A..92T}, this work$^2$  \\
K2-79 & s & $1.066_{-0.070}^{+0.057}$ & $1.265_{-0.027}^{+0.041}$ & $5897\pm118$ & $0.035\pm0.060$ & $6.5\pm1.3$ & 77 & 77 & 1 & 1517 & \refsys{2022AJ....163...41N}  \\
K2-96\,/\,HD\,3167 & m & $0.837_{-0.043}^{+0.053}$ & $0.880_{-0.013}^{+0.012}$ & $5261\pm60$ & $0.04\pm0.05$ & $7.8\pm4.3$ & 471 & 213 & 5 & 1940 & 13, \refsys{2017AJ....154..122C}, \refsys{2017AJ....154..123G}, \refsys{2022A&A...668A..31B}  \\
K2-106 & s & $0.950_{-0.048}^{+0.060}$ & $0.988\pm0.011$ & $5532\pm62$ & $0.11\pm0.07$ & $7.5_{-4.0}^{+3.8}$ & 111 & 44 & 4 & 1231 & \refsys{2017AJ....153..271S}, \refsys{2017A&A...608A..93G}, this work$^2$ \\
K2-110 & s & $0.738\pm0.018$ & $0.713\pm0.020$ & $5010\pm50$ & $-0.34\pm0.03$ & $ 8\pm3 $ & 32 & 15 & 2 & 1470 & \refsys{2017A&A...604A..19O}  \\
K2-111 & s & $0.84\pm0.02$ & $1.25\pm0.02$ & $5775\pm70$ & $-0.46\pm0.05$ & $ 13.5_{-0.9}^{+0.4}$ & 155 & 114 & 1 & 1590 & \refsys{2017A&A...604A..16F}, \refsys{2020MNRAS.499.5004M}  \\
K2-131 & s & $0.803\pm0.034$ & $0.7595_{-0.0089}^{+0.0087}$ & $5120\pm71$ & $-0.04\pm0.07$ & $ 5.3_{-3.7}^{+4.9}$ & 114 & 83 & 2 & 463 & \refsys{2017AJ....154..226D}, this work$^2$ \\
K2-135\,/\,GJ\,9827 & m & $0.606_{-0.014}^{+0.015}$ & $0.602_{-0.004}^{+0.005}$ & $4305\pm49$ & $-0.26\pm0.09$ & $ 10_{-5}^{+3}$ & 127 & 50 & 4 & 3922 & \refsys{2018AJ....155..148T}, \refsys{2018A&A...618A.116P}, \refsys{2019MNRAS.484.3731R} \\
K2-141 & m & $0.708\pm0.028$ & $0.681\pm0.018$ & $4570\pm100$ & $0.00\pm0.06$ & $ 6.3_{-4.7}^{+6.6}$ & 74 & 49 & 2 & 1501 & \refsys{2018AJ....155..107M}, \refsys{2018A&A...612A..95B} \\
K2-167 & s & $1.010_{-0.071}^{+0.081}$ & $1.494_{-0.034}^{+0.036}$ & $6011\pm60$ & $-0.40\pm0.06$ & $ 7.5_{-2.1}^{+2.4}$ & 82 & 82 & 1 & 2208 & this work$^2$ \\
K2-222 & s & $0.989_{-0.065}^{+0.070}$ & $1.115\pm0.029$ & $5942\pm119$ & $-0.315\pm0.060$ & $7.1_{-1.7}^{+1.5}$ & 70 & 70 & 1 & 1599 & 26  \\
K2-262\,/\,Wolf\,503 & s & $0.688_{-0.016}^{+0.023}$ & $0.689_{-0.020}^{+0.021}$ & $4716\pm60$ & $-0.47\pm0.08$ & $11\pm2$ & 84 & 25 & 4 & 1116 & \refsys{2018AJ....156..188P}, \refsys{2021AJ....162..238P}  \\
K2-263 &  s & $0.88\pm0.03$ & $0.85\pm0.02$ & $5368\pm44$ & $-0.08\pm0.05$ & $ 7\pm4$ & 95 & 95 & 1 & 1545 & \refsys{2018MNRAS.481.1839M}  \\
K2-312\,/\,HD\,80653 & s & $1.18\pm0.04$ & $1.22\pm0.01$ & $5959\pm61$ & $0.255\pm0.065$ & $2.67\pm1.20$ & 208 & 208 & 1 & 1068 & \refsys{2020A&A...633A.133F} \\
K2-418\,/\,EPIC-229004835 & s & $0.97\pm0.04$ & $0.999\pm0.008$ & $5868\pm60$ & $-0.12\pm0.05$ & $4.9_{-1.7}^{+1.9}$ & 126 & 126 & 1 & 845 & \refsys{Tronsgaardetalinprep} \\
% & & & & & & & & & & & & \\ 
% & & & & & & & & & & & & \\ 
%\hline
\label{table_system_parameters}
\end{longtable}
%\end{longtab}
%\vspace{-1.0 cm}
%\begin{tablenotes}[flushleft]
\tablefoot{From left to right, the columns report the name of the system, the multiplicity of transiting planets, the stellar parameters (mass, radius, effective temperature, metallicity and isochronological age), the number of radial velocities from all surveys ($N_{\rm RV}$~tot), the number of HARPS-N radial velocities ($N_{\rm RV}$~HN), the number of radial-velocity datasets ($N_{\rm Dat}$), the total duration of the radial-velocity time series, and the literature references for both the stellar parameters and the radial-velocity measurements. 
Table also available at the CDS.}
%\end{tablenotes}
\begin{flushleft}
\footnotemark[1]{ Two datasets were considered for the HARPS-N data, because the replacement of the red side of the HARPS-N CCD 
in late September 2012 resulted in a different zero point for the RVs gathered after that epoch.} 
\footnotemark[2]{ The new system parameters $M_{\rm s}$, $R_{\rm s}$, and age were derived with the public EXOFASTv2 tool by fitting the stellar SED and using the MIST evolutionary tracks. 
Gaussian priors were imposed on the $T_{\rm eff}$ and $\rm [Fe/H]$,
as derived from the analysis of the HARPS-N spectra, and on stellar parallax from Gaia EDR3 (see text for more details).}
\end{flushleft}
\tablebib{
% Kepler-10
(\refsysbib{2014ApJ...789..154D})~\citealt{2014ApJ...789..154D}; 
% Kepler-19
(\refsysbib{2011ApJ...743..200B})~\citealt{2011ApJ...743..200B};
(\refsysbib{2017AJ....153..224M})~\citealt{2017AJ....153..224M};
% Kepler-20
(\refsysbib{2016AJ....152..160B})~\citealt{2016AJ....152..160B}; 
% Kepler-21
(\refsysbib{2012ApJ...746..123H})~\citealt{2012ApJ...746..123H}; (\refsysbib{2016AJ....152..204L})~\citealt{2016AJ....152..204L}; 
% Kepler-22
(\refsysbib{2019AJ....157...52B})~\citealt{2019AJ....157...52B};
% Kepler-37
(\refsysbib{2021MNRAS.507.1847R})~\citealt{2021MNRAS.507.1847R}; 
% Kepler-68
(\refsysbib{2019AJ....157..145M})~\citealt{2019AJ....157..145M}; (\refsysbib{Marginietalinprep})~\citealt{Marginietalinprep};
% Kepler-78
(\refsysbib{2013Natur.503..377P})~\citealt{2013Natur.503..377P}; (\refsysbib{2013Natur.503..381H})~\citealt{2013Natur.503..381H};
(\refsysbib{2019ApJ...883...79D})~\citealt{2019ApJ...883...79D};
% Kepler-102
(\refsysbib{2015ApJ...800..135D})~\citealt{2015ApJ...800..135D}; (\refsysbib{2014ApJS..210...20M})~\citealt{2014ApJS..210...20M};
% Kepler-102
(\refsysbib{2023AJ....165...74B})~\citealt{2023AJ....165...74B};
%Kepler-103
(\refsysbib{2019MNRAS.490.5103D})~\citealt{2019MNRAS.490.5103D};
%Kepler-107
(\refsysbib{2019NatAs...3..416B})~\citealt{2019NatAs...3..416B};
%Kepler-454
(\refsysbib{2016ApJ...816...95G})~\citealt{2016ApJ...816...95G};
%Kepler-538
(\refsysbib{2019AJ....158..165M})~\citealt{2019AJ....158..165M};
% Kepler-1655
(\refsysbib{2018AJ....155..203H})~\citealt{2018AJ....155..203H};
% Kepler-1876
% K2-2
(\refsysbib{2015ApJ...800...59V})~\citealt{2015ApJ...800...59V};
% K2-3
(\refsysbib{2018A&A...615A..69D})~\citealt{2018A&A...615A..69D};
% K2-36
(\refsysbib{2019A&A...624A..38D})~\citealt{2019A&A...624A..38D};
% K2-38
(\refsysbib{2020A&A...641A..92T})~\citealt{2020A&A...641A..92T}; 
% K2-79
(\refsysbib{2022AJ....163...41N})~\citealt{2022AJ....163...41N};
% K2-96/HD3167
(\refsysbib{2017AJ....154..122C})~\citealt{2017AJ....154..122C}; (\refsysbib{2017AJ....154..123G})~\citealt{2017AJ....154..123G}; 
(\refsysbib{2022A&A...668A..31B})~\citealt{2022A&A...668A..31B}; 
% K2-106
(\refsysbib{2017AJ....153..271S})~\citealt{2017AJ....153..271S}; (\refsysbib{2017A&A...608A..93G})~\citealt{2017A&A...608A..93G}; 
% K2-110
(\refsysbib{2017A&A...604A..19O})~\citealt{2017A&A...604A..19O};
% K2-111
(\refsysbib{2017A&A...604A..16F})~\citealt{2017A&A...604A..16F}; (\refsysbib{2020MNRAS.499.5004M})~\citealt{2020MNRAS.499.5004M}; 
% K2-131
(\refsysbib{2017AJ....154..226D})~\citealt{2017AJ....154..226D};
% K2-135/GJ9827
(\refsysbib{2018AJ....155..148T})~\citealt{2018AJ....155..148T}; (\refsysbib{2018A&A...618A.116P})~\citealt{2018A&A...618A.116P}; 
(\refsysbib{2019MNRAS.484.3731R})~\citealt{2019MNRAS.484.3731R}; 
% K2-141
(\refsysbib{2018AJ....155..107M})~\citealt{2018AJ....155..107M}; (\refsysbib{2018A&A...612A..95B})~\citealt{2018A&A...612A..95B}; 
% K2-167
% K2-222
% K2-262
(\refsysbib{2018AJ....156..188P})~\citealt{2018AJ....156..188P}; (\refsysbib{2021AJ....162..238P})~\citealt{2021AJ....162..238P}; 
% K2-263
(\refsysbib{2018MNRAS.481.1839M})~\citealt{2018MNRAS.481.1839M};
% K2-312
(\refsysbib{2020A&A...633A.133F})~\citealt{2020A&A...633A.133F};
% EPIC-229004835
(\refsysbib{Tronsgaardetalinprep})~\citealt{Tronsgaardetalinprep}.
}

\renewcommand{\arraystretch}{1.0}

%%%%%%%%%%%%%%%%%%%%%%%%%%%%%%%%%%%%%%%
%% END OF TABLE OF STELLAR PARAMETERS AND RV DATASETS%%%%%%%%
%%%%%%%%%%%%%%%%%%%%%%%%%%%%%%%%%%%%%%%

%%%%%%%%%%%%%%%%%%%%%%%%%%%%%%%%%%%%%%%
%%%%%% TABLE OF RV DATA %%%%%%%%%%%%%%%%%%%%%%%%
%%%%%%%%%%%%%%%%%%%%%%%%%%%%%%%%%%%%%%%

\vspace{1.5 cm}
\begin{table}[h!]
\centering
\caption{HARPS-N measurements of radial velocity and activity indicators.}
%HARPS-N measurements of the radial velocities (RV) and activity indicators Full Width at Half Maximum (FWHM), 
%Contrast (C) and Bisector span (BIS) of the cross-correlation function, and the CaII H\&K Mount Wilson S-index ($S_{\rm MW}$)) and $\log{R^{'}_{HK}}$.}
\begin{tabular}{l c c c c c c c c c c c c c} 
\hline
System & Time  & RV & $\sigma_{\rm RV}$ & FWHM & $\sigma_{\rm FWHM}$ & C & $\sigma_{\rm C}$ & BIS & $\sigma_{\rm BIS}$ & 
$S_{\rm MW}$ & $\sigma_{\rm S_{\rm MW}}$ & $\log{R^{'}_{\rm HK}}$ & $\sigma_{\log{R^{'}_{\rm HK}}}$  \\
             & [$\rm BJD_{TDB}-2.45\mbox{\sc{e}06}$] & [\ms] & [\ms] & [\ms] & [\ms] & [\%] & [\%] & [\ms] & [\ms] & & & dex & dex \\
\hline
Kepler-10 & 6072.682383 & -98740.32 & 1.76 & 6762.45 & 9.56 & 45.65 & 0.05 & -23.88 & 2.49 & 0.163 & 0.004 & -4.99 & 0.02 \\
Kepler-10 & 6072.704768 & -98740.14 & 1.86 & 6753.98 & 9.55 & 45.70 & 0.05 & -27.56 & 2.64 & 0.170 & 0.005 & -4.96 & 0.02 \\
... & ... & ... & ... & ... & ... & ... & ... & ... & ... & ... & ... & ... & ... \\
\hline
\end{tabular}
\tablefoot{From left to right, che columns report the name of the system, the epoch of the observation, the radial velocity and its uncertainty, 
the activity indicators full width at half maximum (FWHM), contrast (C), and bisector span (BIS) of the cross-correlation function,
and the CaII H\&K Mount Wilson $S$ index ($S_{\rm MW}$) and $\log{R^{'}_{\rm HK}}$ along with their uncertainties. Data are available at both the CDS and DACE databases. A portion is shown here for guidance regarding its form and content.}
%\end{tablenotes}
\label{table_RV_data}
\end{table}

%%%%%%%%%%%%%%%%%%%%%%%%%%%%%%%%%%%%%%%
%%%%%% TABLE OF PRIORS ON GP HYPER-PARAMETERS%%%%%%%%%%
%%%%%%%%%%%%%%%%%%%%%%%%%%%%%%%%%%%%%%%

%\clearpage

\normalsize

\begin{table}[b!]
\centering 
\caption{Priors imposed on the hyper-parameters of the Gaussian process regression for the stars 
showing significant magnetic activity variations.} 
%The hyper-parameters of the covariance function are the radial-velocity semi-amplitude $h$, 
%the exponential decay time $\lambda_{1}$, the inverse harmonic complexity term $\lambda_{2}$, and the rotation period $P_ {\rm rot}$. 
%The second column indicate the use of either Squared Exponential (SE) or Quasi-Periodic (QP) kernel for the covariance function, 
%the former having just two hyper-parameters, i.e. $h$ and $\lambda_{1}$.
%$U$ stands for uniform (uninformative) prior.} 
\begin{tabular}{l c c c c c}
%\renewcommand{\footnoterule}{}      
%\centering
\hline %\hline
System & Adopted GP kernel & $h$     & $\lambda_{1}$ & $\lambda_{2}$ & $P_ {\rm rot}$  \\
             & SE or QP                 & [\ms]   & [day]                & -                       & [day]                  \\
\hline 
Kepler-21 & QP & $U[0, +\infty[$ & $U[0, 300]$ & $U[0.1, 5]$ & $U[10, 15]$ \\
Kepler-78 & QP & $U[0, +\infty[$ & $U[0, 300]$ & $U[0.1, 5]$ & $U[0, 20]$ \\
Kepler-93 & SE & $U[0, +\infty[$ & $U[0, 500]$ & - & - \\
Kepler-102 & QP &  $U[0, +\infty[$ & $\lambda_{1} > P_{\rm rot}/2 $ & $U[0.1, 5]$ & $U[15, 40]$ \\
K2-2\,/\,HIP116454 & SE & $U[0, +\infty[$ & $U[0, 50]$ & - & - \\
K2-3 & QP & $U[0, +\infty[$ & $U[20, 60]$ & $U[0.1, 5]$ & $U[35, 43]$ \\
K2-36 & QP & $U[0, +\infty[$ & $U[0, 300]$ & $U[0.1, 5]$ & $U[15, 20]$ \\
K2-131 & QP & $U[0, +\infty[$ & $U[0, 300]$ & $U[0.1, 5]$ & $U[5, 15]$ \\
K2-135\,/\,GJ\,9827 & QP & $U[0, +\infty[$ & $U[0, 300]$ & $U[0.1, 5]$ & $U[25, 35]$ \\
K2-141 & QP & $U[0, +\infty[$ & $\lambda_{1} > P_{\rm rot}/2 $ & $U[0.1, 5]$ & $U[10, 20]$ \\
K2-312\,/\,HD\,80653 & QP & $U[0, +\infty[$ & $\lambda_{1} > P_{\rm rot}/2 $ & $U[0.1, 5]$ & $U[15, 25]$ \\
% & & & & & & & & & & & & \\ 
% & & & & & & & & & & & & \\ 
\hline
\end{tabular}
%\begin{tablenotes}[flushleft]
%\item {From left to right the columns report the name of the system, the adopted kernel for the covariance function 
%-- either Squared Exponential (SE) or Quasi-Periodic (QP) kernel -- and the priors on the hyper-parameters of the covariance function, 
%namely the radial-velocity semi-amplitude $h$, the exponential decay time $\lambda_{1}$, 
%the inverse harmonic complexity term $\lambda_{2}$, and the rotation period $P_ {\rm rot}$ 
%(the SE kernel has only the two hyper-parameters $h$ and $\lambda_{1}$). $U$ stands for uniform (uninformative) prior.}
%\end{tablenotes}
\tablefoot{From left to right, the columns report the name of the system, the adopted kernel for the covariance function 
-- either squared exponential (SE) or quasi-periodic (QP) kernel -- and the priors on the hyper-parameters of the covariance function, 
namely the radial-velocity semi-amplitude, $h$, the exponential decay time, $\lambda_{1}$, 
the inverse harmonic complexity term, $\lambda_{2}$, and the rotation period, $P_ {\rm rot}$ 
(the SE kernel has only the two hyper-parameters $h$ and $\lambda_{1}$). $U$ stands for uniform (uninformative) prior.}
\label{table_priors_GP_hyperparameters}
\end{table}

%%%%%%%%%%%%%%%%%%%%%%%%%%%%%%%%%%%%%%%
%% END OF TABLE OF PRIORS ON GP HYPER-PARAMETERS%%%%%%%%%%
%%%%%%%%%%%%%%%%%%%%%%%%%%%%%%%%%%%%%%%

\end{landscape}

\newpage

%%%%%%%%%%%%%%%%%%%%%%%%%%%%%%%%%%%%%%%%%%%%%
%% TABLE OF RESULTS ON STELLAR JITTERS, GP HYPER-PARAMETERS, ACCELERATIONS%%
%%%%%%%%%%%%%%%%%%%%%%%%%%%%%%%%%%%%%%%%%%%%%

\tiny

\begin{landscape}

%\vspace{-0.25 cm}
\setlength{\tabcolsep}{3pt}
\centering 
\begin{longtable}{l c c c c c c c c c c c}
\caption{HARPS-N systemic radial velocities and jitter terms, Gaussian process hyper-parameters, and linear accelerations from the 
DE-MCMC radial-velocity modeling.}
\\
\hline
System & $\gamma_{\rm HN-1}$  & $\gamma_{\rm HN-2}$ & $\sigma_{\rm jit, HN-1}$ & $\sigma_{\rm jit, HN-2}$ & $h$     & $\lambda_{1}$ & $\lambda_{2}$ & $P_ {\rm rot}$ & $\dot\gamma$                    &  $\Delta BIC>10$  &  Comment \\
             & [\ms]                               &  [\ms]                            &  [\ms]                             &  [\ms]                             &  [\ms]  & [day]                & -                       & [day]               &   [$ \rm m\,s^{-1}\,day^{-1}$]                 &                               &        \\
\hline 
%
%
%
%
%
%\hline
%
%
Kepler-10 & $-98737.12\pm0.47$ & $0.09\pm0.17$$^1$ & $2.52_{-0.41}^{+0.44}$ & $2.13\pm0.15$ &  -  & - & - & - & $-2.7~\mbox{\sc{e}-04}\pm1.6~\mbox{\sc{e}-04}$  & - &  \\
Kepler-19 & $-10614.72\pm1.31$  & $-10762.08 _{-0.74}^{+0.76}$ & $4.78_{-0.96}^{+1.20}$ & $ 3.51_{-0.42}^{+0.47}$ & - & - & - & - & $1.7~\mbox{\sc{e}-03}\pm0.9~\mbox{\sc{e}-03}$ & - &  \\
Kepler-20 & - & $-21070.23\pm 0.95$ & - & $ 4.06_{-0.41}^{+0.45}$ & - & - & - & - & $0.7~\mbox{\sc{e}-03}\pm1.2~\mbox{\sc{e}-03}$ & - &  \\
Kepler-21 & - & $-19322.87_{-1.41}^{+1.38}$ & - & $2.61 \pm 0.44$ & $4.28_{-0.89}^{+1.45}$ & $46_{-21}^{+90}$ & $0.88_{-0.25}^{+0.47}$ & $13.42_{-0.47}^{+0.24}$ & $-5.6_{-1.7}^{+1.9}~\mbox{\sc{e}-03}$ & - &  \\
Kepler-22 & $-4076.15_{-1.81}^{+1.91}$ & $-4221.01\pm0.99$ & $5.09_{-1.45}^{+1.97}$ & $2.47_{-0.58}^{+0.62}$  &  -  & - & - & - & $1.8~\mbox{\sc{e}-04}\pm8.0~\mbox{\sc{e}-04}$ & - &  \\
Kepler-37 & - & $-30820.12\pm0.56$ & - & $2.59_{-0.19}^{+0.21}$  &  -  & - & - & - & $6.9~\mbox{\sc{e}-04}\pm7.4~\mbox{\sc{e}-04}$ & - &  \\
Kepler-68 & - & $-20903.57_{-0.36}^{+0.45}$ & - & $1.81\pm 0.17$  &  -  & - & - & - & $-1.2_{-0.48}^{+0.55}~\mbox{\sc{e}-03}$ & - & 2 CJs (ro) \\
Kepler-78 & - & $-3530.11\pm3.21$ & - & $1.65 \pm 0.33$ & $8.90_{-1.49}^{+2.13}$ & $26.7_{-5.9}^{+6.8}$ & $0.475_{-0.076}^{+0.086}$ & $12.92\pm0.12$ & $-1.6~\mbox{\sc{e}-03}\pm3.2~\mbox{\sc{e}-03}$ & - &  \\
Kepler-93 & - & $27245.19_{-1.12}^{+1.23}$ & - & $1.25 \pm 0.16$ & $3.23_{-0.73}^{+1.07}$ & $130_{-34}^{+46}$ & - & - & \boldmath{$3.92~\mbox{\sc{e}-02}\pm0.11~\mbox{\sc{e}-02}$} & y & BD/LMS  \\
Kepler-102 & $-28152.22\pm1.50$ & $-28163.26\pm1.06$ & $<2.20$ & $<0.72$ & $3.26_{-0.50}^{+0.54}$ & $21.3_{-4.6}^{+8.9}$ & $0.73_{-0.16}^{+0.20}$ & $27.95_{-1.67}^{+3.64}$ & $-7.3~\mbox{\sc{e}-04}\pm7.0~\mbox{\sc{e}-04}$ & - &  \\
Kepler-103 & - & $-28609.58\pm0.71$ & - & $2.26_{-1.15}^{+0.98}$ &  -  & - & - & - & $-2.3~\mbox{\sc{e}-03}\pm1.8~\mbox{\sc{e}-03}$ & - &  \\
Kepler-107 & - & $5536.53\pm0.87$ & - & $<2.34$  &  -  & - & - & - & $-1.2~\mbox{\sc{e}-03}\pm1.4~\mbox{\sc{e}-03}$ & - &  \\
Kepler-109 & - & $-37129.88 \pm1.53$ & - & $2.85_{-1.19}^{+1.06}$ &  -  & - & - & - & $-2.9~\mbox{\sc{e}-03}\pm2.0~\mbox{\sc{e}-03}$ & - &  \\
Kepler-323 & - & $13349.22_{-1.64}^{+1.61}$ & - & $< 2.86$ &  -  & - & - & - & $3.2~\mbox{\sc{e}-03}\pm1.4~\mbox{\sc{e}-03}$ & - &  \\
Kepler-409 & - & $ -28541.49 \pm 0.57$ & - & $2.39_{-0.29}^{+0.36}$  &  -  & - & - & - & $0.0~\mbox{\sc{e}-03}\pm0.4~\mbox{\sc{e}-03}$ & - &  \\
Kepler-454 & - & $ -71463.30_{-2.2}^{+1.2}$ & - & $2.26_{-0.33}^{+0.35}$  &  -  & - & - & - & \boldmath{$9.15_{-1.3}^{+2.6}~\mbox{\sc{e}-03}$} & y & CJ/BD/LMS + 2 CJs (ro)  \\
Kepler-538 & - & $-37467.97_{-0.85}^{+0.80}$ & - & $2.34_{-0.29}^{+0.32}$  &  -  & - & - & - & $-0.2~\mbox{\sc{e}-03}\pm1.1~\mbox{\sc{e}-03}$ & - &   \\
Kepler-1655 & - & $-40769.17 \pm 0.64$ & - & $<0.91$  &  -  & - & - & - & $1.8~\mbox{\sc{e}-03}\pm1.0~\mbox{\sc{e}-03}$ & - & \\
Kepler-1876 & - & $7681.39\pm0.45$ & - & $1.87_{-0.36}^{+0.39}$  &  -  & - & - & - & $7.8~\mbox{\sc{e}-03}\pm3.2~\mbox{\sc{e}-03}$ & - &  \\
K2-2\,/\,HIP\,116454 &  - & $-3299.73\pm0.61$$^2$ & - & $0.97 \pm 0.45$ & $2.72_{-0.42}^{+0.47}$ & $4.15_{-1.05}^{+2.44}$ & - & - & $0.2~\mbox{\sc{e}-03}\pm1.4~\mbox{\sc{e}-03}$ & - &  \\
K2-3 & - & $-0.06\pm0.55$$^3$ & - & $0.98_{-0.31}^{+0.24}$ & $2.60_{-0.28}^{+0.32}$ & $41.6_{-8.8}^{+9.8}$ & $0.42_{-0.12}^{+0.21}$ & $40.71_{-1.36}^{+0.87}$ & $-0.4~\mbox{\sc{e}-03}\pm1.7~\mbox{\sc{e}-03}$ & - & \\
K2-12 & - & $17000.18 \pm 0.99$ & - & $4.36_{-0.96}^{+1.02}$  &  -  & - & - & - & \boldmath{$1.884~\mbox{\sc{e}-02}\pm0.096~\mbox{\sc{e}-02}$} & y & CJ/BD/LMS  \\
K2-36 & - & $13625.38 \pm 4.25$ & - & $2.64 \pm 1.06$ & $15.5_{-2.6}^{+3.8}$ & $99_{-29}^{+41}$ & $0.67_{-0.12}^{+0.13}$ & $17.13_{-0.10}^{+0.07}$ & $3.7~\mbox{\sc{e}-03}\pm7.3~\mbox{\sc{e}-03}$ & - & \\
K2-38 & - & $-36040.15\pm0.56$ & - & $2.07\pm0.67$  &  -  & - & - & - & $-8.2~\mbox{\sc{e}-03}\pm3.1~\mbox{\sc{e}-03}$ & - & \\
K2-79 & - & $-10314.99\pm0.76$ & - & $<1.74$  &  -  & - & - & - & $1.8~\mbox{\sc{e}-03}\pm1.3~\mbox{\sc{e}-03}$ & - & \\
K2-96\,/\,HD\,3167 & - & $19371.36 \pm0.24$ & - & $1.43_{-0.09}^{+0.10}$  &  -  & - & - & - & \boldmath{$1.32_{-0.30}^{+0.28}~\mbox{\sc{e}-03}$} & y & Activity \\
K2-106 & - & $-15735.71\pm1.12$ & - & $< 1.17$  &  -  & - & - & - & $-1.4~\mbox{\sc{e}-03}\pm2.1~\mbox{\sc{e}-03}$ & - &   \\
K2-110 & - & $-21798.17  \pm0.93$ & - & $<1.91$  &  -  & - & - & - & $0.7~\mbox{\sc{e}-03}\pm1.5~\mbox{\sc{e}-03}$ & - &    \\
K2-111 & - & $-16291.29\pm0.26$ & - & $<0.93$  &  -  & - & - & - & $-1.8~\mbox{\sc{e}-03}\pm0.5~\mbox{\sc{e}-03}$ & n &  \\
K2-131 &  - & $6855.45\pm6.17$ & - & $< 1.59$ & $21.38_{-3.35}^{+4.68}$ & $31.9_{-7.3}^{+9.8}$ & $0.435\pm0.074$ & $9.194\pm0.065$ & $3.8~\mbox{\sc{e}-02}\pm3.3~\mbox{\sc{e}-02}$ & - & \\
K2-135\,/\,GJ\,9827 &  - & $ 31940.16_{-1.33}^{+1.38}$ & - & $<0.95$ & $3.83_{-0.65}^{+0.82}$ & $54_{-29}^{+48}$ & $0.61_{-0.14}^{+0.17}$ & $28.90_{-0.70}^{+1.09}$ & $-0.6~\mbox{\sc{e}-03}\pm1.3~\mbox{\sc{e}-03}$ & - & \\
K2-141 & - & $-3408.80\pm4.92$ & - & $< 1.17$ & $12.75_{-3.13}^{+6.11}$ & $16.4_{-6.5}^{+16.2}$ & $0.94_{-0.23}^{+0.36}$ & $15.17_{-2.36}^{+3.63}$ & $-1.0_{-7.4}^{+7.1}~\mbox{\sc{e}-03}$ & - & \\
K2-167 & - & $-17550.44 \pm0.50$ & - & $2.36_{-0.24}^{+0.20}$  &  -  & - & - & - & $1.2~\mbox{\sc{e}-03}\pm0.6~\mbox{\sc{e}-03}$ & - & \\
K2-222 & - & $1681.65\pm0.46$ & - & $1.79_{-0.31}^{+0.34}$  &  -  & - & - & - & $5.8~\mbox{\sc{e}-04}\pm7.1~\mbox{\sc{e}-04}$ & - & \\
K2-262\,/\,Wolf\,503 &  - & $-46781.43\pm0.46$ & - & $1.97_{-0.34}^{+0.42}$  &  -  & - & - & - & \boldmath$6.6~\mbox{\sc{e}-03}\pm1.0~\mbox{\sc{e}-03}$ & y & Activity \\
K2-263 &  - & $29837.72\pm0.29$ & - & $1.50\pm0.37$  &  -  & - & - & - & $4.1~\mbox{\sc{e}-04}\pm4.9~\mbox{\sc{e}-04}$ & - &  \\
K2-312\,/\,HD\,80653 & - & $8323.37_{-0.71}^{+0.76}$ & - & $0.64\pm0.30$ & $2.86_{-0.25}^{+0.28}$ & $18.8_{-3.5}^{+4.1}$ & $0.46_{-0.06}^{+0.07}$ & $19.55_{-0.50}^{+0.57}$ & $0.0\pm1.4$ & - & 1 CJ (ro) \\
K2-418\,/\,EPIC-229004835 &  - & $-31487.88\pm0.30$ & - & $2.48\pm0.24$  &  -  & - & - & - & $-2.5~\mbox{\sc{e}-03}\pm1.1~\mbox{\sc{e}-03}$ & - &  \\
% & & & & & & & & & & & & \\ 
% & & & & & & & & & & & & \\ 
\hline
%\end{tabular}
\label{table_accelerations}
\end{longtable}
%\tiny
%\vspace{-0.50 cm}
\tablefoot{From left to right, the columns report the name of the system, the systemic radial velocity ($\gamma$), 
the jitter terms ($\sigma_{\rm jit}$), the Gaussian process hyper-parameters -- namely 
radial-velocity semi-amplitude ($h$), exponential decay time ($\lambda_{1}$), inverse harmonic complexity term ($\lambda_{2}$), and 
rotation period ($P_ {\rm rot}$) -- , the linear acceleration ($\dot\gamma$) and the $\Delta BIC$ in favor of the model with acceleration.
Even though, for simplicity, $\gamma$ and $\sigma_{\rm jit}$ are reported 
for the HARPS-N spectrograph only, the DE-MCMC radial-velocity modeling was carried out by including the data gathered with other spectrographs, 
when available. Accelerations $\dot\gamma$ evaluated as significant, from both a confidence level higher than $3\sigma$  and 
an odds ratio greater than 10 from the $\Delta BIC$, are highlighted in boldface. 
HN-1 and HN-2 refer to the HARPS-N data collected before and after September 2012, respectively. The last column reports comments 
about the origin of the detected trends, such as possible cold Jupiter (CJ), brown dwarf (BD), or low-mass star (LMS) companions or stellar magnetic activity, and about the presence of cold Jupiters in resolved orbits (ro), which were modeled with Keplerians and are thus not responsible for the trends.}
%\vspace{-0.25 cm}
\begin{flushleft}
\footnotemark[1]{The radial-velocity zero point is close to zero because the radial velocities were extracted with the Yarara-v2 tool \citep{2022A&A...659A..68C} from the HARPS-N spectra reduced with the DRS-v2.3.5 (see \citealt{Bonomoetalinprep}).}
\footnotemark[2]{The last observing season was fitted as an independent dataset for the presence of an offset in the RVs of very likely activity origin. The systemic velocity and jitter of this season were found to be $\gamma=-3292.51_{-1.23}^{+1.26}$~\ms and $\sigma_{\rm jit}=1.70_{-1.13}^{+1.43}$~\ms.} 
\footnotemark[3]{The radial-velocity zero point is close to zero because the radial velocities were extracted with the TERRA software (see text for details). The systemic velocity is $30149.2_{-0.7}^{+0.6}$~\ms \citep{2018A&A...615A..69D}.}
\end{flushleft}
\clearpage

%%%%%%%%%%%%%%%%%%%%%%%%%%%%%%%%%%%%%%%%%%%%%%%%%
%% END OF TABLE OF RESULTS ON STELLAR JITTERS, GP HYPER-PARAMETERS, ACCELERATIONS%%
%%%%%%%%%%%%%%%%%%%%%%%%%%%%%%%%%%%%%%%%%%%%%%%%%

%\end{landscape}

%\clearpage
%\newpage

%\begin{landscape}

\normalsize

\begin{table}
\centering
\caption{Orbital parameters and minimum masses of the companions causing the long-term trends observed in the Kepler-93, Kepler-454, and K2-12 systems.}
\begin{tabular}{l c c c c} 
\hline
System & $\rm P$  &  $\rm a$  & $ \rm K$  & $M_{\rm p} \sin{i}$  \\
             & [yr]          &    [AU]     & [\ms]                  & [$\rm M_{Jup}$]  \\
\hline
Kepler-93   & $>48.6$    &  $>13.0$     & $>174$     &  $>21$ \\
Kepler-454 & $>48.2$    &  $>13.4$     & $>40.3$    &  $>5.3$ \\
K2-12         & $>27.8$    &  $>9.1$       & $>47.8$    &  $>5.0$ \\
\hline
\end{tabular}
\tablefoot{From left to right, the columns report the name of the system and the lower limits to orbital period, semi-major axis, RV semi-amplitude and minimum mass of the companions.}
\label{table_companion_trends}
\end{table}

\begin{table} 
\centering
\caption{Parameters of the long-period non-transiting giant planets with resolved orbit.}
\begin{tabular}{l c c c c c c c} 
\hline
Name    & $T_{\rm c}$                                               & $P $ & $a$  &  $e$ & $\omega$  &$K$        &  $M_{\rm p} \sin{i}$  \\ 
             & [$\rm BJD_{TDB}-2.45\mbox{\sc{e}06}$] & [day] & [AU] &          & [deg]         & [\ms]        &  $[\rm M_{Jup}]$ \\
\hline
Kepler-68d & $5892.0 \pm 5.4 $  &  $632.62 \pm 1.03$   & $ 1.469 \pm 0.010$ & $0.102\pm0.016$ & $256.4\pm10.4$  & $17.20 \pm 0.29$ & $0.749 \pm 0.017$\\
Kepler-68e &  $7791_{-126}^{+134}$ & $3455_{-169}^{+348}$ & $ 4.60_{-0.16}^{+0.32}$ & $0.33\pm0.11$ & $196_{-20}^
{+17}$ & $3.73\pm0.39$&  $0.272 \pm 0.032 $\\
Kepler-454c & $7151.87 \pm 0.69$ & $524.19 \pm 0.20$ & $1.287 \pm 0.017$ & $<0.0053$ & -  & $111.58 \pm 0.55$ & $4.51 \pm 0.12$\\
Kepler-454d & $8433\pm26$ & $4073_{-186}^{+399}$  & $5.10_{-0.19}^{+0.34}$ & $0.089_{-0.027}^{+0.036}$ & $117_{-21}^{+25}$ & $28.4_{-1.3}^{+2.4}$ &  $2.31_{-0.16}^{+0.27}$\\
K2-312c\,/\,HD\,80653c &$8968.80 \pm 0.41$ & $921.2\pm10.8$ & $1.961 \pm 0.027$&  $0.853\pm0.011$ & $41.6\pm2.2$ &  $194_{-13}^{+17}$& $5.41_{-0.44}^{+0.52}$\\
\hline
\end{tabular}
\tablefoot{From left to right, the columns report the name of the system, the inferior conjunction time, the orbital period, the semi-major axis,
the eccentricity, the argument of periastron, the radial-velocity semi-amplitude, and the planet minimum mass.}
\label{table_giant_planets}
\end{table}

%\tiny

\begin{table}
\centering
\caption{Completeness (recovery rate) and fraction of stars hosting both short-period small planets and cold Jupiters
for different intervals of orbital separation.}
\begin{tabular}{c c c c c}
\hline
Orbital separation [AU] & Completeness [$\%$] & $N_{\star, \rm eff}$ & $N_{\star, \rm CJ}$ & $f_{\rm CJ|SP} [\%]$ \\
\hline
$1-2$   & 96.9 & 35.9 & 2 & $5.6^{+6.5}_{-1.8}$ \\
$2-4$   & 92.2  & 34.1 & 2 & $5.9^{+6.8}_{-1.9}$ \\
$4-10$ & 77.4  & 28.6 & 0 & $<3.80$ \\
$1-10$ & 87.9  & 32.4 & 3 & $9.3^{+7.7}_{-2.9}$ \\
\hline
\end{tabular}
\tablefoot{From left to right, the columns report the interval of semi-major axis, the survey completeness, 
the effective number of stars ($N_{\star, \rm eff} \leq 37$), the number of systems with detected cold Jupiters ($N_{\star, \rm CJ}$) in the sample, 
and the fraction of stars hosting both short-period small planets and cold Jupiters ($f_{\rm CJ|SP}$).}
\label{completeness_frequency}
\end{table}

\begin{table}
\renewcommand{\arraystretch}{1.20}
\vspace{-0.1cm}
\centering
\caption{Fraction of stars with cold Jupiters in our sample compared with those from previous works.}
\begin{tabular}{c c c c c c } 
\hline
Planetary Mass [$\rm M_{Jup}$]  & Orbital separation [AU]       & $f_{\rm CJ|SP} [\%]$              &  $f_{\rm CJ|SP} [\%]$               & $f_{\rm CJ} [\%]$  & $f_{\rm CJ|SP} [\%]$~$^1$  \\
                                                        &                                               & from Keplerians                      &  from Keplerians and trends      & \citep{2020MNRAS.492..377W}          & \citep{2019AJ....157...52B}  \\
\hline
$0.3-13$  &       $1-2$     & $5.6^{+6.5}_{-1.8}$  & -                                   & $8.0^{+3.7}_{-2.2}$       & -\\
$0.3-13$  &       $2-4$     & $5.9^{+6.8}_{-1.9}$  & -                                   & $5.3^{+2.8}_{-1.5}$       & -  \\
$0.3-13$  &       $4-10$   & $<3.8$                      & -                                   & $6.9^{+4.2}_{-2.1}$ & - \\
$0.3-13$  &       $1-10$   & $9.3^{+7.7}_{-2.9}$  &  $12.3^{+8.1}_{-3.7}$  & $20.2^{+6.3}_{-3.4}$    & - \\
\hline
$0.5-13$  &       $1-10$   & $8.8^{+7.4}_{-2.8}$  &  $11.8^{+7.7}_{-3.5}$  & -                               & $36^{+7}_{-6}$ \\
$0.5-13$  &       $1-20$   & $8.3^{+7.0}_{-2.6}$  &  $11.1^{+7.4}_{-3.3}$  & -                               & $41^{+8}_{-7}$ \\
\hline
$0.5-20$  &       $1-10$   & $8.7^{+7.3}_{-2.7}$  &  $11.3^{+7.7}_{-3.4}$  & -                               & $38 \pm 7$ \\
$0.5-20$  &       $1-20$   & $9.6^{+7.9}_{-3.0}$  &  $12.7^{+8.3}_{-3.9}$  & -                               & $39 \pm 7$ \\
\hline
\end{tabular}
\tablefoot{From left to right, the columns report the intervals of planetary mass and semi-major axis; the fraction of stars with cold Jupiters ($f_{\rm CJ|SP}$) in our sample by considering Keplerian signals only (third column) as well as including the long-term trend of K2-12 (fourth column), because it is still compatible with a planetary companion; the occurrence rates of cold Jupiters derived by \citet{2020MNRAS.492..377W} and \citet{2019AJ....157...52B}.}
\begin{flushleft}
\tiny
\footnotemark[1]{$f_{\rm CJ|SP}$ for $1<M_{\rm p}<10~\rm M_\oplus$ instead of the wider range $1<M_{\rm p}<20~\rm M_\oplus$ used in 
both this work and \citet{2018AJ....156...92Z}.} \\
\end{flushleft}
\label{frequency_comparison}
\end{table}

\renewcommand{\arraystretch}{1.0}

\begin{table} 
\centering
\caption{Parameters of the non-transiting small planets.}
\begin{tabular}{ l c c c  c c c } 
\hline
Name    & $T_{\rm c}$                                               & $P $ & $a$  &  $e$ & $K$        &  $M_{\rm p} \sin{i}$  \\ 
             & [$\rm BJD_{TDB}-2.45\mbox{\sc{e}06}$] & [day] & [AU] &         & [\ms]        &  $[\rm M_{\oplus}]$ \\
\hline
Kepler-10d & $ 7165.4_{-5.3}^{+4.7}$   & $151.04\pm0.45$               & $0.5379 \pm 0.0044$  & $<0.26$ & $1.68\pm0.28$ & $12.68\pm2.24$\\  
Kepler-1876c$^1$ & $ 6988.65_{-1.8}^{+1.0}$& $15.76_{-0.16}^{+0.10}$ & $0.1302 \pm 0.0033 $& $<0.10$ &  $2.52\pm0.59$&   $11\pm2.7$\\ 
K2-96d\,/\,HD\,3167d & $ 7745.37 \pm 0.18$ & $8.4112\pm0.0052$ & $0.0763\pm0.0015$     & $<0.12$ &  $1.55\pm0.15$     &   $4.33 \pm 0.45$\\
K2-96e\,/\,HD\,3167e & $7738.43 \pm 2.67$ & $96.63\pm0.29$$^2$ & $0.3885\pm0.0079$ & $<0.15$ &  $1.33\pm0.16$  &   $8.41 \pm 1.02$\\
K2-111c & $8119.62 \pm 0.33$ & $15.6805\pm 0.0064$ & $0.11569\pm0.00094$  & $< 0.071$ & $3.19\pm0.29$ &   $11.08 \pm 1.03$\\
\hline
\end{tabular}
\tablefoot{From left to right, the columns report the name of the planet, the inferior conjunction time, the orbital period, the semi-major axis, the eccentricity, the radial-velocity semi-amplitude, and the planet minimum mass. The planet candidate Kepler-1876c is also included.}
\begin{flushleft}
\tiny
\footnotemark[1]{Planet candidate.} 
\footnotemark[2]{Other orbital periods are possible due to aliasing effects (see also \citealt{2022A&A...668A..31B}).}\\
\end{flushleft}
\label{table_nontransiting_planets}
\end{table}

\end{landscape}

\clearpage
%\newpage

%\twocolumn

%\twocolumn

\begin{appendix}

%\onecolumn

\tiny

\begin{landscape}

\section{Orbital and physical parameters of the 64 transiting Kepler and K2 planets.}

\setlength{\tabcolsep}{2pt}
\renewcommand{\arraystretch}{1.35}

\vspace{-0.5 cm}
\begin{longtable}{l c c c c c c c c c c c c c}
\caption{Orbital and physical parameters of the 64 transiting Kepler and K2 planets.}\\
\hline %\hline
Name & $T_{\rm c}$                                                  & $P$     & $R_{\rm p}$           &   $i$      & Ref.           & $e$ & $K$   & $ M_{\rm p}$   & $\rho_{\rm p}$                 & $\log{g_{\rm p}}$ & $a$   & $T_{\rm eq}$ & $F_{\rm p}$ \\
          & [$\rm BJD_{TDB}-2.45\mbox{\sc{e}06}$]  & [day]    & [$\rm R_{\oplus}$] &   [deg]   & transits     &           & [\ms]  & [$\rm M_{\oplus}$] &  [$\rm g\;cm^{-3}$] &  [cgs]                  & [AU] &  [K]        & [$\rm F_{\oplus}$] \\
\hline
\endfirsthead
\caption{continued.}\\
\hline 
Name & $T_{\rm c}$                                                  & $P$     & $R_{\rm p}$           &   $i$      & Ref.           & $e$ &  $K$   & $ M_{\rm p}$   & $\rho_{\rm p}$                 & $\log{g_{\rm p}}$ & $a$   & $T_{\rm eq}$ & $F_{\rm p}$ \\
          & [$\rm BJD_{TDB}-2.45\mbox{\sc{e}06}$]  & [day]    & [$\rm R_{\oplus}$] &   [deg]   & transits     &           & [\ms]  & [$\rm M_{\oplus}$] &  [$\rm g\;cm^{-3}$] &  [cgs]                  & [AU] &  [K]        & [$\rm F_{\oplus}$] \\
\hline
\endhead
\hline
\endfoot
%
%
%\hline
Kepler-10b & $5034.08687(18)$ & $0.83749070(20)$ & $1.470_{-0.020}^{+0.030}$ & $84.8_{-3.9}^{+3.2}$ & \reftr{2014ApJ...789..154D} &0(fixed) & $2.34\pm0.21$ & $3.26\pm0.30$ & $5.57_{-0.59}^{+0.61}$ & $3.166_{-0.045}^{+0.042}$ & $0.01685\pm0.00013$ & $2188\pm16$ & $3820\pm120$ \\ 
Kepler-10c & $5062.26648(81)$ & $45.294301(48)$ & $2.355\pm0.022$ & $89.623\pm0.011$ &  1, \reftr{Bonomoetalinprep}  &$0.130\pm0.050$ & $2.19\pm0.24$ & $11.4\pm1.3$ & $4.78_{-0.54}^{+0.56}$ & $3.302_{-0.051}^{+0.047}$ & $0.2410\pm0.0019$ & $578\pm4$ & $18.68\pm0.57$ \\ 
Kepler-19b & $4959.7074(14)$ & $9.2869900(00)$ & $2.209\pm0.048$ & $89.940_{-0.44}^{+0.060}$ & \reftr{2011ApJ...743..200B}, \reftr{2017AJ....153..224M}  & $<0.41$ & $2.13_{-0.81}^{+0.90}$ & $6.1_{-2.7}^{+2.8}$ & $3.1\pm1.4$ & $3.09_{-0.25}^{+0.16}$ & $0.0846\pm0.0012$ & $851\pm14$ & $87.6\pm6.0$ \\ 
Kepler-20b & $4967.50225(33)$ & $3.6961049(16)$ & $1.773_{-0.030}^{+0.053}$ & $87.36_{-1.6}^{+0.22}$ & \reftr{2016AJ....152..160B}, \reftr{2019RAA....19...41G}   &$<0.083$ & $4.23\pm0.54$ & $9.7\pm1.3$ & $9.4_{-1.4}^{+1.5}$ & $3.474_{-0.066}^{+0.060}$ & $0.04565\pm0.00089$ & $1187\pm16$ & $331\pm19$ \\ 
 Kepler-20c & $4971.60886(16)$ & $10.8540774(21)$ & $2.894_{-0.033}^{+0.036}$ & $89.815_{-0.63}^{+0.036}$ & 5, 6  &$<0.076$ & $3.38\pm0.61$ & $11.1\pm2.1$ & $2.51_{-0.47}^{+0.48}$ & $3.113_{-0.090}^{+0.075}$ & $0.0936\pm0.0018$ & $828\pm11$ & $78.7\pm4.6$ \\ 
Kepler-20d & $4997.7296(11)$ & $77.611455(96)$ & $2.606_{-0.039}^{+0.053}$ & $89.708_{-0.053}^{+0.17}$ & 5, 6  &$<0.082$ & $2.12\pm0.57$ & $13.4_{-3.6}^{+3.7}$ & $4.1_{-1.1}^{+1.2}$ & $3.28_{-0.14}^{+0.11}$ & $0.3474\pm0.0067$ & $430\pm6$ & $5.71\pm0.33$ \\ 
Kepler-20e & $4968.93956(34)$ & $6.0984882(99)$ & $0.821\pm0.022$ & $87.63_{-0.13}^{+1.1}$ & 5, 6 &$<0.092$ & $<0.28$ & $<0.76$ & $<7.5$ & $<3.0$ & $0.0637\pm0.0012$ & $1004\pm14$ & $169.8\pm9.8$ \\ 
Kepler-20f & $4968.19883(57)$ & $19.578328(48)$ & $0.952_{-0.087}^{+0.047}$ & $88.788_{-0.072}^{+0.43}$ & 5, 6 &$<0.094$ & $<0.35$ & $<1.4$ & $<8.4$ & $<3.2$ & $0.1387\pm0.0027$ & $681\pm9$ & $35.9\pm2.1$ \\ 
Kepler-21b & $5093.83716(85)$ & $2.7858212(32)$ & $1.639_{-0.015}^{+0.019}$ & $83.20_{-0.26}^{+0.28}$ &  \reftr{2016AJ....152..204L}  &0(fixed) & $2.70\pm0.46$ & $7.5\pm1.3$ & $9.3\pm1.6$ & $3.435_{-0.081}^{+0.069}$ & $0.04340_{-0.00032}^{+0.00021}$ & $2015\pm16$ & $2749\pm90$ \\ 
Kepler-22b & $4966.7001(68)$ & $289.863876(13)$ & $2.10\pm0.12$ & $89.764_{-0.042}^{+0.025}$ &  \reftr{2012ApJ...745..120B}, \reftr{2016ApJS..225....9H}, t.w.$^1$  &$<0.72$ & $<1.6$ & $<9.1$ & $<5.2$ & $<3.3$ & $0.812_{-0.013}^{+0.011}$ & $279\pm4$ & $1.013\pm0.060$ \\ 
Kepler-37b & $5017.0473(37)$ & $13.367020(60)$ & $0.3098_{-0.0076}^{+0.0059}$ & $88.63_{-0.53}^{+0.30}$ & 6, \reftr{2013Natur.494..452B},  t.w.$^1$ &$<0.098$ & $<0.25$ & $<0.79$ & $<140$ & $<3.9$ & $0.1019\pm0.0014$ & $718\pm10$ & $44.5\pm2.8$ \\ 
Kepler-37c & $5024.83997(87)$ & $21.301848(18)$ & $0.755_{-0.055}^{+0.033}$ & $89.07_{-0.33}^{+0.19}$ & 6, 10, t.w.$^1$   &$<0.099$ & $<0.34$ & $<1.3$ & $<15$ & $<3.3$ & $0.1390\pm0.0020$ & $615\pm9$ & $23.9\pm1.5$ \\ 
Kepler-37d & $5008.24982(13)$ & $39.7922622(65)$ & $2.030_{-0.039}^{+0.030}$ & $89.335_{-0.047}^{+0.043}$ & 6, 10, t.w.$^1$ &$<0.10$ & $<0.44$ & $<2.0$ & $<1.3$ & $<2.7$ & $0.2109\pm0.0030$ & $499\pm7$ & $10.39\pm0.65$ \\ 
Kepler-68b & $5006.858780(76)$ & $5.39875259(52)$ & $2.357\pm0.023$ & $87.23_{-0.17}^{+0.22}$ &  6,  \reftr{Marginietalinprep} &$<0.090$ & $2.83\pm0.23$ & $8.03\pm0.67$ & $3.37_{-0.29}^{+0.30}$ & $3.151_{-0.038}^{+0.036}$ & $0.06135\pm0.00043$ & $1275\pm17$ & $441\pm25$ \\ 
Kepler-68c & $4969.3821(11)$ & $9.605027(13)$ & $0.979\pm0.019$ & $87.071_{-0.094}^{+0.087}$ & 6, 11 &$<0.099$ & $<0.37$ & $<1.3$ & $<7.5$ & $<3.1$ & $0.09008\pm0.00063$ & $1052\pm14$ & $204\pm12$ \\ 
Kepler-78b & $4953.95984(15)$ & $0.355007450(80)$ & $1.201\pm0.028$ & $75.2_{-2.1}^{+2.6}$ & \reftr{2013Natur.503..381H}, t.w.$^1$ &0(fixed) & $1.75\pm0.28$ & $1.68\pm0.27$ & $5.33_{-0.93}^{+0.97}$ & $3.058_{-0.080}^{+0.069}$ & $0.00901_{-0.00019}^{+0.00012}$ & $2223\pm32$ & $4070\pm240$ \\ 
Kepler-93b & $4944.29227(13)$ & $4.72673978(97)$ & $1.478\pm0.019$ & $89.183\pm0.044$ & \reftr{2014ApJ...790...12B}   &0(fixed) & $1.89\pm0.21$ & $4.66\pm0.53$ & $7.93_{-0.94}^{+0.96}$ & $3.320_{-0.053}^{+0.048}$ & $0.05343\pm0.00065$ & $1133\pm17$ & $275\pm18$ \\ 
Kepler-102b & $4968.8696(11)$ & $5.286965(12)$ & $0.460\pm0.026$ & $89.78\pm0.22$ & 6, \reftr{2014ApJS..210...20M}, t.w.$^1$  &$<0.100$ & $<0.47$ & $<1.1$ & $<62$ & $<3.7$ & $0.05521\pm0.00049$ & $857\pm20$ & $90.0\pm9.0$ \\ 
Kepler-102c & $4972.9746(24)$ & $7.071392(22)$ & $0.567\pm0.028$ & $89.82\pm0.15$ &  6, 14, t.w.$^1$  &$<0.094$ & $<0.66$ & $<1.7$ & $<52$ & $<3.7$ & $0.06702\pm0.00059$ & $777\pm18$ & $61.0\pm6.1$ \\ 
Kepler-102d & $4967.091280(00)$ & $10.3117670(41)$ & $1.154\pm0.058$ & $89.49\pm0.11$ & 6, 14, t.w.$^1$   &$<0.092$ & $1.02\pm0.44$ & $3.0\pm1.3$ & $10.7_{-4.6}^{+5.1}$ & $3.34_{-0.24}^{+0.16}$ & $0.08618\pm0.00076$ & $686\pm16$ & $36.9\pm3.7$ \\ 
Kepler-102e & $4967.75370(12)$ & $16.1456994(22)$ & $2.17\pm0.11$ & $89.488\pm0.051$ & 6, 14, t.w.$^1$   &$<0.089$ & $1.37\pm0.53$ & $4.7\pm1.8$ & $2.50_{-0.98}^{+1.1}$ & $2.99_{-0.21}^{+0.15}$ & $0.1162\pm0.0010$ & $590\pm14$ & $20.3\pm2.0$ \\ 
Kepler-102f & $4978.0276(16)$ & $27.453592(60)$ & $0.861\pm0.022$ & $89.320\pm0.037$ & 6, 14, t.w.$^1$   &$<0.10$ & $<1.1$ & $<4.3$ & $<37$ & $<3.8$ & $0.1656\pm0.0015$ & $495\pm11$ & $10.0\pm1.0$ \\ 
Kepler-103b & $5677.65243(28)$ & $15.9653287(92)$ & $3.486_{-0.054}^{+0.057}$ & $87.914_{-0.072}^{+0.073}$ &  \reftr{2019MNRAS.490.5103D}  &$<0.093$ & $2.21\pm0.88$ & $9.8\pm3.9$ & $1.28_{-0.50}^{+0.52}$ & $2.90_{-0.22}^{+0.15}$ & $0.13223_{-0.0013}^{+0.00085}$ & $973\pm13$ & $149.9\pm8.6$ \\ 
Kepler-103c & $5667.15973(44)$ & $179.60978(20)$ & $5.45\pm0.18$ & $87.704_{-0.055}^{+0.12}$ & 15  &$<0.095$ & $3.8\pm1.7$ & $38_{-16}^{+17}$ & $1.29_{-0.55}^{+0.58}$ & $3.10_{-0.24}^{+0.16}$ & $0.6639_{-0.0063}^{+0.0043}$ & $434.0\pm6.0$ & $5.95\pm0.34$ \\ 
Kepler-107b & $5701.08414(37)$ & $3.1800218(29)$ & $1.536\pm0.025$ & $89.05\pm0.67$ &  \reftr{2019NatAs...3..416B}  &$<0.10$ & $1.44\pm0.68$ & $3.8_{-1.7}^{+1.8}$ & $5.8_{-2.6}^{+2.7}$ & $3.20_{-0.26}^{+0.17}$ & $0.04544\pm0.00036$ & $1592\pm19$ & $1073\pm53$ \\ 
Kepler-107c & $5697.01829(79)$ & $4.9014520(00)$ & $1.597\pm0.026$ & $89.49_{-0.44}^{+0.34}$ & 16   &$<0.080$ & $3.29\pm0.66$ & $10.0\pm2.0$ & $13.5_{-2.8}^{+2.9}$ & $3.586_{-0.099}^{+0.081}$ & $0.06064\pm0.00048$ & $1378\pm16$ & $602\pm30$ \\ 
Kepler-107d & $5702.9547(60)$ & $7.95839(12)$ & $0.860\pm0.060$ & $87.55_{-0.48}^{+0.64}$ & 16   &$<0.11$ & $<2.1$ & $<7.7$ & $<67$ & $<4.0$ & $0.08377\pm0.00065$ & $1173\pm14$ & $315\pm16$ \\ 
Kepler-107e & $5694.48550(46)$ & $14.749143(19)$ & $2.903\pm0.035$ & $89.67\pm0.22$ &  16  &$<0.10$ & $3.22\pm0.74$ & $14.1\pm3.3$ & $3.18_{-0.74}^{+0.75}$ & $3.216_{-0.11}^{+0.091}$ & $0.12638\pm0.00099$ & $955\pm11$ & $138.7\pm6.8$ \\ 
Kepler-109b & $4955.97791(60)$ & $6.4816307(48)$ & $2.49\pm0.11$ & $87.06\pm0.11$ & 6, 14, t.w.$^1$   &$<0.11$ & $<1.6$ & $<4.9$ & $<1.7$ & $<2.9$ & $0.0701\pm0.0019$ & $1276\pm23$ & $442\pm34$ \\ 
Kepler-109c & $4970.5722(10)$ & $21.222650(27)$ & $2.65\pm0.12$ & $89.63\pm0.19$ &  6, 14, t.w.$^1$  &$<0.098$ & $<1.8$ & $<8.2$ & $<2.4$ & $<3.1$ & $0.1546\pm0.0042$ & $859\pm16$ & $91.0\pm7.1$ \\ 
Kepler-323b & $4953.9478(14)$ & $1.6783280(15)$ & $1.381\pm0.035$ & $88.3_{-3.9}^{+1.5}$ & 9,  \reftr{2015ApJS..217...16R}, t.w.$^1$ &$<0.095$ & $<2.1$ & $<3.9$ & $<8.2$ & $<3.3$ & $0.02778\pm0.00068$ & $1838\pm35$ & $1900\pm150$ \\ 
Kepler-323c & $4956.9823(20)$ & $3.5538229(22)$ & $1.570\pm0.040$ & $88.7\pm1.3$ &  9, 17, t.w.$^1$  &$<0.095$ & $2.8\pm1.4$ & $6.8_{-3.2}^{+3.4}$ & $9.6_{-4.6}^{+5.0}$ & $3.43_{-0.28}^{+0.18}$ & $0.0458\pm0.0011$ & $1431\pm27$ & $700\pm56$ \\ 
Kepler-409b & $5012.0829(34)$ & $68.9583216(39)$ & $1.199\pm0.043$ & $86.30\pm0.13$ & 9, 14, t.w.$^1$  &$<0.69$ & $<1.6$ & $<6.0$ & $<19$ & $<3.6$ & $0.3192\pm0.0060$ & $438\pm7$ & $6.15\pm0.42$ \\ 
Kepler-454b & $5008.06758(77)$ & $10.5737534(78)$ & $2.37\pm0.13$ & $87.90\pm0.20$ &  \reftr{2016ApJ...816...95G}  &$<0.32$ & $1.64\pm0.41$ & $5.4\pm1.4$ & $2.23_{-0.64}^{+0.75}$ & $2.98_{-0.14}^{+0.11}$ & $0.09528_{-0.00091}^{+0.0013}$ & $916\pm10$ & $117.7\pm5.6$ \\ 
Kepler-538b & $5044.6789(11)$ & $81.73778(13)$ & $2.215_{-0.034}^{+0.040}$ & $89.730_{-0.060}^{+0.14}$ &  \reftr{2019AJ....158..165M}  &$<0.21$ & $2.09\pm0.47$ & $12.9\pm2.9$ & $6.5\pm1.5$ & $3.411_{-0.11}^{+0.089}$ & $0.3554_{-0.0047}^{+0.0068}$ & $417\pm5$ & $5.07\pm0.29$ \\ 
Kepler-1655b & $5013.89795(69)$ & $11.8728787(85)$ & $2.213\pm0.082$ & $87.62\pm0.55$ &   \reftr{2018AJ....155..203H}  &$<0.19$ & $1.51\pm0.48$ & $5.4\pm1.7$ & $2.72_{-0.90}^{+0.97}$ & $3.03_{-0.17}^{+0.13}$ & $0.1029\pm0.0014$ & $938\pm15$ & $129.1\pm8.8$ \\ 
Kepler-1876b & $4971.0510(50)$ & $6.992050(38)$ & $0.853_{-0.026}^{+0.039}$ & $89.13_{-2.0}^{+0.75}$ &  \reftr{2016ApJS..224...12C}, t.w.$^1$  &$<0.098$ & $<0.71$ & $<2.4$ & $<19$ & $<3.5$ & $0.0758\pm0.0019$ & $1299\pm22$ & $475\pm34$ \\ 
K2-2b\,/\,HIP\,116454b & $6907.887(29)$ & $9.0949(26)$~$^2$ & $2.53\pm0.18$ & $88.43\pm0.40$ &  \reftr{2015ApJ...800...59V}  &$<0.089$ & $3.73\pm0.42$ & $10.1_{-1.1}^{+1.2}$ & $3.41_{-0.72}^{+0.94}$ & $3.187_{-0.080}^{+0.079}$ & $0.07765\pm0.00093$ & $745\pm15$ & $51.4\pm4.3$ \\ 
K2-3b & $6813.41843(39)$ & $10.0546260(00)$ & $2.25\pm0.23$ & $89.588_{-0.100}^{+0.12}$ &  \reftr{2019AJ....157...97K}, t.w.$^1$  &$<0.094$ & $2.66\pm0.37$ & $6.47_{-0.99}^{+1.0}$ & $3.12_{-0.90}^{+1.3}$ & $3.10\pm0.11$ & $0.0778\pm0.0026$ & $513\pm29$ & $11.6_{-2.4}^{+2.8}$ \\ 
K2-3c & $6812.28013(95)$ & $24.646582(39)$ & $1.69\pm0.17$ & $89.905_{-0.088}^{+0.066}$ & 23, t.w.$^1$   &$<0.095$ & $1.01\pm0.35$ & $3.3\pm1.2$ & $3.7_{-1.5}^{+2.1}$ & $3.05_{-0.20}^{+0.16}$ & $0.1414\pm0.0047$ & $381\pm21$ & $3.51_{-0.73}^{+0.85}$ \\ 
K2-3d & $6826.22347(53)$ & $44.556456(97)$ & $1.62\pm0.18$ & $89.788_{-0.029}^{+0.033}$ & 23, t.w.$^1$   &$<0.097$ & $<0.39$ & $<1.6$ & $<2.1$ & $<2.8$ & $0.2097\pm0.0070$ & $312\pm17$ & $1.60_{-0.33}^{+0.38}$ \\ 
K2-12b & $6815.3701(29)$ & $8.28246(52)$ & $2.44_{-0.14}^{+0.30}$ & $87.6_{-5.4}^{+1.8}$ & \reftr{2018AJ....155..136M}, t.w.$^1$   &$<0.50$ & $<2.1$ & $<5.2$ & $<1.6$ & $<2.9$ & $0.0792_{-0.0013}^{+0.0016}$ & $1026\pm15$ & $185\pm12$ \\ 
K2-36b & $6810.8916(13)$ & $1.422614(38)$ & $1.430\pm0.080$ & $84.45_{-0.48}^{+0.78}$ & \reftr{2019A&A...624A..38D}   &$<0.093$ & $2.85\pm0.92$ & $4.3\pm1.4$ & $8.0_{-2.7}^{+3.1}$ & $3.31_{-0.17}^{+0.13}$ & $0.02288\pm0.00010$ & $1328\pm12$ & $519\pm20$ \\ 
K2-36c & $6812.84001(71)$ & $5.340888(86)$ & $3.20\pm0.30$ & $86.917_{-0.056}^{+0.066}$ &  25  &$<0.089$ & $3.4\pm1.2$ & $7.9\pm2.8$ & $1.30_{-0.51}^{+0.71}$ & $2.88_{-0.20}^{+0.16}$ & $0.05528\pm0.00023$ & $854\pm8$ & $89.0\pm3.4$ \\ 
K2-38b & $6896.8786(54)$ & $4.01593(50)$ & $1.655\pm0.096$ & $88.36_{-0.15}^{+0.17}$ &  \reftr{2016ApJ...827...78S}, \reftr{2020A&A...641A..92T}, t.w.$^1$  &$<0.11$ & $3.02\pm0.43$ & $7.7_{-1.1}^{+1.2}$ & $9.3_{-1.9}^{+2.4}$ & $3.440_{-0.085}^{+0.080}$ & $0.0503\pm0.0011$ & $1310\pm22$ & $490\pm35$ \\ 
K2-38c & $6900.4752(33)$ & $10.56103(90)$ & $2.49\pm0.22$ & $87.68_{-0.28}^{+0.31}$ & 26, 27, t.w.$^1$   &$<0.086$ & $2.11\pm0.37$ & $7.4_{-1.3}^{+1.4}$ & $2.63_{-0.72}^{+1.0}$ & $3.07\pm0.11$ & $0.0959\pm0.0022$ & $949\pm16$ & $135.3\pm9.5$ \\ 
K2-79b & $7103.22750(84)$ & $10.99470(47)$ & $4.09_{-0.12}^{+0.17}$ & $88.44\pm0.44$ & \reftr{2022AJ....163...41N} &$<0.23$ & $2.63\pm0.69$ & $9.2\pm2.4$ & $0.73_{-0.20}^{+0.21}$ & $2.73_{-0.14}^{+0.11}$ & $0.0988_{-0.0023}^{+0.0017}$ & $1022\pm20$ & $181\pm15$ \\ 
K2-96b\,/\,HD\,3167b & $7394.37454(43)$ & $0.959641(11)$ & $1.670_{-0.100}^{+0.17}$ & $83.4_{-7.7}^{+4.6}$ &  \reftr{2017AJ....154..122C}, t.w.$^1$  &0(fixed) & $3.56\pm0.15$ & $4.97_{-0.23}^{+0.24}$ & $5.6_{-1.3}^{+1.2}$ & $3.227_{-0.072}^{+0.057}$ & $0.01796_{-0.00031}^{+0.00037}$ & $1774\pm29$ & $1650\pm110$ \\ 
K2-96c\,/\,HD\,3167c & $7394.9788(12)$ & $29.8454(12)$ & $3.00_{-0.21}^{+0.45}$ & $89.30_{-1.0}^{+0.50}$ & 29, t.w.$^1$ &$<0.060$ & $2.58\pm0.15$ & $11.13_{-0.74}^{+0.78}$ & $2.11_{-0.69}^{+0.51}$ & $3.060_{-0.11}^{+0.064}$ & $0.1776_{-0.0031}^{+0.0037}$ & $564\pm9$ & $16.9\pm1.1$ \\ 
K2-106b & $7394.01140(00)$ & $0.571292(12)$ & $1.725\pm0.039$ & $86.4_{-4.1}^{+2.5}$ & \reftr{2017A&A...608A..93G}, t.w.$^1$  &0(fixed) & $6.50\pm0.52$ & $8.21_{-0.74}^{+0.76}$ & $8.77_{-0.94}^{+1.0}$ & $3.432_{-0.045}^{+0.044}$ & $0.01326_{-0.00023}^{+0.00028}$ & $2300\pm36$ & $4670\pm300$ \\ 
K2-106c & $7405.7316(44)$ & $13.33970(96)$ & $2.836_{-0.073}^{+0.086}$ & $89.35_{-0.46}^{+0.43}$ & 30, t.w.$^1$  &$<0.13$ & $2.48\pm0.66$ & $8.9\pm2.4$ & $2.12_{-0.57}^{+0.58}$ & $3.03_{-0.13}^{+0.10}$ & $0.1083_{-0.0018}^{+0.0023}$ & $805\pm12$ & $70.0\pm4.5$ \\ 
K2-110b & $7275.32992(61)$ & $13.86375(26)$ & $2.592_{-0.097}^{+0.098}$ & $89.35_{-0.24}^{+0.41}$ &  \reftr{2017A&A...604A..19OG}  &$<0.13$ & $5.25\pm0.89$ & $15.9\pm2.7$ & $5.02_{-0.98}^{+1.1}$ & $3.366_{-0.087}^{+0.077}$ & $0.10207\pm0.00083$ & $638\pm11$ & $27.7\pm2.0$ \\ 
K2-111b & $7100.0768(19)$ & $5.35180(40)$ & $1.820_{-0.090}^{+0.11}$ & $86.43_{-0.21}^{+0.37}$ &  \reftr{2020MNRAS.499.5004M}  &$<0.100$ & $2.30\pm0.30$ & $5.58_{-0.73}^{+0.74}$ & $5.0_{-1.0}^{+1.1}$ & $3.212_{-0.079}^{+0.073}$ & $0.05650\pm0.00045$ & $1309\pm19$ & $490\pm30$ \\ 
K2-131b & $7582.9360(11)$ & $0.3693038(91)$ & $1.690_{-0.058}^{+0.085}$ & $85.0_{-10.}^{+9.0}$ & \reftr{2017AJ....154..226D}, t.w.$^1$   &0(fixed) & $8.0\pm1.3$ & $7.9\pm1.3$ & $8.8_{-1.7}^{+1.9}$ & $3.426_{-0.084}^{+0.075}$ & $0.00936\pm0.00014$ & $2223\pm37$ & $4070\pm280$ \\ 
K2-135b\,/\,GJ\,9827b & $7738.82586(26)$ & $1.2089819(71)$ & $1.577_{-0.031}^{+0.027}$ & $86.07_{-0.34}^{+0.41}$ &   \reftr{2019MNRAS.484.3731R}  &$<0.063$ & $4.31\pm0.39$ & $5.14\pm0.47$ & $7.19_{-0.76}^{+0.81}$ & $3.306_{-0.045}^{+0.042}$ & $0.01880\pm0.00016$ & $1175\pm14$ & $318\pm16$ \\ 
K2-135c\,/\,GJ\,9827c & $7742.19930(73)$ & $3.648096(63)$ & $1.241_{-0.026}^{+0.024}$ & $88.19_{-0.18}^{+0.21}$ & 34  &$<0.094$ & $<0.75$ & $<1.3$ & $<3.7$ & $<2.9$ & $0.03925\pm0.00033$ & $813\pm10$ & $73.0\pm3.7$ \\ 
K2-135d\,/\,GJ\,9827d & $7740.96115(45)$ & $6.201470(63)$ & $2.022_{-0.043}^{+0.046}$ & $87.443\pm0.045$ & 34   &$<0.13$ & $1.73\pm0.43$ & $3.53_{-0.88}^{+0.87}$ & $2.34_{-0.59}^{+0.62}$ & $2.926_{-0.13}^{+0.099}$ & $0.05590\pm0.00046$ & $681\pm8$ & $36.0\pm1.9$ \\ 
K2-141b & $7744.07160(22)$ & $0.2803244(15)$ & $1.510\pm0.050$ & $86.3_{-3.6}^{+2.7}$ &  \reftr{2018AJ....155..107M}  &0(fixed) & $6.10\pm0.39$ & $4.97_{-0.34}^{+0.35}$ & $7.93_{-0.91}^{+1.0}$ & $3.330_{-0.042}^{+0.041}$ & $0.00747\pm0.00010$ & $2103\pm56$ & $3260_{-330}^{+370}$ \\ 
K2-141c & $7751.15460(00)$ & $7.74850(22)$ & $7.0_{-2.8}^{+4.6}$ & $87.2_{-2.0}^{+1.6}$ &  35  &$<0.092$ & $<3.2$ & $<8.0$ & $<0.081$ & $<1.9$ & $0.06830\pm0.00091$ & $695\pm18$ & $39.1\pm4.3$ \\ 
K2-167b & $6979.9368(25)$ & $9.97748(00)$ & $2.30_{-0.14}^{+0.21}$ & $88.6_{-2.0}^{+1.0}$ &  24, t.w.$^1$  &$<0.47$ & $1.97_{-0.55}^{+1.1}$ & $6.5_{-1.5}^{+1.6}$ & $2.87_{-0.36}^{+0.24}$ & $3.073_{-0.071}^{+0.050}$ & $0.0910\pm0.0025$ & $1174\pm24$ & $317\pm27$ \\ 
K2-222b & $7399.0595(16)$ & $15.38857(88)$ & $2.350_{-0.070}^{+0.080}$ & $89.12_{-0.41}^{+0.55}$ &  28  &$<0.16$ & $2.29\pm0.43$ & $8.7\pm1.7$ & $3.68_{-0.77}^{+0.86}$ & $3.188_{-0.097}^{+0.085}$ & $0.1206\pm0.0029$ & $871\pm23$ & $96.0_{-9.7}^{+11}$ \\ 
K2-262b\,/\,Wolf\,503b & $8191.36145(11)$ & $6.001270(21)$ & $2.043\pm0.069$ & $89.87\pm0.13$ & \reftr{2021AJ....162..238P}   &$0.409\pm0.085$ & $3.11\pm0.39$ & $6.27_{-0.84}^{+0.85}$ & $4.03_{-0.64}^{+0.72}$ & $3.168_{-0.069}^{+0.063}$ & $0.05712_{-0.00045}^{+0.00063}$ & $789\pm16$ & $64.7\pm5.5$ \\ 
K2-263b & $8111.1274(12)$ & $50.818947(94)$ & $2.41\pm0.12$ & $89.240_{-0.070}^{+0.050}$ & \reftr{2018MNRAS.481.1839M}   &$<0.15$ & $2.83\pm0.39$ & $14.9\pm2.1$ & $5.8_{-1.1}^{+1.3}$ & $3.399_{-0.078}^{+0.073}$ & $0.2573\pm0.0030$ & $470\pm7$ & $8.17\pm0.52$ \\ 
K2-312b\,/\,HD\,80653b & $8134.42440(70)$ & $0.719573(21)$ & $1.613\pm0.071$ & $82.1\pm2.4$ &  \reftr{2020A&A...633A.133F}  &0(fixed) & $3.62\pm0.21$ & $5.72_{-0.35}^{+0.36}$ & $7.47_{-1.00}^{+1.2}$ & $3.333_{-0.046}^{+0.048}$ & $0.01661\pm0.00019$ & $2463\pm30$ & $6130\pm310$ \\ 
K2-418b\,/\,EPIC-22...35b & $7920.44584(80)$ & $16.141132(19)$ & $2.332_{-0.094}^{+0.080}$ & $88.08_{-0.24}^{+0.26}$ & \reftr{Tronsgaardetalinprep} & $<0.23$ & $2.76\pm0.39$ & $10.4_{-1.5}^{+1.6}$ & $4.49_{-0.81}^{+0.92}$ & $3.272_{-0.077}^{+0.070}$ & $0.1237\pm0.0017$ & $804\pm10$ & $69.7\pm3.7$ \\ 
\label{table_planet_parameters}
\end{longtable}
\tablefoot{From left to right, the columns report the planet name, the transit mid-time, the orbital period, the planetary radius, the orbital inclination, 
the references for the transit parameters, the orbital eccentricity, the radial-velocity semi-amplitude, 
the planet mass, density and surface gravity, the semi-major axis, the equilibrium temperature by considering 
a null Bond albedo and full heat redistribution from the day to the night side, and the stellar incident flux. Table also available at the CDS.}
\begin{flushleft}
\footnotemark[1]{This work: the planet radius was newly determined from the $R_{\rm p}/R_{\rm s}$ transit parameter in the literature and 
the stellar radius $R_{\rm s}$ as reported in Table~\ref{table_system_parameters}.} 
\footnotemark[2]{The orbital period of K2-2b comes from the RVs by imposing a Gaussian prior on $T_{\rm c}$ only, and slightly differs from the value reported in \citet{2015ApJ...800...59V}, which is affected by systematics in the photometric data of the MOST satellite (A. Vanderburg, private communication).} 
\end{flushleft}
\tablebib{
% Kepler-10
(\reftrbib{2014ApJ...789..154D})~\citealt{2014ApJ...789..154D};
(\reftrbib{Bonomoetalinprep})~\citealt{Bonomoetalinprep};
% Kepler-19
(\reftrbib{2011ApJ...743..200B})~\citealt{2011ApJ...743..200B};
(\reftrbib{2017AJ....153..224M})~\citealt{2017AJ....153..224M};
% Kepler-20
(\reftrbib{2016AJ....152..160B})~\citealt{2016AJ....152..160B}; 
(\reftrbib{2019RAA....19...41G})~\citealt{2019RAA....19...41G}
% Kepler-21
(\reftrbib{2016AJ....152..204L})~\citealt{2016AJ....152..204L}; 
% Kepler-22
(\reftrbib{2012ApJ...745..120B})~\citealt{2012ApJ...745..120B};
(\reftrbib{2016ApJS..225....9H})~\citealt{2016ApJS..225....9H};
% Kepler-37
(\reftrbib{2013Natur.494..452B})~\citealt{2013Natur.494..452B}; 
% Kepler-68
(\reftrbib{Marginietalinprep})~\citealt{Marginietalinprep}; 
% Kepler-78
(\reftrbib{2013Natur.503..381H})~\citealt{2013Natur.503..381H};
% Kepler-93
(\reftrbib{2014ApJ...790...12B})~\citealt{2014ApJ...790...12B}; 
% Kepler-102
(\reftrbib{2014ApJS..210...20M})~\citealt{2014ApJS..210...20M};
%Kepler-103
(\reftrbib{2019MNRAS.490.5103D})~\citealt{2019MNRAS.490.5103D};
%Kepler-107
(\reftrbib{2019NatAs...3..416B})~\citealt{2019NatAs...3..416B};
%Kepler-323
(\reftrbib{2015ApJS..217...16R})~\citealt{2015ApJS..217...16R};
%Kepler-454
(\reftrbib{2016ApJ...816...95G})~\citealt{2016ApJ...816...95G};
%Kepler-538
(\reftrbib{2019AJ....158..165M})~\citealt{2019AJ....158..165M};
% Kepler-1655
(\reftrbib{2018AJ....155..203H})~\citealt{2018AJ....155..203H};
% Kepler-1876
(\reftrbib{2016ApJS..224...12C})~\citealt{2016ApJS..224...12C};
% K2-2
(\reftrbib{2015ApJ...800...59V})~\citealt{2015ApJ...800...59V};
% K2-3
(\reftrbib{2019AJ....157...97K})~\citealt{2019AJ....157...97K};
% K2-12
(\reftrbib{2018AJ....155..136M})~\citealt{2018AJ....155..136M};
% K2-36
(\reftrbib{2019A&A...624A..38D})~\citealt{2019A&A...624A..38D};
% K2-38
(\reftrbib{2016ApJ...827...78S})~\citealt{2016ApJ...827...78S}; 
(\reftrbib{2020A&A...641A..92T})~\citealt{2020A&A...641A..92T}; 
% K2-79
(\reftrbib{2022AJ....163...41N})~\citealt{2022AJ....163...41N};
% K2-96/HD3167
(\reftrbib{2017AJ....154..122C})~\citealt{2017AJ....154..122C}; 
% K2-106
(\reftrbib{2017A&A...608A..93G})~\citealt{2017A&A...608A..93G}; 
% K2-110
(\reftrbib{2017A&A...604A..19O})~\citealt{2017A&A...604A..19O};
% K2-111
(\reftrbib{2020MNRAS.499.5004M})~\citealt{2020MNRAS.499.5004M}; 
% K2-131
(\reftrbib{2017AJ....154..226D})~\citealt{2017AJ....154..226D};
% K2-135/GJ9827
(\reftrbib{2019MNRAS.484.3731R})~\citealt{2019MNRAS.484.3731R};
% K2-141
(\reftrbib{2018AJ....155..107M})~\citealt{2018AJ....155..107M}; 
% K2-167
% K2-222
% K2-262
(\reftrbib{2021AJ....162..238P})~\citealt{2021AJ....162..238P}; 
% K2-263
(\reftrbib{2018MNRAS.481.1839M})~\citealt{2018MNRAS.481.1839M};
% K2-312
(\reftrbib{2020A&A...633A.133F})~\citealt{2020A&A...633A.133F}; 
% EPIC-229004835
(\reftrbib{Tronsgaardetalinprep})~\citealt{Tronsgaardetalinprep};
}

\end{landscape}

\end{appendix}
\end{document}